\documentclass[12pt,preprint]{aastex}

\usepackage[bookmarks=true]{hyperref}
\usepackage{hyperref}
\hypersetup{
colorlinks=true,
pdfborder={0 0 0},
linkcolor=red,
citecolor=blue
 }
\usepackage{epsfig}
\usepackage{amsmath, amssymb}
\usepackage{morefloats}

\newcommand \beq {\begin{equation}}
\newcommand \enq {\end{equation}}
\newcommand \omgb {\Omega_{\rm bin}}
\newcommand \omgk {\Omega_{\rm K}}
\newcommand \rp {r_{\rm p}}
\newcommand \sigmap {\Sigma_{\rm p}}
\newcommand \md {M_{\rm d}}

\newcommand \lan {\langle}
\newcommand \ran {\rangle}
\newcommand \ebin {e_{\rm bin}}
\newcommand \cost {\lan\cos\theta\ran}
\makeatletter

\newcommand{\Rmnum}[1]{\expandafter\@slowromancap\romannumeral #1@}
\makeatother

\slugcomment{}

\shorttitle{3D MHD Simulation of Circumbinary Disk}
\shortauthors{Shi, Krolik, Lubow \& Hawley}

\begin{document}


\title{Three Dimensional MHD Simulation of Circumbinary Accretion Disks: Disk
Structures and Angular Momentum Transport }



\author{Ji-Ming Shi\altaffilmark{1,2} and Julian H. Krolik\altaffilmark{1}}
\and
\author{Stephen H. Lubow\altaffilmark{3}}
\and
\author{John F. Hawley\altaffilmark{4}}

\altaffiltext{1}{Department of Physics and Astronomy, Johns Hopkins
University, Baltimore, MD 21218}
\altaffiltext{2}{Center for Integrative Planetary Sciences, Astronomy Department, University of
California at Berkeley, Berkeley, CA94720; jmshi@astro.berkeley.edu}
\altaffiltext{3}{Space Telescope Science Institute, Baltimore, MD 21218 }
\altaffiltext{4}{Department of Astronomy, University of Virginia, Charlottesville, VA 22903 }

\begin{abstract}
We present the first three-dimensional magnetohydrodynamic (MHD) simulations of a circumbinary disk surrounding an equal mass binary.  The binary maintains a fixed circular orbit of separation $a$.  As in previous hydrodynamical simulations, strong torques by the binary can maintain a gap of radius $\simeq 2a$.  Streams curve inward from $r \simeq 2a$ toward the binary; some of their mass passes through the inner boundary, while the remainder swings back out to the disk.   However, we also find that near its inner edge the disk develops both a strong $m=1$ asymmetry and growing orbital eccentricity.  Because the MHD stresses introduce more matter into the gap, the total torque per unit disk mass is $\simeq 14$ times larger than found previously.  The inner boundary accretion rate per unit disk mass is $\simeq 40$ times greater than found from previous hydrodynamical calculations.  The implied binary shrinkage rate is determined by a balance between the rate at which the binary gains angular momentum by accretion and loses it by gravitational torque.  The large accretion rate brings these two rates nearly into balance, but in net, we find that $\dot a/a < 0$, and its magnitude is about $2.7$ times larger than predicted by the earlier hydrodynamic simulations.  If the binary comprises two massive black holes, the accretion rate may be great enough for one or both to be AGN, with consequences for the physical state of the gas both in the disk body and in its inner gap.

\end{abstract}



\section{INTRODUCTION \label{sec:intr}}
Various types of astronomical binary systems can be embedded in gaseous disks, from
young binary stars to stars with growing planets to binary black holes. Such disks have
been observed directly in nearby star-forming regions. One of the best resolved (around
a young binary) is in GG~Tau \citep[e.g.,][]{dgs94,krist05}. To date, however, only a few
possible disk-planet systems have been directly imaged \citep[e.g.,][]{greaves08,kalas08,hash11}.
Although there is no direct evidence for the existence of circumbinary disks involving binary
black hole systems, it is generally believed that such configurations should exist near the
centers of galaxies after a galaxy merger
\citep[e.g.,][]{bbr80,ivan99,mm05,escala04,escala05,mayer07,dotti09}.

Tidal forces exerted by the binary can sometimes clear a gap in the disk.
When the binary mass ratio $q\equiv M_2/M_1\ll 1$, where $M_1$ denotes the mass of the
star and $M_2$ the mass of the planet, the gap that is formed is an annular ring
around the primary through which the secondary travels.  Whether such a gap opens
depends on whether the secondary's mass is sufficiently large to overcome the
gap closing effects of internal stresses.  Independent of whether such a gap
exists, the secondary can exchange angular momentum with the gas via gravitational torques.
Inward orbital migration of the secondary may occur on the disk inflow timescale if the
disk mass is large compared to the secondary mass \citep{lp86,lp93}. Otherwise the migration
will be slower \citep{ivan99, an2002}.  There has been extensive theoretical study of this situation,
using both analytic and numerical methods
\citep[e.g.,][]{gt80,lp86,lp93,bryden99,ivan99,bate03,nelson03,winter03}. 

When $q$ is closer to unity, the gap can include the entire binary itself.  In this case,
the resulting configuration can contain as many as three disks: one around each member of the
binary and one that orbits outside the binary, called the circumbinary disk.  Observational evidence of
such large gap clearing and circumbinary disks has been found in several young stellar binaries
\citep{mathieu94}. Numerical simulations have also been applied to study $q\lesssim 1$ binary systems 
with a circumbinary disk \citep[e.g.,][]{al94,bb97,gk02,escala05,mm08,haya07,cuadra09,hoa10}.
These studies have found
that the radius of the disk inner edge depends on several factors, including the binary
separation, mass ratio, eccentricity, and the strength of the disk turbulence \citep{al94}.

In a one-dimensional model (one that considers only radial dependences), a circumbinary disk
would in principle behave as a ``decretion" disk' \citep{Pringle91}:
the binary loses angular momentum to the surrounding disk so that the disk is repelled. However, the
one-dimensional model neglects any non-axisymmetric properties of the disk.
For instance, there could be some directions where the binary torque becomes weak enough that
matter could leak through the disk inner edge.  The gas then penetrates the gap and may accrete
onto the binary.  Indeed, this penetration effect has been witnessed in numerous two-- or
three--dimensional simulations of disks with various binary mass ratios using either
smoothed particle hydrodynamics \citep[SPH;][]{al96,escala05,cuadra09,dotti09} or grid-based
methods \citep[][]{bryden99,gk02,mm08,deVal11}.  Such work has found that in the low-density gap
there are gas flows from the disk inner edge to the binary in form of narrow, high-velocity spiral
streams. The flow rate through the gap depends on the binary and the disk properties. Compared to
the accretion rate near the disk center that would be expected in the absence of binary torque,
the accretion rate appears to be reduced \citep[e.g.,][hereafter
MM08]{lubow99,ld06,mm08}. However, \citet{rl97} found no reduction at all.

Disk eccentricity is another potential non-axisymmetric property. A disk can become eccentric
as a result of its interaction with the binary.  Circumbinary disks can, of course, gain
eccentricity through direct driving by an eccentric binary \citep{arty91,pnm01, roe11}.
However, there is also evidence that disk eccentricity can arise even when the orbit of the binary
is circular \citep[e.g.,][MM08]{pnm01}. Simulations of protoplanetary disks have shown that
these disks are subject to a resonant mode coupling instability \citep{kd06,dlb06}. Through this
instability, the disk's eccentricity can grow although the planet is on a fixed circular orbit. This
instability follows the same tidal resonance mechanism found for 
eccentrically unstable circumstellar disks in superhump binaries \citep[][]{lubow91b, kpo08}.
For circumstellar disks, \citet{lubow94} found that orbiting secondaries can drive eccentricity
by stream impacts if they are strongly modulated in time.  Recently, MM08 found that the disk around an
equal mass binary on a circular orbit also became eccentric after a large number of binary orbits.
They suggested an eccentricity generation mechanism that involved
the action of the binary torque on the gas within the low density gap.  Conversely, disk
eccentricity might also excite eccentricity in the binary \citep[e.g.,][]{pnm01}. 
For coalescing massive black holes, the residue eccentricity in the emitted gravitational
waves might be detected by proposed instruments like the Laser Interferometer Space Antenna
(LISA), perhaps signaling gas-driven evolution \citep{an2005, kc11}. 

Almost all previous studies of circumbinary disks have adopted the $\alpha$-prescription to describe
internal stresses and treat either as evolving through diffusion (in one dimension) or as a result
of ``viscous" stresses (in two or three dimensions).  In these efforts, the internal stress to pressure
ratio $\alpha$ was taken to be constant everywhere and at all times. Although that might be a reasonable
approximation for vertically-integrated and time-averaged conditions in the main body of a disk, it
becomes unrealistic for highly time-dependent turbulent accretion flows and low density regions outside
the disk body\citep{hk01}.  Since the exchange of angular momentum between the binary and the
disk is crucial for both the circumbinary disk and the binary, we need more a realistic description
of the underlying internal stresses.  It is now generally recognized that whenever the material of
the disk is sufficiently ionized so as to be well-coupled to any embedded magnetic field, the principal
mechanism of angular momentum transport is MHD turbulence induced by the magnetorotational instability
(MRI).  It is therefore necessary to study circumbinary disks using MHD simulations in which internal
stress arises self-consistently from turbulence generated by the MRI. 
To date, this has been done only for extreme mass ratio star-planet systems, 
using either unstratified \citep[][]{nelson03, winter03, bfnm11} or stratified \citep[][]{ukfh11}
MHD simulations.

It is the goal of this project to construct the first three-dimensional (3D) MHD simulation of
a circumbinary disk around an equal mass binary. To simplify our model, we assume the disk and
binary to be coplanar. The binary orbits on a fixed circular orbit.  On the basis of this simulation,
we will try to answer the following questions: (1) What is the inner disk structure? Is the disk
truncated or not?  (2) How is angular momentum transported within the disk?  Can the internal stress
balance the binary torque and therefore allow the disk to achieve a quasi-steady state? (3) Is there
any eccentricity growth of the disk? If so, what is the cause? (4) How does the accretion rate onto
a binary compare with the rate onto single point mass?

We organize this paper as follows: In $\S$~\ref{sec:numerical}, we describe the
physical model and numerical procedures of our circumbinary disk simulations.
In $\S$~\ref{sec:result}, we present our simulation results.  We then discuss the binary
torque and binary contraction in $\S$~\ref{sec:discussion_torq} and \ref{sec:discussion_shrink}.
In $\S$~\ref{sec:discussion_ecc}, we discuss possible mechanisms for disk eccentricity growth.
We explain the formation of an asymmetric density concentration near the disk inner edge in 
$\S$~\ref{sec:discussion_lump}.
Finally, in $\S$~\ref{sec:conclusion} we summarize our conclusions.

\section{NUMERICAL SIMULATION\label{sec:numerical}}
In this section, we discuss in detail the numerical procedure of this work. The code used is a
modern version of the 3D, time-explicit Eulerian finite-differencing ZEUS code for
MHD \citep{stone92a,stone92b,hs95}. We modified the code to cope with the time-dependent
binary potential.
\subsection{Physical Model\label{sec:model}}

We construct our physical disk model in the inertial frame in which the center of mass of the binary
is at rest at the origin. Treating the simplest case first, we assume an equal mass circular
restricted binary system, i.e. the binary orbits circularly in the disk plane, and we neglect binary
evolution as the disk inflow time scale is usually much smaller than the shrinkage timescale of the
binary, and certainly more than the duration of the simulation (see estimate in
$\S$~\ref{sec:conclusion}). We set the gravitational constant $G$, total binary mass $M$ and the
binary separation $a$ to be unity, and therefore the binary frequency $\omgb=\sqrt{GM/a^3}$ is
unity as well. 
We also assume the circumbinary disk to be cold and thin. As we are mainly concerned with
the orbital dynamics of the flow, we choose a simple global isothermal sound speed $c_s=0.05$. 
The disk flares at larger radii because the ratio of height to
radius $H/R=c_s/R\omgk=0.05(R/a)^{1/2}$, where $H$ denotes the scale height of the disk, $R$ is
the disk radius in cylindrical coordinates, and $\omgk=\sqrt{GM/R^3}$ is the Keplerian frequency.
As the fluid is well coupled to the magnetic field even in a cold disk where the
ionization fraction is far below unity\citep[see, e.g., ][]{sghb2000}, we assume ideal MHD.
For this reason, we include no explicit diffusivity except the von Neumann-Richtmyer bulk
viscosity in compressive regions that ensures the right jump conditions for shock waves.
The effect on
damping the angular momentum is negligible compared to other transport mechanisms.
Because we are most concerned with the inner part of the disk, not accretion onto the binary,
we excise a central region.  This cut-out must be well
inside the inner edge of the disk so that it does not affect fully resolving the
inner disk and any gas leakage from the inner edge. We choose to cut out the area within $0.8a$.
This region is beyond the main extent of the interior disks that surround each binary member because
these disks are each tidally truncated at $\sim 0.3a$ from their central objects \citep{p77}. 
Once the disk reaches the quasi-steady stage, we find that the inner edge of the disk is
located at $r \simeq 2a$, which is about a factor of $2.5$ outside the cut-out.
We approximate the time-dependent binary potential in the disk region using Newtonian
dynamics because the disk region we are interested in is far from the
gravitational radii of the individual binary components. The disk mass is assumed to be much
smaller than the binary mass, so that the Toomre parameter satisfies $Q\approx
(H/R)(M/M_{\rm d})\gg 1$, where $M_{\rm d}$ denotes the disk mass.  We therefore neglect the
contribution of the disk self-gravity to the potential.

\subsection{Grid Scheme and Boundary Conditions\label{sec:grid}}
The properties of circumbinary disks require us to resolve three different
length scales when setting up the grid.
The first is the half disk thickness $H$. The computational domain has to contain at least several
($\lesssim 4$) scale heights on each side of the midplane, and for each $H$, several tens of cells
are needed. The second is the maximum growth-rate wavelength of the MRI,
$\lambda_{\rm MRI}\equiv 8\pi/\sqrt{15}v_A/\Omega(R)$, where $v_{A}$ is
the Alfven speed, and $\Omega(R)$ is the disk rotational frequency.  We require
$\lambda_{\rm MRI}$ to be resolved by at least six grid elements. The last one is the spiral density
wavelength, $\lambda_{\rm d}\sim 2\pi c_{s}/\omgb$.  There should be many cells across this
wavelength (MM08); we require at least six.

In order to satisfy the above requirements, we follow the scheme proposed in \citet{N10} to construct
our grid in spherical coordinates ($r$,$\theta$,$\phi$). We adopt a logarithmic grid in
the radial direction, which provides a constant $\Delta r/ r$. The vertical grid is derived by
mapping a simple linear function $y(x) = x$ for $x\in[0,1]$ through a polynomial transformation
(see equation~(6) in \citet{N10}):
\beq
\theta(y) =
\frac{\pi}{2}\left[1+(1-\xi)(2y-1)+\left(\xi-\frac{2\theta_c}{\pi}\right)(2y-1)^n\right],
\label{eq:theta}
\enq
where $\xi$, $\theta_c$ and $n$ are parameters that define the shape of the polynomial. Note that
$y=0.5$ is exactly mapped to the midplane.  The merit of this $\theta$-grid is that for $n>1$ it
ensures dense and nearly uniform cell elements close to the disk
midplane. The azimuthal grid is evenly spaced and covers all $2\pi$. Theoretically speaking, the
aspect ratios of the cell shape should be as isotropic as possible, but the azimuthal cell
widths can be a factor of a few longer than the other two, as the shearing tends to draw out
features in this direction.

The grid resolution used in the present simulation is $400\times160\times540$ in $(r,\theta,\phi)$,
with a computational domain covering $[r_{\rm in},r_{\rm out}]$ radially, $[\theta_c,\pi-\theta_c]$
meridionally and $[0,2\pi]$ azimuthally (see also in Table \ref{tab1}), where $r_{\rm in} =0.8a$,
$r_{\rm out}=16a$, and $\theta_c=0.2$ is the width of the cut-out around the polar axis. The other
parameters used in equation~(\ref{eq:theta}) are $\xi=0.9$ and $n=9$. Using this grid, we are able
to resolve one disk scale height with $20$ cells at the inner boundary of the computational domain.
The resolution grows to $\sim 40$ cells per scale height at radius $\sim 3a$.  Within two scale
heights of the midplane, $\lambda_{\rm MRI}$ is resolved by more than six cells if the plasma
$\beta$ is no greater than $\sim 100$. The grid also resolves $\lambda_{\rm d}$ with at least
six cells in the $r$--$\phi$ plane for $r\lesssim 7a$.

We choose outflow boundary conditions in the radial and meridional directions for the gas, which
permit only flows going outward; any inward velocities are set to zero. As we cover all
$2\pi$ of $\phi$, the boundary conditions for $\phi$--grid are simply periodic.
For the magnetic field in the radial and meridional directions, we set the transverse components of
the field to be zero in the ghost zones.  The components normal to the boundaries are
calculated by imposing the divergence-free constraint.

\subsection{Initial Setup\label{initial}}
We begin with a prograde disk orbiting in the binary plane. Between $r_{\rm min}=3a$ and
$r_{\rm max}=6a$, its density is constant in the
midplane. The initial disk is axisymmetric, and the polar angle density distribution
is $\rho=\rho_0 \exp{[-(\theta-\pi/2)^2/(\sqrt{2}H/r)^2]}$, in which
$\rho_0=1$ is the unit of disk density. This form provides initial hydrostatic balance
vertically for a point mass potential and zero radial
pressure gradient along the midplane. For a first order approximation, the difference between a binary
potential and a point-mass in the midplane can be described by the temporally and
azimuthally averaged quadrupole moment of the binary potential. We therefore modify the angular
frequency of the initial disk to account for the quadrupole contribution:
\beq
\Omega(r)^2 \approx \omgk^2\left[1+\frac{3}{4}\left(\frac{a}{r}\right)^2\frac{q}{(1+q)^2}\right],
\label{eq:omega}
\enq
where $q=1$ is the mass ratio of the binary. Here we replace $R$ with $r$ as
$r=R\sin{\theta}\approx R$ for regions near the midplane. We also add $1/(r\Sigma)dP/dr$ to the
right hand side of equation~(\ref{eq:omega}) to compensate for the small radial gradient of the
vertically integrated pressure $P=\Sigma c_s^2$.

The initial magnetic field is a single poloidal loop within the main body of the disk. It
is subthermal with plasma $\beta=100$ on average.
The field is constructed from the vector potential $\mathbf{A} = (0,0,A_{\phi})$ and we define
$A_{\phi}$ by
$A_{\phi}= A_0\sqrt{\rho}\sin(2\pi r/kH)(r/r_{\rm min}-1)(1-r/r_{\rm max})-\sqrt{\rho_{\rm cut}}$
if $A_{\phi}>0$ and zero otherwise, where $k = 2\omgb a/c_{s}$, $\rho_{\rm cut}=10^{-3}\rho_0$,
and $A_0$ is a constant determined by the constraint on the averaged $\beta$. 

\subsection{Diagnostics\label{diag}}
Three-dimensional MHD simulations usually produce a large amount of data, which poses
challenges for data storage, transport and post-simulation analysis. In order to facilitate
the study of the spatial and temporal properties of the disk, we write out spatially averaged history
data at a frequency of one dump per time unit and write out 3D snapshots every five units of time.

We use two different types of spatially averaged history data.  The first is defined by either
an integration or an average over radial shells.  Shell-averaging for variable $X$ is defined by
\beq
\lan X\ran (r,t) \equiv \frac{\int X r^2\sin\theta d\theta d\phi}{\int r^2\sin\theta d\theta d\phi},
\label{eq:shell_int}
\enq
where $ \int r^2\sin\theta d\theta d\phi \equiv A_{\rm x}(r)$ is the shell surface area.
With this definition, the density-weighted shell average is
$\lan X\ran_{\rho}\equiv \lan \rho X\ran /\lan \rho\ran$. For example, the net disk
accretion rate is
\beq
\dot M(r,t)\equiv \int \rho v_{\rm r} r^2 \sin\theta d\theta d\phi =A_{\rm x}\lan\rho
v_r\ran,
\label{eq:mdot}
\enq
and the average specific angular momentum $l=\lan v_{ \phi}r\ran_{\rho}=\lan\rho
v_{\phi} r\ran/\lan \rho\ran$. The surface density is vertically integrated and azimuthally
averaged quantity:
\beq
\Sigma(r,t) \equiv \frac{1}{2\pi}\int\rho r \sin\theta d\theta d\phi .
\enq

The second type of average is two-dimensional, either an azimuthal average of poloidal slices
or a vertical average referred to the equatorial plane.
We define the vertical average by
\beq
\lan Y\ran_z (r,\phi,t) \equiv \frac{\int Y r\sin\theta d\theta}{\int r\sin\theta d\theta},
\label{eq:vert_int}
\enq
and the density weighted vertical average by
\beq
\lan Y\ran_{z,\rho}(r,\phi,t)\equiv \frac{\int \rho Y r\sin\theta d\theta}{\int \rho r\sin\theta d\theta} =
\frac{\lan \rho Y\ran_z}{\lan\rho\ran_z}.
\label{eq:vert_int_den}
\enq
Similarly we have the azimuthal average is $\lan Y\ran_{\phi}(r,\theta,t)\equiv
\frac{1}{2\pi}\int Y d\phi$.

We need to be very careful about the definition of the $r$--$\phi$ component of the internal stress
in the present simulation. We follow \citet{hk01} and define the stress as
\beq
w_{r\phi} (r,\theta,\phi,t)= t_{r\phi} + r_{r\phi} = -\frac{B_rB_\phi}{4\pi} + \rho \delta v_r\delta v_{\phi},
\label{eq:stress_tot}
\enq
where $t_{r\phi}$ is the Maxwell stress and $r_{r\phi}$ is the Reynolds stress. The
perturbed velocities $\delta v_r$ and $\delta v_{\phi}$ are calculated from
\begin{eqnarray}
\delta v_r (r,\theta,\phi,t)     &=& v_r - \lan v_r\ran_{\rho} \nonumber \\
\delta v_{\phi}(r,\theta,\phi,t) &=& v_{\phi} - \lan v_{\phi}\ran_{\rho}.
\label{eq:delta_vel}
\end{eqnarray}
The vertically integrated and azimuthally averaged total stress can then be described by
\begin{eqnarray}
W_{r\phi}&\equiv& \frac{1}{2\pi}\int w_{r\phi}r\sin\theta d\theta d\phi = L_z\lan w_{r\phi}\ran = T_{r\phi} +
R_{r\phi} ,\nonumber \\
T_{r\phi}& =& L_z\left\lan\frac{-B_rB_{\phi}}{4\pi}\right\ran,\nonumber\\
R_{r\phi}& =&
L_z\left[\lan\rho v_r v_{\phi}\ran - \frac{\lan\rho v_r\ran\lan\rho
v_{\phi}\ran}{\lan\rho\ran}\right],
\label{eq:stress_tot_1d}
\end{eqnarray}
where $T_{r\phi}$ and $R_{r\phi}$ are the average Maxwell stress and Reynolds stress respectively,
and $L_z \equiv A_{\rm x}/(2\pi r)$ is the vertical integral length, which in our case equals the
height of the computational domain at a given radius $r$.
In a similar way, we can obtain the vertically averaged stress $\lan w_{r\phi}\ran_z$ and
azimuthally averaged stress $\lan w_{r\phi}\ran_{\phi}$ as well.

\subsection{Hydrodynamic Simulations\label{sec:hd_run}}
To study the mechanism for disk eccentricity growth, we also carried out a set of
numerical experiments using two-dimensional viscous hydrodynamic simulations. Their purpose was both
to distinguish hydrodynamic from MHD effects and to test the effect of where the inner boundary is
placed.

The hydrodynamic simulations in this paper used the $\alpha$-disk prescription. We chose $\alpha=0.1$
throughout the disk because that is roughly the ratio of stress to pressure in the disk body of our
MHD simulation (note MM08 used $\alpha=0.01$).
The hydrodynamic disks are evolved in polar $(r,\phi)$ coordinates in the
inertial frame. Following the grid scheme of the MHD simulation, we set the radial grid to be
logarithmically spaced, while the azimuthal grid is spaced uniformly. 

Among these simulations, B2D.rin=0.8 serves as the control run. The parameters which
describe the physical properties of the hydrodynamic disk and the binary are kept the same as in the
MHD case. The resolution of the control run is $512\times1024$ for $r\times\phi$.  Its grid covers
the same physical extent in radial and azimuthal directions as the MHD one. The initial surface
density of the two-dimensional disk is simply taken from a vertical integration of the initial
condition of the MHD disk. 
Other hydrodynamic simulations are reruns of B2D.rin=0.8 at $t=500$ with various locations of
the inner boundary. The initial disks of the reruns are obtained by truncating the restart data of
B2D.rin=0.8 to the desired radius while keeping $\Delta r/r$ and $\Delta\phi$ fixed. 
The properties of both the reruns and the control run can be found in Table~\ref{tab1}.

\section{RESULTS \label{sec:result}}
We present one 3D MHD simulation, called B3D, in this paper.\footnote{\small{We also performed a
short duration rerun for $t=300$--$322$ with higher dump rates: ten history dumps and
one 3D dump per time unit. They are used when high time resolution is required, e.g. when we try to
investigate the angular momentum budget and the stream dynamics.}}
This simulation was terminated after the inner disk ($r<3a$) reached a quasi-steady stage for several
hundred time units. Longer simulations require better resolution to resolve the MRI in the growing
density concentration region (see $\S$~\ref{sec:ecc_lump}).

B3D ran for $\simeq 480$ code units, which corresponds to $\simeq 77$ binary orbits.
%
The gas quickly fills in the initially empty region within several binary orbital periods, and
after $\sim 100$ units of time, the disk becomes fully turbulent. The binary torque then is able
to maintain a low density gap, and the inner part of the disk finally reaches a quasi-steady state
after $t~\sim 200$.   
We note that this single simulation
consumed $\sim 720$K CPU hours on the Kraken Cray XT5 system.

In the following subsections, we first describe the overall evolution of the circumbinary
disk and its quasi-steady state. We then try to discuss how the angular momentum is
transported in the inner disk in $\S$ \ref{sec:budget}. We will mainly discuss the characteristic
disk structures in $\S$ \ref{sec:ecc} and \ref{sec:stream}.   The field structure will be
considered in $\S$~\ref{sec:bfield} and the temporal properties of the accretion in the
last subsection.

\subsection{A Secularly Evolving Quasi-steady State \label{sec:evolution}}
\begin{figure}
\centering{
\includegraphics[width=7cm]{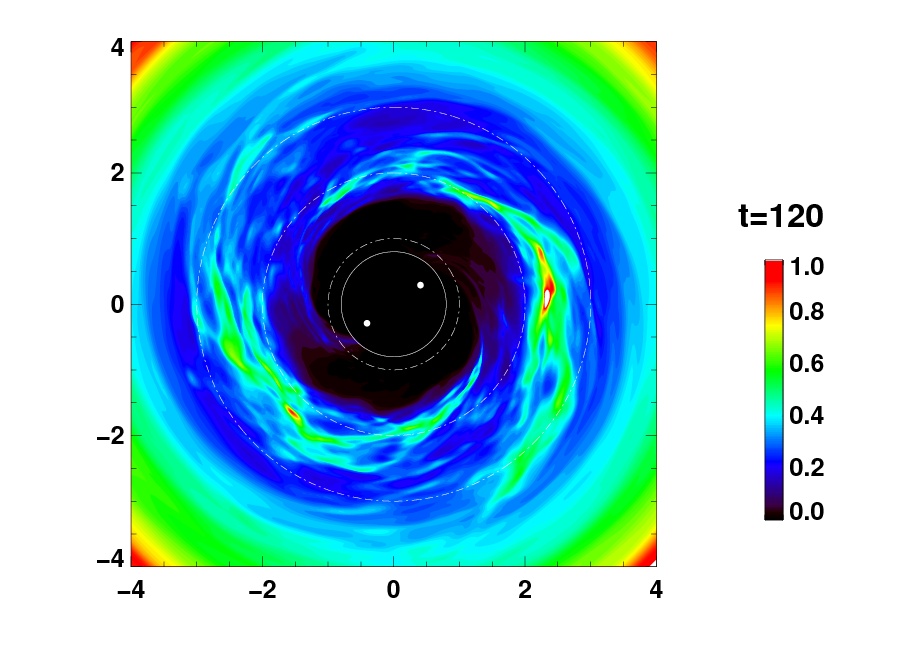}
\includegraphics[width=7cm]{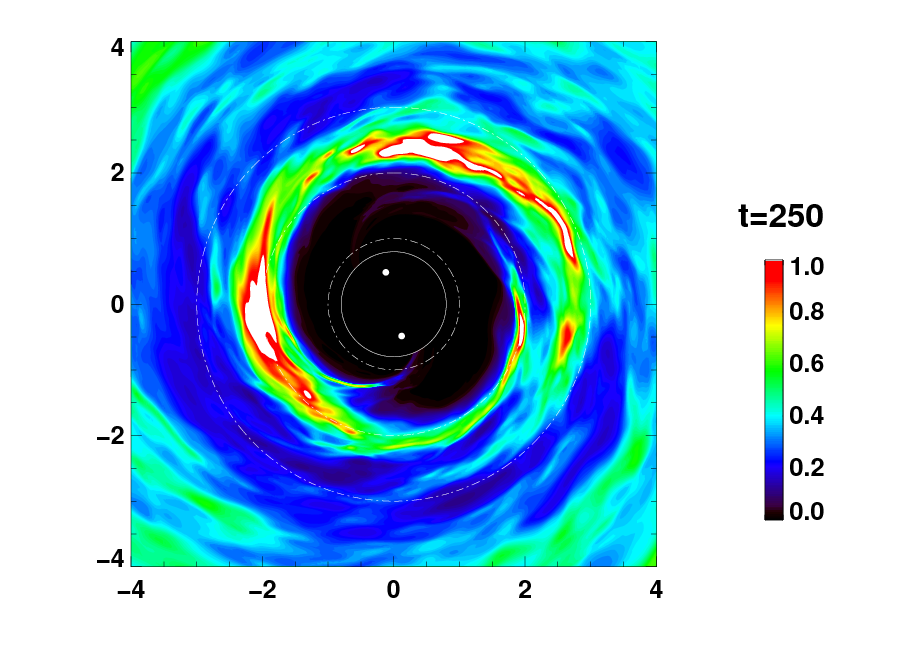}
}
\centering{
\includegraphics[width=7cm]{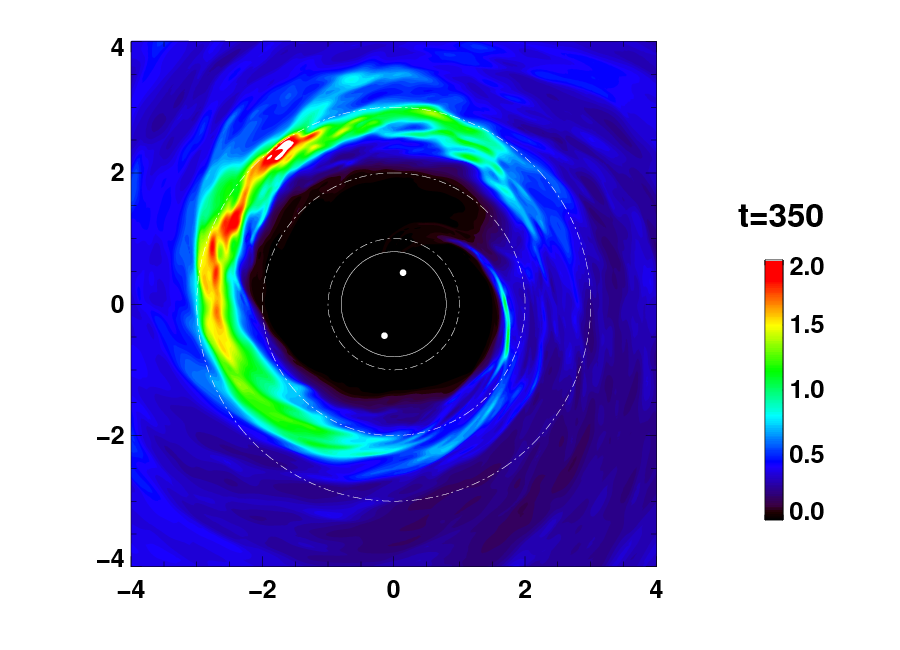}
\includegraphics[width=7cm]{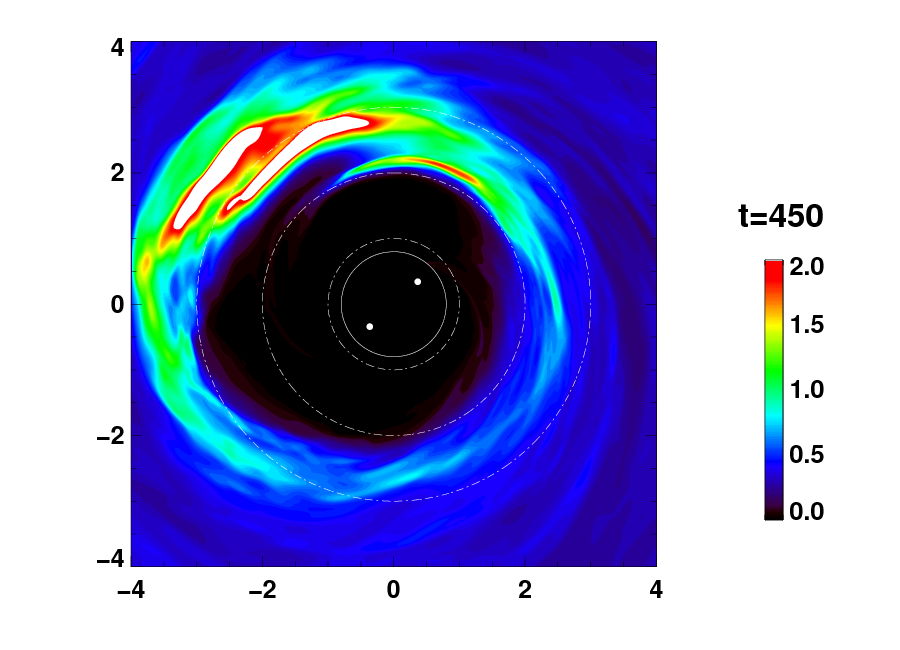}
}
\caption{\small{A series of snapshots of the disk surface density at different times.
Density contours are on a linear scale.  The color scale encoding the density (see
the color bar for each panel) has twice the range in the bottom two panels as in the top two;
the white regions are density peaks which are off the scale.
White dots show the position of
the binary; the faint white solid circle shows the boundary of the central cut-out; the
white dash-dotted circles represent the radii $r=1$, $2$ and $3a$.}}\label{fig:4snapshots}
\end{figure}

After the first $100$ time units, the binary torque starts to clear out a low
density gap between the inner boundary and $r\sim 2a$. In the top-left panel of
Figure \ref{fig:4snapshots}, we show the vertically integrated surface density of the
circumbinary disk at $t=120$ in the $x-y$ plane. In that panel, we can clearly see the disk
is truncated at around twice the binary separation.  We also find two streams emanating from
the disk edge toward the binary components. From the other three panels in
Figure~\ref{fig:4snapshots}, we find that as the simulation continues the gap
persists, the surface density gradually increases outside the gap, and finally an incomplete
ring of dense gas forms due to the combined effects of mass accretion and
gravitational torque.  The stream also persists, but only one arm at a time.  We will
study the properties and effects of the transient stream in \S~\ref{sec:stream} and
\ref{sec:discussion}.  At $t=250$ the disk appears to be elliptical and slightly off center,
and at $t=350$ and $450$ the disk obviously becomes eccentric. We also observe a growing
azimuthal density asymmetry.  We will discuss the eccentricity growth and density
asymmetry in \S~\ref{sec:ecc} and \ref{sec:discussion}.  Despite the slow growth
of both disk eccentricity and asymmetry, the principal disk structures, the gap and stream,
are sufficiently time-steady to suggest that in a qualitative sense the simulation has
achieved a statistically stationary state.  An analysis based on angle-averaged
quantities will further demonstrate that.

\begin{figure}
\plotone{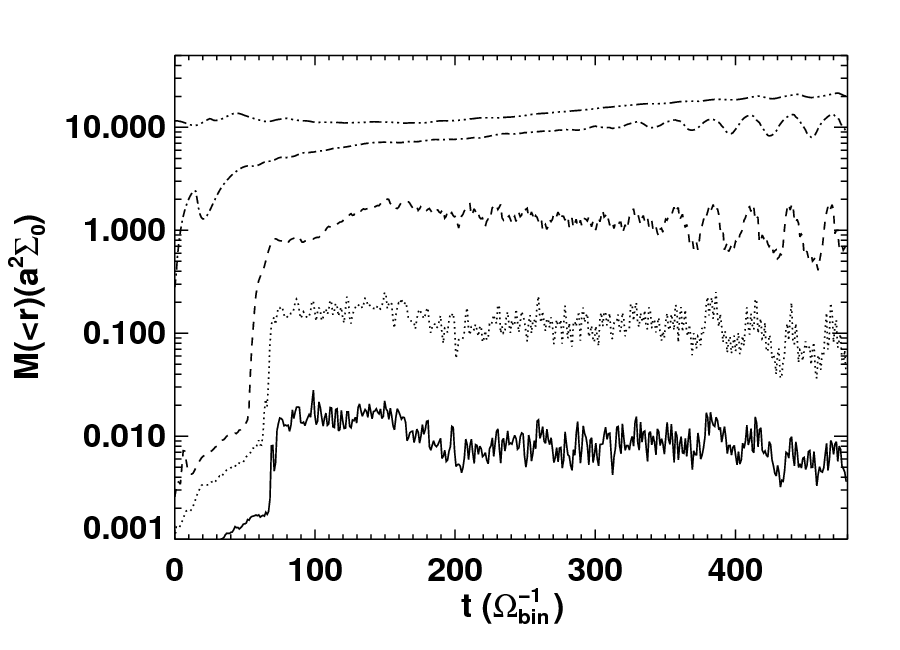}
\caption{\small{Disk mass profile history.  The mass interior to $r=1.0a$ is given by
the solid curve, $1.5a$ the dotted curve, $2.0a$ the dashed curve, 
$3a$ the dash-dotted curve, and $4a$ the dash-triple-dotted curve.}}\label{fig:mdisk}
\end{figure}

%
We show the enclosed mass (integrated disk mass interior to a given radius) inside $r=a$, $1.5a$,
$2a$, $3a$ and $4a$ as a function of time in Figure~\ref{fig:mdisk}. We find that the mass
at $r<3a$ (i.e., inside the initial disk's inner edge) undergoes dramatic growth during the
first $\sim 100$ time units as the initial disk fills in regions with $<r_{\rm min}$.
The radial pressure gradient at the edge of our initial disk leads to an inflow
during the first several
orbits, and once toroidal field develops near the disk inner edge, the Maxwell stress
quickly removes the angular momentum of the low density flow and drives inflow.
As this transient phase passes, the mass in the gap levels off and stays quasi-steady
(falling very slowly) after $t \sim 200$.
We therefore define the quasi-steady stage as the period between $t=200$
and the end of the simulation.  On the other hand, the mass inside $r=3a$ and $4a$ rises
by a factor of two after $t=100$, indicating that this region does not achieve
inflow equilibrium. The mass-interior profiles also possess high frequency fluctuations
(typical time scale $\sim 3$ time units) due to the orbital modulation of mass contained
in the streams.  In addition, after $\sim 350$ time units, there are slower oscillations
(time scale $\simeq 20$--$30$ time units) of the enclosed mass caused by
the eccentrically orbiting density lump.

\begin{figure}
\centering{
\includegraphics[scale=0.4]{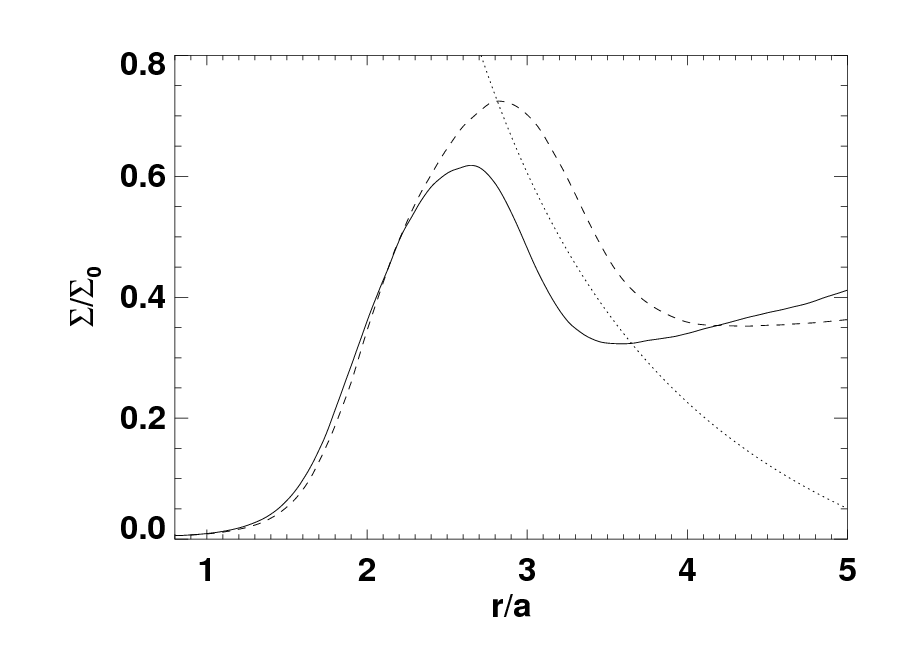} 
\includegraphics[scale=0.4]{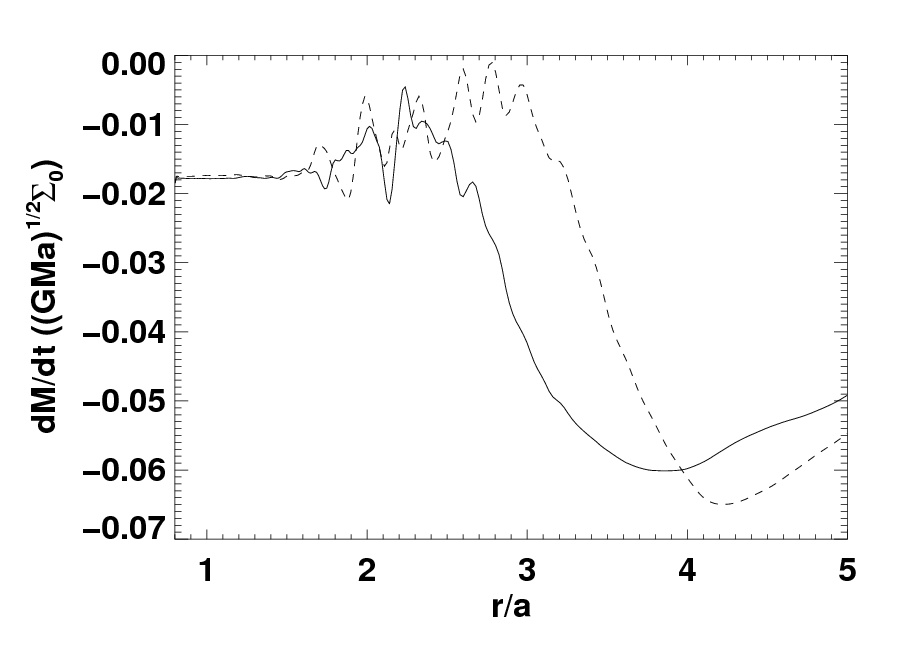} 
}
\caption{\small{ Time- and shell-integrated surface density (top) and accretion rate (bottom)
averaged over $t=250$--$350$ ($\Delta T_1$, solid) and $t=350$--$450$ ($\Delta T_2$, dashed).
The dotted line in the surface density plot displays the $\propto r^{-2}$ density profile of
a `decretion' disk.}}\label{fig:steady}
\end{figure}

In Figure~\ref{fig:steady}, we display the time-averaged surface density and accretion rate
over two intervals: $t=250$--$350$ ($\Delta T_1$) and $t=350$--$450$ ($\Delta T_2$).
We find there is only minor change in the averaged surface density and accretion rate of
the disk region $<2.5a$ over $\Delta T_1$ and $\Delta T_2$.  If we define the unit of
surface density $\Sigma_0 \equiv \rho_0 a$, the peak density grows in
time, but only slowly, rising from $\simeq 0.6\Sigma_0$ to $\simeq 0.7\Sigma_0$.  Its
position $\rp$ also changes, but similarly slowly, gradually moving from $\simeq 2.5a$ to
$\simeq 3a$.  The shape of the surface density profile likewise hardly changes in time.
On the inside, the disk is truncated exponentially,
$\lan\Sigma(r<2a)\ran_t\sim \Sigma_0\exp(3.8r/a-8.4)$ for both intervals,
where $\lan~\ran_{t}$ represents an average over time.  Toward larger radii, the surface
density is $\propto r^{-2}$, the predicted profile of a `decretion disk' \citep{Pringle91}.
All that changes is that the region where $\Sigma(r) \propto r^{-2}$ extends farther
outward at later times.

Averaged over the quasi-steady period, the accretion rate at the inner boundary is
$\dot{M}\simeq 0.018(GMa)^{1/2}\Sigma_0$.  However, the disk evolves throughout the
simulation at $r>3a$.  We see gas inflow rates as large as
$\sim 0.06 (GMa)^{1/2}\Sigma_0$ outside the region $\sim 3a$ during both time intervals.
As a result, the surface density continuously build up around this radius.

Many properties of these disks can be expected to scale with the mass near the surface
density peak.  It is therefore convenient to define a `disk mass'
$\md \equiv \pi \rp^2 \lan\Sigma_p\ran_t$, where $\rp =3a$ and
$\lan \Sigma_{\rm p}\ran_t = 0.65\Sigma_0$.  In terms of this mass (and the natural unit of
time for the simulation), the mean accretion rate through the inner boundary can be
rewritten as
\begin{equation}
\langle \dot M \rangle = 1.0 \times 10^{-3}\md \omgb.
\label{eq:accretion_rate}
\end{equation}
The surface density of such disks is very uncertain.  A sense of scale, however, can be
gleaned from translating eqn.~\ref{eq:accretion_rate} into the luminosity that would be produced
if the accretion were converted into radiation at customary black hole efficiency (10\%);
in Eddington units, it is $L/L_E = 0.036 M_8^{-1/2} a_{0.1}^{1/2}\tau_p$, where we have scaled to a
total black hole mass of $10^8 M_{\odot}$, a binary separation of 0.1~pc, and a disk
column density whose Thomson optical depth is $\tau_p$.

\subsection{Angular Momentum Budget \label{sec:budget}}
Unlike a steady accretion disk in a point-mass potential, where the internal stress simply
transports angular momentum outward and thereby drives an inflow, the
binary consistently interacts gravitationally with the circumbinary disk by torquing the
surrounding gas.  The angular momentum delivered through these torques is transported
by two mechanisms: MHD stresses, mostly due to MRI-driven turbulence; and Reynolds stresses
associated with coherent gas motions in the gap region.  We have checked that the code's numerical
shear viscosity is negligible compared to the transport mechanisms discussed here.
In this section, we first
investigate the mechanisms of angular momentum transport and the properties of the binary
torque, and then discuss the balance between binary torques and the stresses.

\subsubsection{Maxwell and Reynolds Stresses \label{sec:stress}}

\begin{figure}
\centering{
\includegraphics[scale=0.4]{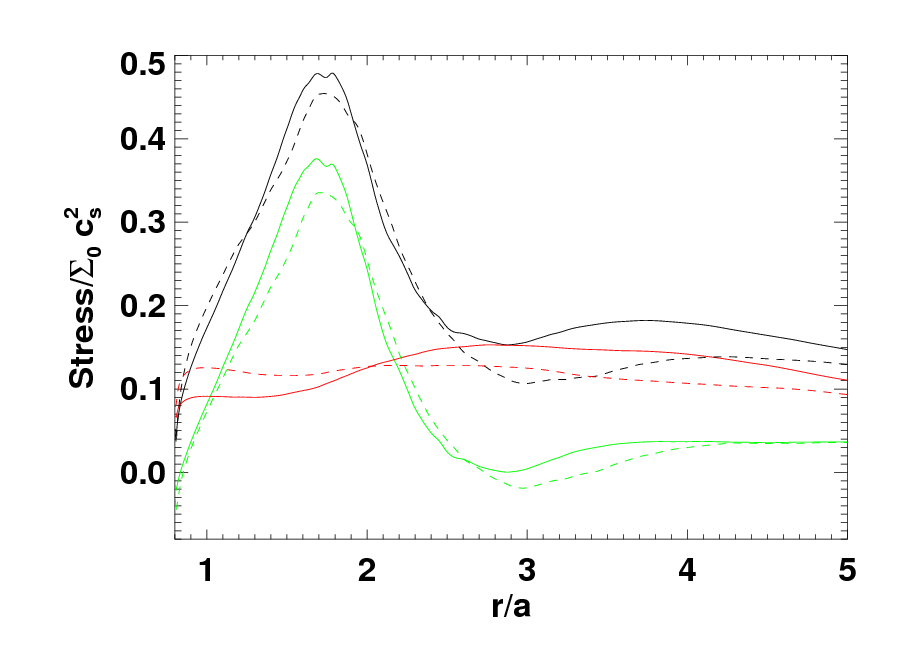}
\includegraphics[scale=0.4]{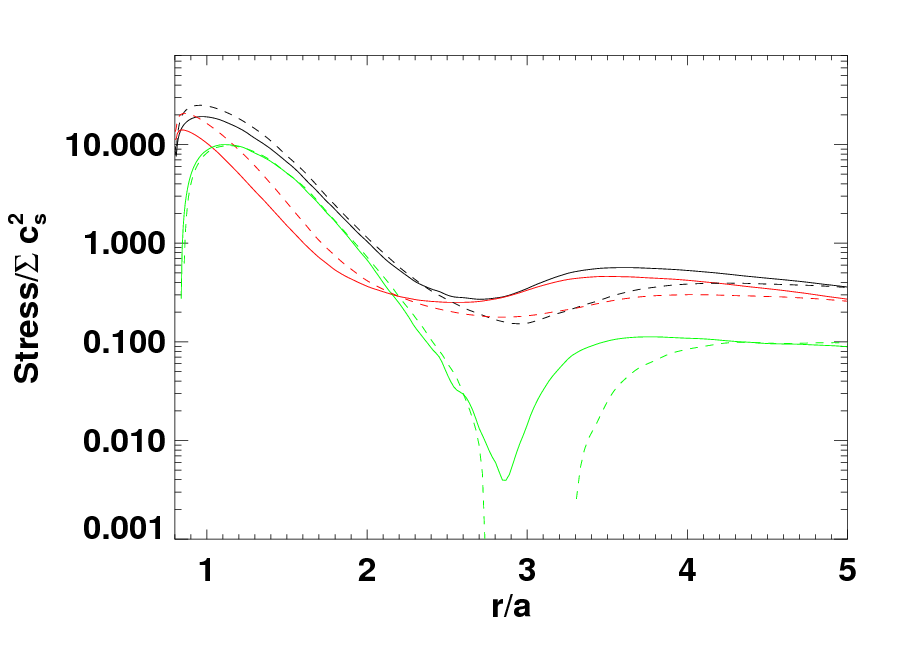}
}
\caption{\small{Top: Time-averaged Reynolds stress (green), Maxwell stress (red) and the total
stress (black) as functions of radius, solid for $\Delta T_1$ and dashed for $\Delta T_2$.
Bottom: Same stresses but normalized with the local pressure.}\label{fig:stress_1d}}
\end{figure}

In the first panel of Figure~\ref{fig:stress_1d}, we show the vertically-integrated and
time- and azimuthally-averaged stresses and their sum as a function of radius.
Within $r <3a$, the character of these stresses is very similar in our two averaging periods,
$\Delta T_1$ and $\Delta T_2$.  The Reynolds stress shows big fluctuations radially. Its
first peak, at $r \simeq 1.7a$, is $\sim 0.35 \Sigma_0 c_s^2$. It dips at around $3a$ to
nearly zero and then rises up again slowly. The ratio of the first peak to the second peak
is $\sim 10$. However, the Maxwell stress has a flatter radial profile. It is roughly
$\sim 0.1\Sigma_0 c_s^2$ on average and varies only slowly over a range of at most a
factor of two.  It diminishes within the gap and decreases smoothly towards larger distance.
Comparing the two, we find the Reynolds stress exceeds the Maxwell stress by a factor
$\sim 4$ at $r\sim 1$--$2a$.  On the other hand, beyond the gap region, the MHD stress always
dominates the Reynolds stress by a factor $\gtrsim 3$. Within one binary separation, the
Maxwell stress also exceeds the Reynolds stress, mainly because the Reynolds
stress falls with the lower gas density near the boundary.
The total internal stress therefore follows the Reynolds stress at $r\lesssim 2a$, while
it is close to the magnetic stress outside that region.
Similar results have been reported in previous studies on gap formation in protoplanetary disks
with an embedded planet using MHD simulations \citep{nelson03,winter03}. It is not
a surprising outcome as the disk with a planet is just a special case of a circumbinary
with an extreme mass ratio.


In the language of the $\alpha_{\rm SS}$ parameter of \citet{SS73}, we can also
measure the stresses in units of the pressure.  In the right-hand panel of
Figure~\ref{fig:stress_1d}, we present the time averaged stresses in these units.
%
%
The most significant distinction compared with the absolute stresses is that the stress ratios
increase steeply inward due to the low surface density in the gap. The total stress
ratio peaks at $\alpha_{\rm SS}\simeq 11$ at $ r=1.1a$. Its value in the disk body is almost
two orders of magnitude smaller.

\begin{figure}
\epsscale{0.8}\plotone{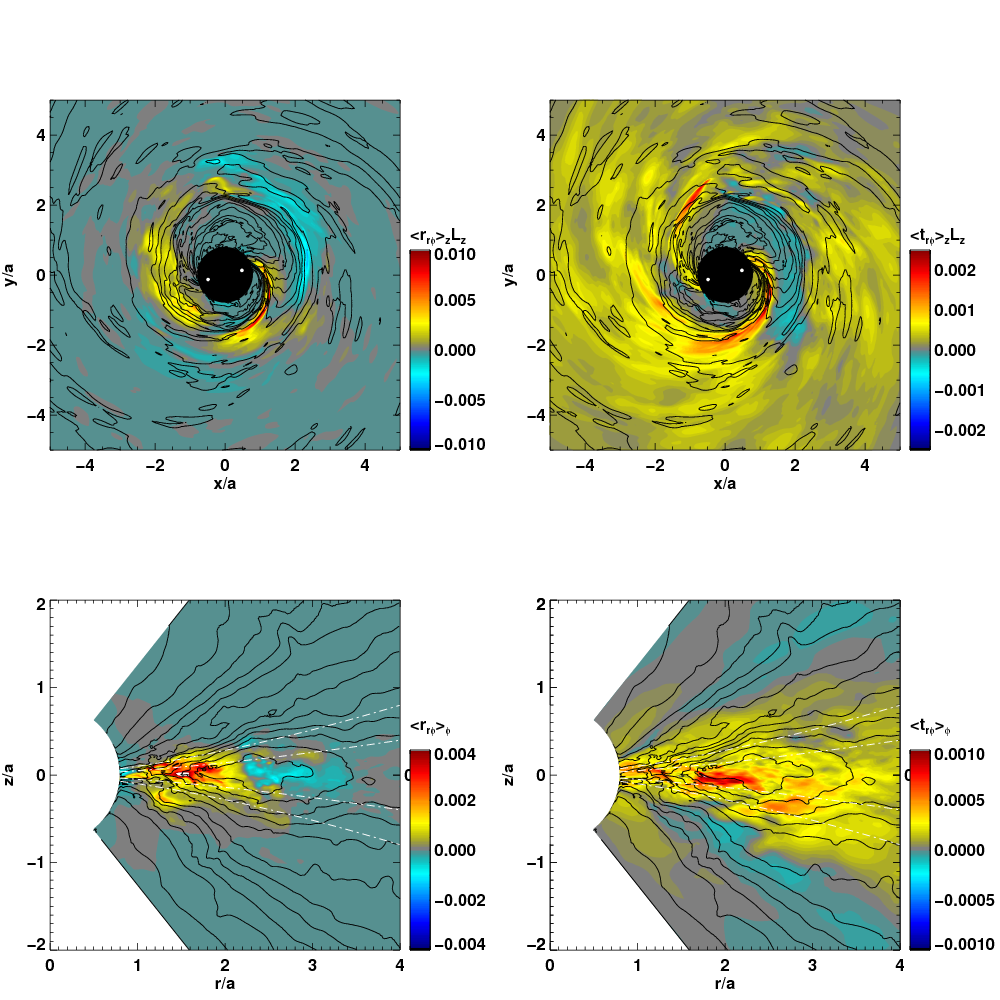}
\caption{\small{Snapshots of Reynolds stress (left column) and Maxwell stress (right column) at
$t=305$ averaged either vertically (top row) or azimuthally (bottom row). The $z=\pm H$ and $\pm 2H$
levels are plotted as white dash-dotted lines.}\label{fig:stress_2d}}
\end{figure}

Our simulation shows that the circumbinary disk possesses highly time-dependent structures
in the horizontal plane, making it important to look at the instantaneous spatial distribution
of the stresses in addition to the time-averaged values.  Vertically-integrated snapshots
of both Reynolds stress (top left) and Maxwell stress (top right)
taken at $t=305$ time units are shown in Figure \ref{fig:stress_2d}. On top of each
plot, we also draw the surface density with $15$ contour lines, logarithmically spaced
from the density floor value to the density maximum. Nearly all the Reynolds stress is
confined within $r<3a$. The most interesting point is that we find relatively large stress
in a single gas stream extending from $\sim 1$--$2a$ (see the red stripe in the top left panel),
where the gas gains angular momentum ($\delta v_{\phi}>0$) and is being pushed out by the
binary.  The gas being kicked thus goes on an eccentric orbit, creating negative stress at
$r\sim 2$--$3a$ and $-\pi/4<\phi<\pi/2$ ($\delta v_r>0$ and $\delta v_{\phi}<0$).
In the same range of radii but a different azimuthal location ($\pi/2<\phi<5\pi/4$),
the gas falls back toward the binary, creating positive stress (both $\delta v_r$
and $\delta v_{\phi}<0$).
This bi-symmetric property provides a near cancellation so that only the stress within the
stream contributes significantly to the radial profile of the Reynolds stress.
On the other hand, the Maxwell stress is more evenly distributed in the disk body,
although the typical magnitude is a factor of $4$ less than the typical magnitude of
the Reynolds stress in the inner disk. In the gap, the Maxwell stress is strongly positive
in the streams, where the field lines are collimated by the gas flow, but negative elsewhere.

To illustrate how the stresses depend on the disk height, we present azimuthally-averaged
snapshots at the same time in Figure \ref{fig:stress_2d}. We find the Reynolds stress
is well concentrated in the midplane. Its strength decreases to about half the midplane
value at only one scale height away from the midplane, and about one tenth at two scale
heights. However, the Maxwell stress is less sensitive to the altitude. It appears to be
mainly confined within four scale heights, but also shows positive contributions
at higher altitude near the pole due to field buoyancy.

\subsubsection{Binary Torque \label{sec:torq}}
\begin{figure}
\centering{
\includegraphics[width=7cm]{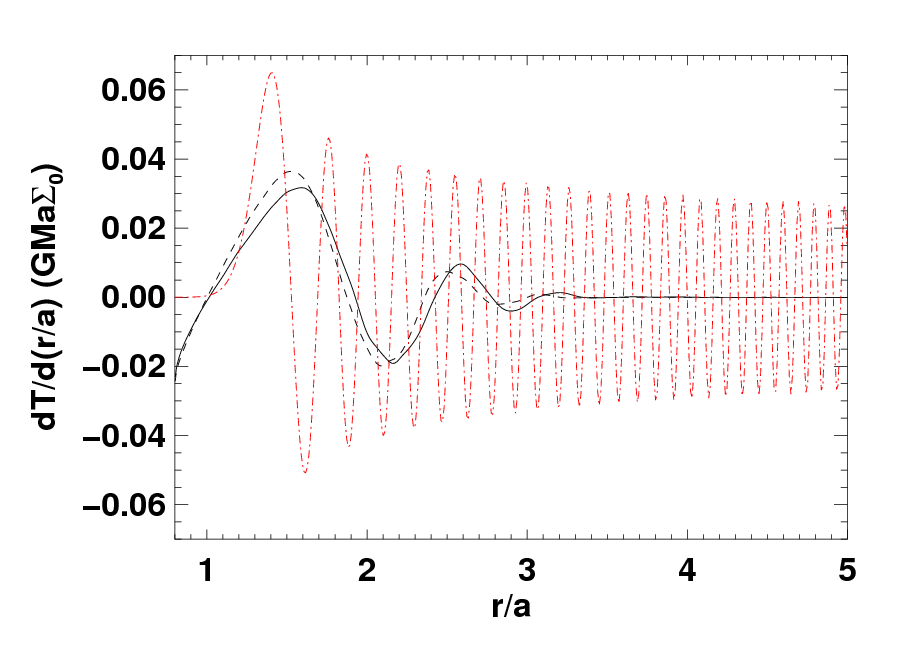}
\includegraphics[width=7cm]{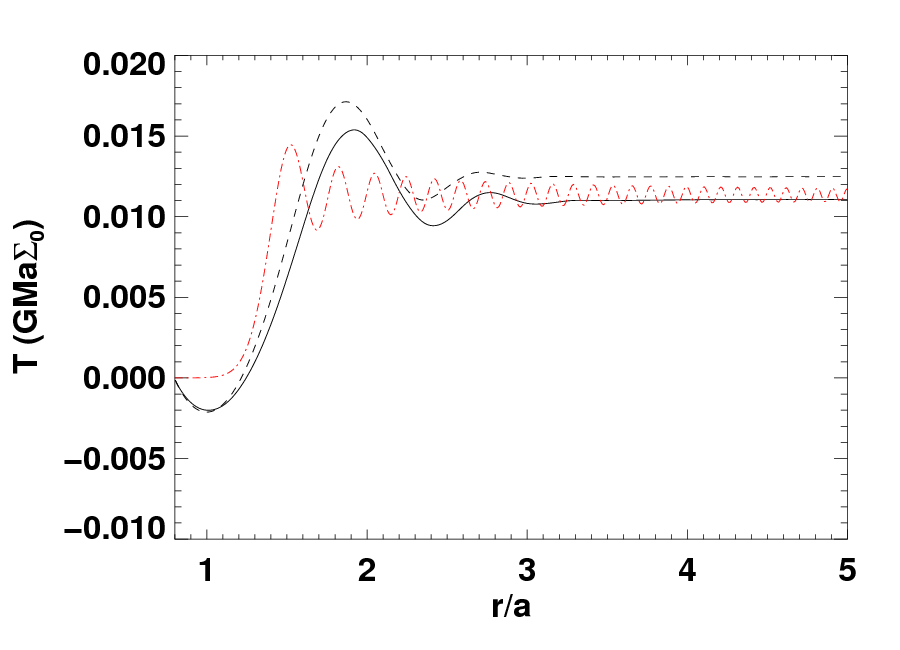} 
}
\centering{
\includegraphics[width=7cm]{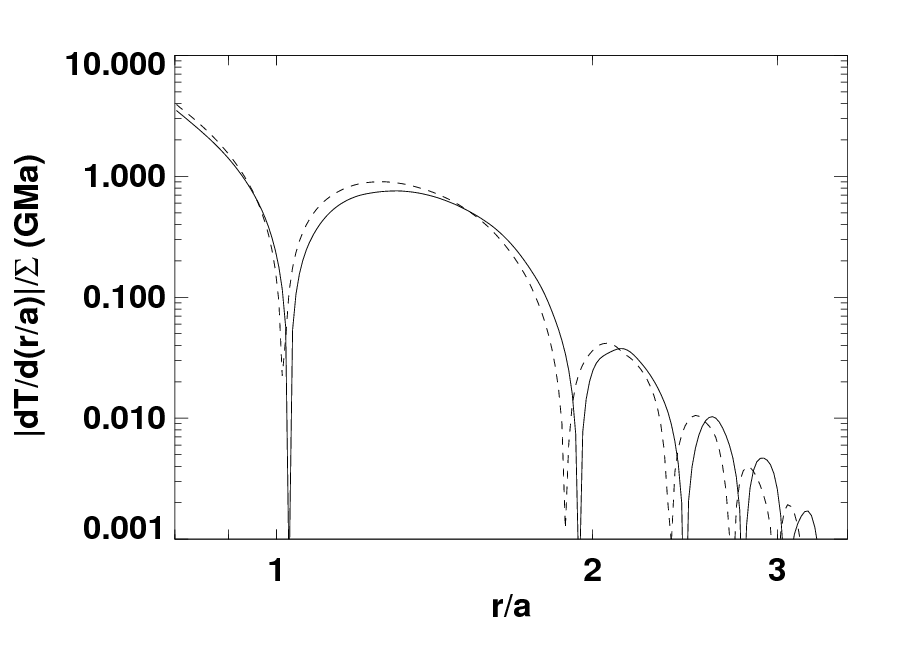}
\includegraphics[width=7cm]{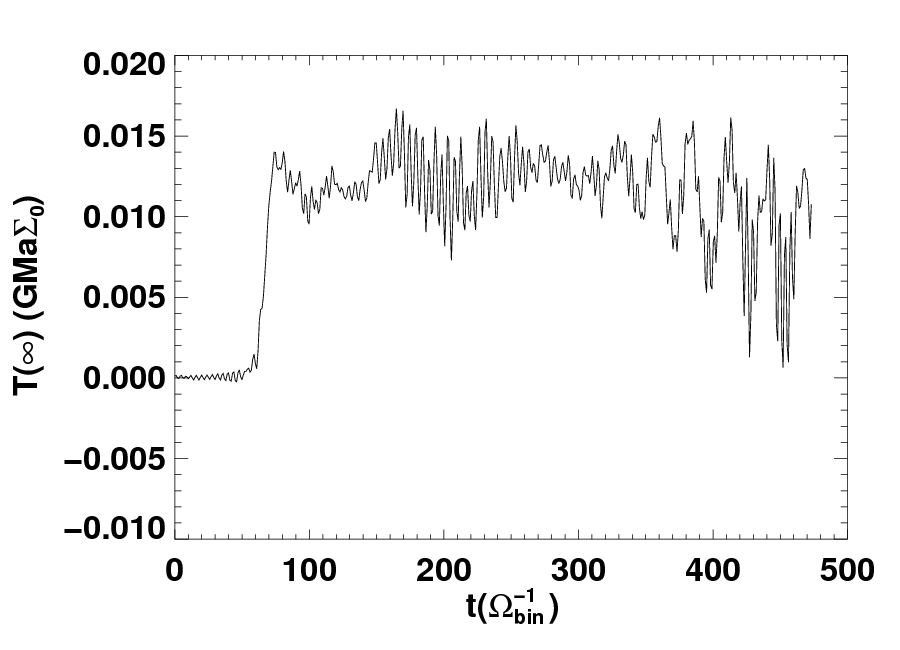}
}
\caption{\small{Time averages (dashed for $\Delta T_1$, solid for $\Delta T_2$) of local (top left),
total(top right), and specific (bottom left) binary torque as functions of radius.  The linear
theory prediction using 1/4 the time-averaged surface density at the $3:2$ resonance is shown
as a red-dotted curve in the upper two panels.  The history of the total torque is shown in
the bottom right panel.  The curve is smoothed by boxcar average whose width is twice the binary
period.\label{fig:torq}}
}
\end{figure}

Having discussed the mechanism for angular momentum transport, let us now consider the
torque exerted on the disk by the binary. We plot the time-averaged torque density and
the integrated torque in Figure~\ref{fig:torq}, where the torque is calculated approximately
using the surface density instead of the local density itself. The definition follows MM08:
\beq
\frac{dT}{dr} \equiv -\int \Sigma(r,\phi)\frac{\partial\Phi}{\partial\phi} rd\phi
\enq
is the local torque, and the integrated torque is
\beq
T(r) \equiv \int^{r}_{r_{\rm in}} \frac{dT}{dr'}dr',
\enq
where $\Phi=\Phi(r,\phi)$ is the binary potential in the midplane.

The first panel of Figure~\ref{fig:torq} shows the averages of the local torques
for the two intervals $\Delta T_1$ and $\Delta T_2$ are very similar, which again indicates that
the inner part of the disk reaches a quasi-steady state.  The local torque density is
positive at $r\simeq 1$--$2a$ and peaks at $r\simeq 1.5$--$1.6a$ with
$\frac{dT}{dr}\simeq 0.03$--$0.04 GM\Sigma_0$ as the peak value.
It then goes negative at $r\simeq 2$--$2.4a$, reaching its first negative maximum at
$r \simeq 2.0$--$2.2a$, but with a smaller magnitude $\simeq -0.02 GM\Sigma_0$.
Similar to what was previously found in MM08 and \citet{cuadra09}, we find the torque
density oscillates around zero but damps quickly toward larger distance.
In addition, the torque density for $r\lesssim a$ is negative
because the gas within that region is advanced in phase (greater angular frequency) with
respect to the binary, an effect that would appear in purely hydrodynamic treatments as
well as MHD.

The integrated binary torque is displayed in the top right panel of Figure~\ref{fig:torq}.
The total binary torque exerted on the disk is $T(\infty) \simeq 0.011$--$0.013 GMa\Sigma_0$.
In order to make a comparison with the torque found in MM08, some normalization is needed.
%
%
First, we normalize the torque to the surface density at the radius where the local
binary torque peaks. We denote that radius as $r_{\rm torq}$.  In our MHD simulation, the first
positive peak of the torque takes place at $r_{\rm torq}\simeq 1.6a$  while it is $\sim 1.7a$
in MM08.  In terms of this normalization, $T(\infty)/\Sigma(r_{\rm torq})\sim 0.07 GMa$ in
the hydrodynamic simulation of MM08.  In our MHD simulation, we find that it is
$\simeq 0.12 GMa$ (averaged in the quasi-steady period), a
factor of $1.7$ greater than in the hydrodynamic and $\alpha$-viscosity treatment.

However, as the peak of the torque is located in the gap region, where the surface density
rises rapidly with increasing radius, this means of normalization is unreliable.  We
therefore turn to a different method, normalizing the binary torque to the disk mass.
With this method, the normalized torque is
$T(\infty)/\md \simeq 6.5 \times 10^{-4} GMa^{-1}$ in
our MHD simulation. However it is only $\sim 4.5\times 10^{-5} GMa^{-1}$ in MM08, $\sim 14$
times smaller.  Just as for the mass accretion rate, there is a natural set of units for
the total torque: it is convenient to describe it in terms of how rapidly the torque
removes the binary's orbital angular momentum.  In that language,
\begin{equation}
T({\infty}) = 2.6 \times 10^{-3} j_{\rm bin}\md\omgb,
\end{equation}
where $j_{\rm bin} = (GMa)^{1/2}/4$ is the specific angular momentum of the binary.

Alternatively, as $\rp$ is roughly around $\lesssim 3a$ for both
MHD and hydrodynamic simulations, we can write the binary torque as some factor times
$GMa\sigmap$, and then we have $T(\infty)\simeq 1.9\times 10^{-2} GMa\sigmap$ in our present
simulation, while MM08 gives $\sim 1.4\times 10^{-3}GMa\sigmap$, $\sim 14$ times smaller.
The largest part of this contrast can be explained
by the fact that in the MHD treatment $\Sigma(r_{\rm torq})/\sigmap\simeq 0.15$, a factor of
$\sim 8$ greater than that of MM08.

This contrast can also be understood in terms of the angular momentum budget.  The
gravitational torque per unit mass approximately balances the torque per unit mass
due to internal stress near the disk inner edge. Therefore, the binary torque should
increase linearly with the effective $\alpha$. In our case,
$\alpha \sim \alpha_{\rm SS}\gtrsim 0.2$ in the disk body at $r \simeq 2$--$5a$, a factor of
$\gtrsim 20$ greater than the constant $\alpha=0.01$ assumed by MM08.

In order to show how the local binary torque evolves with time, we also plot in
Figure~\ref{fig:torq} the specific torque density (bottom left), i.e., the absolute value
of the ratio of torque density to surface density.  By dividing by the surface density,
we remove the effects due to redistribution of the surface density.  We find the specific
torque slightly shifts to larger radius and diminishes its amplitude
from $\Delta T_1$ to $\Delta T_2$. In the bottom right panel, we plot the history of the
total torque smoothed over a short time span
$\sim 2T_{\rm bin}$, where $T_{\rm bin}=2\pi/\omgb$ is the binary period. 
Before smoothing, the total torque fluctuates very
rapidly ($\sim 1.5\omgb$) between $\sim -0.05 GMa\Sigma_0$ and $0.06 GMa\Sigma_0$ in the
quasi-steady period.  The power spectrum of this fluctuation is dominated by a peak at
$\sim1.5\Omega_{\rm bin}$, approximately the beat frequency between the orbital frequency of the
matter being torqued most strongly (at the peak of $dT/dr$, i.e. $r \sim 1.6$ --$ 1.7a$) and
$2\Omega_{\rm bin}$.  After smoothing, we find the binary torque has a sudden rise
around $t=60$ which is due to the initial disk filling the small radius region.  The torque
then levels out between $t=100$--$200$. Once in the
quasi-steady stage, the torque shows a slowly decreasing trend, $\lesssim 30\%$ in
fractional terms over the final $\sim 300$ time units. It also appears to
oscillate at the disk orbital frequency ($r \sim 2 $--$3a$) during the late time, a fluctuation
caused by the eccentric movement of the density lump. As both the disk eccentricity
and the density of the lump grow with time, the torque normalized by the time-dependent
$\sigmap \rp^2$ simply reflects the increasing $\sigmap$ and $\rp$.

\subsubsection{Torque Balance\label{sec:budget3}}

\begin{figure}
\plotone{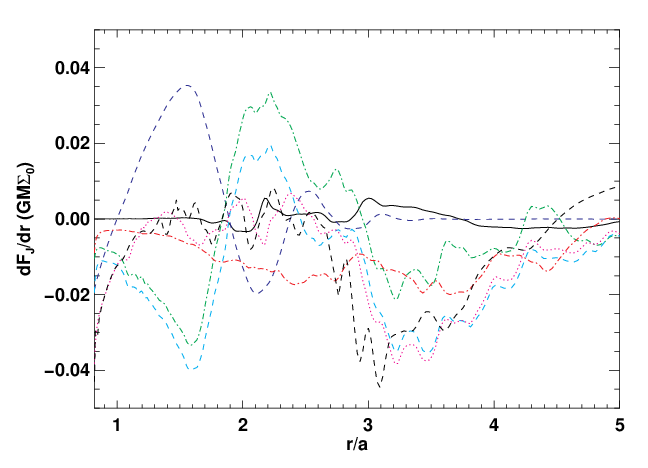}
\caption{\small{Radial derivatives of the different sorts of time-averaged and shell-integrated
angular momentum flux ($dF_J/dr$) appearing in 
equation~(\ref{eq:L3_conserve}) as functions of radius. The averaging period is $t=300$--$320$.  
Net rate of change of the local angular momentum
(solid black curve); binary torque density (blue dashed curve); Reynolds stress (green
dash-dotted curve); Maxwell stress (red dash-dotted curve); the sum of Reynolds and Maxwell stresses
(cyan dashed curve); angular momentum flux due
to gas advection (black dashed curve; the inferred flux
according to conservation law is shown with the magenta dotted curve for comparison).}}  \label{fig:dflux}
\end{figure}

For a steady circumbinary disk, the angular momentum at a given radius should not fluctuate too
much over time. Therefore the angular momentum deposition from the binary must either be
transported outward by Maxwell and Reynolds stresses or advected towards the hole. In order to
test this idea, we perform an estimate of the differential angular momentum flux averaged over
a short time interval between $t=300$--$320$ using the high time-resolution rerun.
Based on the conservation of angular momentum, we write the angular momentum budget for a ring of
disk between ($r-\Delta r$, $r+\Delta r$) as
\begin{eqnarray}
\frac{\partial J}{\partial t}&+& \frac{\pi}{\Delta r}[(r+\Delta
r)^2F(r+\Delta r)-(r-\Delta r)^2F(r-\Delta r)] \nonumber \\
&=& \frac{\pi}{\Delta r}[(r-\Delta r)^2W_{r\phi}(r-\Delta r)-(r+\Delta r)^2 W_{r\phi}(r+\Delta r)] + \frac{dT}{dr},
\label{eq:L3_conserve}
\end{eqnarray}
where $J(r,t) = 2\pi r\lan\rho v_{\phi}r\ran L_z$ is the shell-integrated angular
momentum, $F(r,t)=\lan\rho v_r\ran\lan\rho v_{\phi}\ran L_z/\lan\rho\ran$ is the mean angular
momentum flux due to mass motion, and $W_{r\phi}(r,t)=T_{r\phi}+R_{r\phi}$ is the total
internal stress. We write the
conservation law this way so that a negative differential flux due to advection (the second term
on the left hand side of equation~(\ref{eq:L3_conserve})) indicates a net
angular momentum inflow, while a negative flux due to internal stresses (the first term on 
the right hand side) means angular momentum
is removed and transported to larger radii. 
We calculate the time averages of all these terms, and plot them in Figure~\ref{fig:dflux}
as functions of radius. In its $y$--legend, we use $dF_J/dr$ to represent all the terms in the
equation, which are all radial derivatives of one or another variety of shell-integrated
angular momentum flux ($F_J$).

During this time interval, the local angular momentum stays steady, with only very small
temporal variations from the averaged $\partial J/\partial t$ (solid black curve), as expected
for a quasi-steady accretion disk.  The binary torque (blue dashed curve) at $r\sim 1$--$3a$ is
almost balanced by the differential flux due to the sum of Reynolds stress and Maxwell stress
(cyan dashed curve). In other words, the circumbinary disk within that region
reaches a balance between binary torque and internal stresses. As a result, the angular momentum
flux carried by accretion flows oscillates about zero (magenta curves within $1$--$3a$).
Clearly, the internal torque due to the Reynolds stress (green dash-dotted curve)
dominates the Maxwell stress (red dash-dotted curve) at $r\sim 1$--$3a$, which suggests that
most of the angular momentum dumped by the binary at $r\sim 1$--$1.8a$ is transported outward
via the Reynolds stress, while the angular momentum drawn from the binary at $r\sim 1.8$--$2.4a$
is compensated by the torque of the Reynolds stress.  Finally, we comment that torque balance
is not reached at $r\lesssim a$ (where negative binary torque dominates) or $r\gtrsim 3a$
(where internal stress dominates). 
By summing all the other torques, we find the differential angular momentum flux that must be
advected inward in those two regions (magenta curve).  Directly computing the second term on
the left hand side of equation~(\ref{eq:L3_conserve}) yields the (very similar) black dashed curve in
Figure~\ref{fig:dflux}.

\vspace*{0.4cm}
To sum up, we have found three principal conclusions about the angular momentum budget.
First, the Reynolds stress dominates the Maxwell stress in the gap region, while the latter
exceeds the former outside that. Secondly, the distribution of both stresses follows the gas
streams and are vertically confined: within two scale heights for Reynolds stress and
four scale heights for Maxwell stress. Third, we find the radial profile of the binary
torque is similar to that of previous hydrodynamic results, but the amplitude is different:
normalized by the surface density at the location where local torque peaks,
the binary torque in the present MHD simulation is twice as great as in previous
hydrodynamic calculations; however when normalized by the disk mass, our torque is an
order of magnitude greater than that found in MM08.  Lastly, we find that in the quasi-steady
state, the binary torque is roughly balanced by the torque due to the Reynolds and Maxwell
stresses in the inner disk ($r<3a$). Most of the angular momentum dumped by the binary
in the gap region is transported outward by the torque of the
Reynolds stress associated with the gas streams.

\subsection{Disk Eccentricity \label{sec:ecc}}

\subsubsection{Evolution of the Disk Eccentricity\label{sec:ecc_evolution}}


Although we begin our simulation with an axisymmetric disk configuration, which means
there is zero eccentricity at the start, the disk eccentricity grows throughout our simulation
(e.g., compare the disk surface density snapshots in Figure~\ref{fig:4snapshots}).

\begin{figure}
\plotone{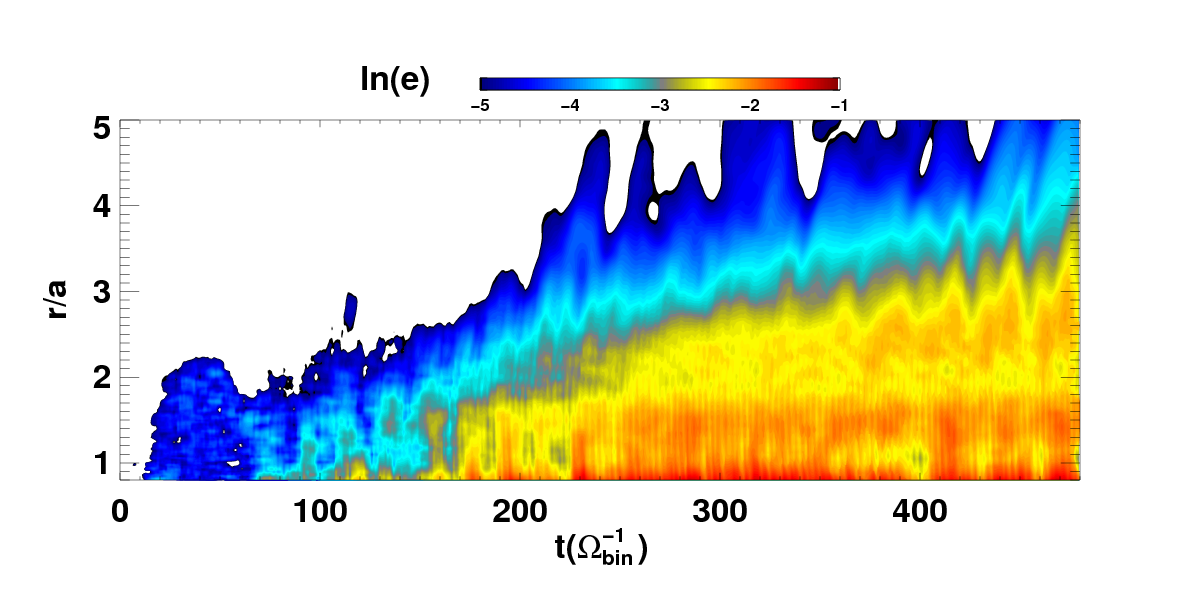}
\plotone{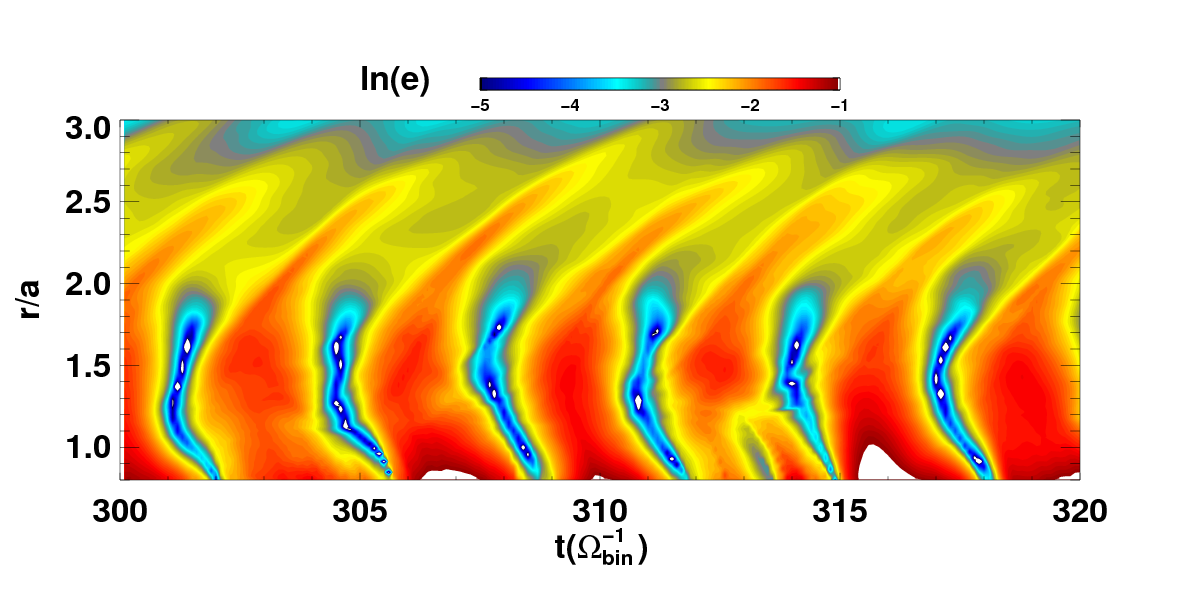}
\caption{\small{Top: Space-time diagram of the local eccentricity (smoothed over $T_{bin}$); color
contours are logarithmic (see color bar).  Bottom: an enlarged section of the
space-time diagram (without time smoothing) to show short timescale behavior in the inner disk.
White regions are eccentricity peaks off the scale.
}}  \label{fig:ecc_diagram}
\end{figure}

To quantify the eccentricity, we define the local eccentricity as
\beq
e(r,t) = \frac{1}{T_{\rm bin}}\int^{t+T_{\rm bin}}_{t}\frac{\left|\left\lan \rho v_r e^{i\phi}\right\ran \right|}
{\left\lan \rho v_{\phi}\right\ran} dt',
\label{eq:ecc_local}
\enq
where $\lan~\ran$ is the shell average defined as in equation~(\ref{eq:shell_int}).
We show this local eccentricity in the space-time diagram in Figure~\ref{fig:ecc_diagram}
(top panel). 
The local eccentricity of the gas in the gap ($r\lesssim 2a$) rises significantly during
the period $t\sim 100$--$200$.  The increasing $e(r)$ in the gap simply reflects the fact
that the dynamics there are strongly non-Keplerian.  The eccentricity in the
main body of the disk ($r>2a$) grows rapidly over the period
$t=100$--$300$. The typical value increases from much less than $0.01$ throughout the disk body
at $t=100$ to $\sim 0.08$ between $r=2$ --$3a$ at $t=300$. It keeps growing
slowly after $300$ time units, and at the end of the simulation a ring of eccentric disk forms
at $2a<r<3.3a$ with typical $e(r)\sim 0.1$.
A more detailed picture of time-variation in the eccentricity is shown in the bottom
panel of Figure~\ref{fig:ecc_diagram}, where we show its evolution from $t=300$ to $320$.
For clarity, we use the data from the high time resolution simulation, and do not take the
temporal smoothing as in equation~(\ref{eq:ecc_local}).
We find strong radial extending structures which connect the gap with the disk body. 
These features bend at $\sim 1.5a$ pointing to later times at both ends, which suggests 
propagation of eccentricity from the gap to both the disk and the binary. The pattern
repeats once every half binary period, indicating a stream-related origin. 
We will discuss the stream effects on the disk eccentricity
in $\S$~\ref{sec:discussion_mechanism1}.

\begin{figure}
\plotone{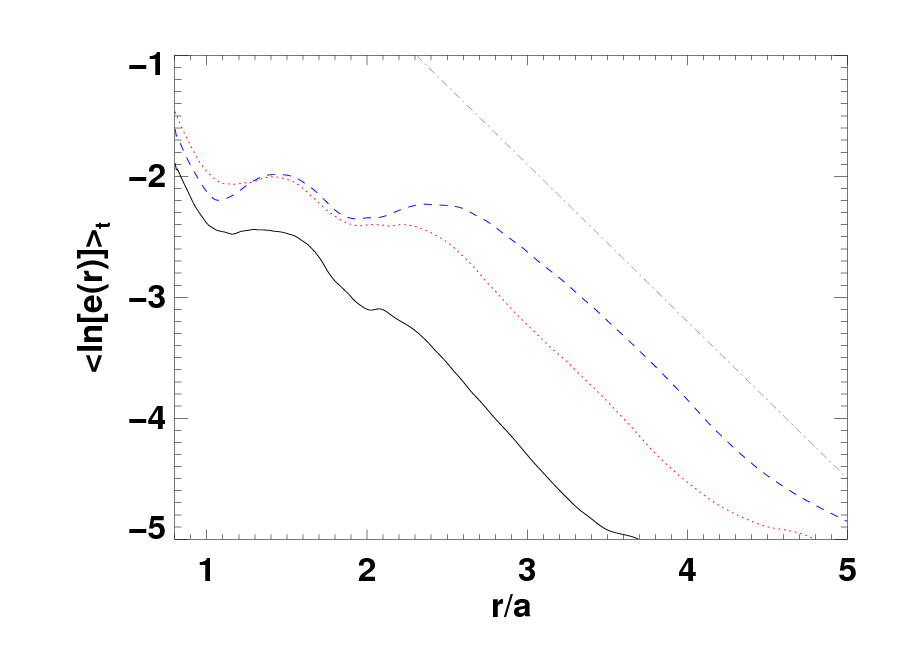}
\caption{\small{Time-averaged disk eccentricity as a function of radius at three epochs:
$t=150$--$250$ (black solid curve), $t=250$--$350$ (red dotted curve) and $t=350$--$450$
(blue dashed curve). The gray dash-dotted line shows a radial dependence
$\propto \exp{[2-1.3(r/a)]}$.}}  \label{fig:ecc_distr}
\end{figure}

In Figure~\ref{fig:ecc_distr}, we present the radial distribution of the disk
eccentricity averaged over three time intervals: $t=150$--$250$(solid black),
$\Delta T_1$ (red dotted) and $\Delta T_2$ (blue dashed).  The distribution curves
in the disk body ($r>2a$) shift toward larger radius and
greater eccentricity as the disk evolves in time. Outside the peaks around $2a$, the slopes of
the distribution curves are constant. The radial dependence of $e(r)$ in the disk body can
be roughly described as $\propto \exp{[-1.3(r/a)]}$. 

Interestingly, the eccentricity distribution found by MM08 was quite similar
to ours in both amplitude and shape, despite that calculation's very different internal
stresses, initial surface density profile, and duration.  In our own 2-d hydrodynamical
simulations, we have likewise found qualitatively similar eccentricity growth.  However,
the eccentricity profile in those simulations matches the MHD simulation (and MM08) only
at a specific time ($t \sim 800\Omega_{\rm bin}^{-1}$).  From these comparisons, it is
clear that the primary mechanism for driving eccentricity must be hydrodynamic response
to gravitational forcing by the binary, but its quantitative development can be affected
both by disk physics (e.g., MHD stresses) and initial conditions (e.g., the initial surface
density profile).  We will discuss specific mechanisms at greater length in $\S$ \ref{sec:edistr}.

\begin{figure}
\plotone{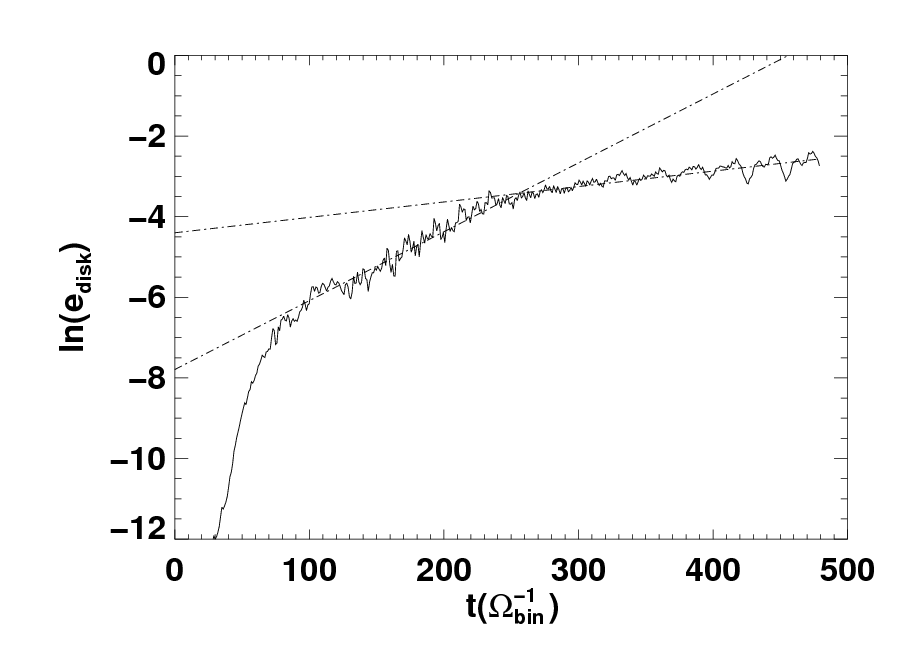}
\caption{\small{Evolution of the disk eccentricity. The dash-dotted lines show linear fits
for the exponential growth phase and the later saturation phase.
}}\label{fig:ms}
\end{figure}

The space-time diagram suggests we may take the radial average of the local eccentricity
for the disk body between $r=2a$ and $r=4a$ as a measure of the disk eccentricity.
In Figure~\ref{fig:ms}, we show the averaged disk eccentricity $e_{\rm disk}$ constructed by
taking volume averages from $r=2a$ to $r=4a$ and time averaging over one binary period:
\beq
e_{\rm disk} =\frac{1}{T_{\rm bin}}\int^{t+T_{\rm bin}}_{t} dt' \frac{\left | \int^{4a}_{2a}dr \int\rho v_{r} e^{i\phi} r^2\sin\theta d\theta d\phi \right
|}{\int^{4a}_{2a} dr\int\rho v_{\phi} r^2\sin\theta d\theta d\phi} .
\label{eq:edisk}
\enq
At early times ($t\lesssim 100$), Figure~\ref{fig:ms} shows the
transient phase of a non-equilibrium disk in a non-Keplerian potential.
During this phase, the noise develops significantly, especially in the low density
region of the initial configuration, and all modes undergo dramatic growth. 
After the transient, the eccentricity of the disk undergoes exponential growth from $t=100$
to $t=250$. By fitting the eccentricity curve during that interval, we find the growth rate
is $\gamma_{\rm e}\simeq 0.018 \omgb$, where the subscript $\rm e$ denotes eccentricity.
The growth rate of disk eccentricity slows by a factor of $4$ after $\sim 250$.
This suggests some nonlinear mechanisms come into play to limit and perhaps eventually
saturate eccentricity growth.

\subsubsection{Precession of the Eccentric Disk\label{sec:ecc_precess}}
\begin{figure}
\plotone{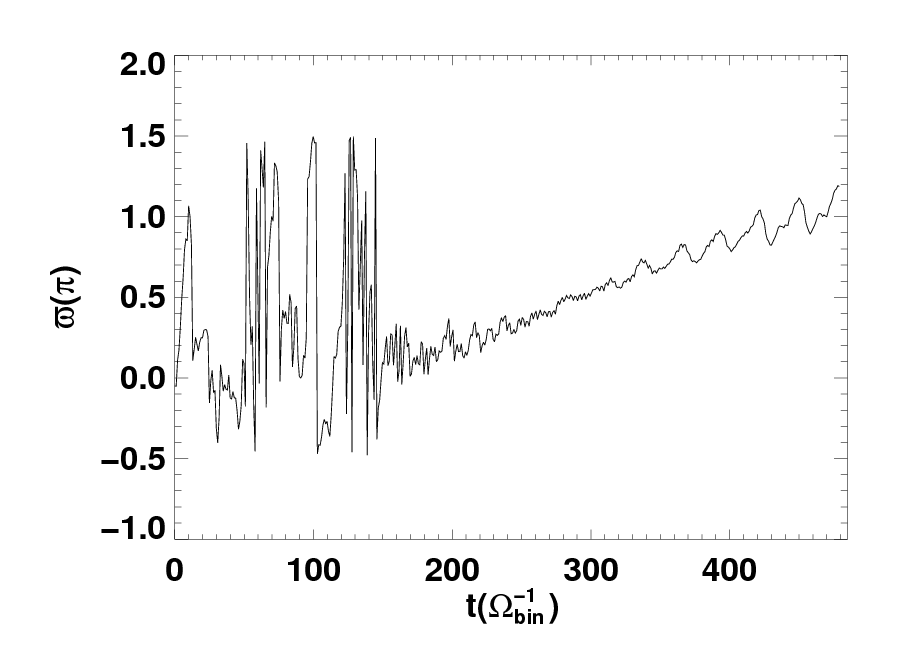}
\caption{\small{Phase angle of the disk eccentricity vector as a function of time.
}}  \label{fig:ecc_precess}
\end{figure}

To follow changes in orientation of the eccentricity, we define the longitude of the apoapse
$\varpi$ in terms of the complex phase angle of the disk eccentricity for $2a \le r \le 4a$.
Its time evolution is shown in Figure~\ref{fig:ecc_precess}.  Here the angle $\varpi$
measures the angular location of the apoapse with respect to the $x$--direction in the
inertial frame.  The disk eccentricity is vigorously perturbed by the motion of the gas
in the gap during the first one hundred units, and the arbitrary orientation in that period
is partly because the eccentricity during this time
interval is very small. The angle $\varpi$ then gradually rotates in a prograde manner. 
By fitting the curve between $t=200$--$480$, we find
$\dot{\varpi} \simeq 3.2\times 10^{-3}\omgb$. Linear perturbation theory predicts the
precession rate for particles around the binary potential is
$\dot{\varpi}\simeq \Omega - \kappa$ for a cold disk, where $\Omega$ is
the angular frequency and $\kappa$ is the epicyclic frequency of the disk. To a first
order approximation,
$\Omega(r) \approx \omgk\left[1+\frac{3}{8}\left(\frac{a}{r}\right)^2\frac{q}{(1+q)^2}\right]$
(also see equation~(\ref{eq:omega})), and
$\kappa(r) \approx \omgk\left[1-\frac{3}{8}\left(\frac{a}{r}\right)^2\frac{q}{(1+q)^2}\right]$.
Thus we find the prediction is
$\dot{\varpi}\simeq 0.19 (a/r)^{7/2}\omgb$ for $q=1$, or $\simeq 4.1 \times 10^{-3}\omgb$
evaluated at $r\simeq 3a$. The precession rates measured in both our MHD simulation and MM08
are close to this prediction. We therefore suggest that the precession is mainly due to the
quadrupolar potential.

\subsubsection{Late Time Lump\label{sec:ecc_lump}}
\begin{figure}
\plotone{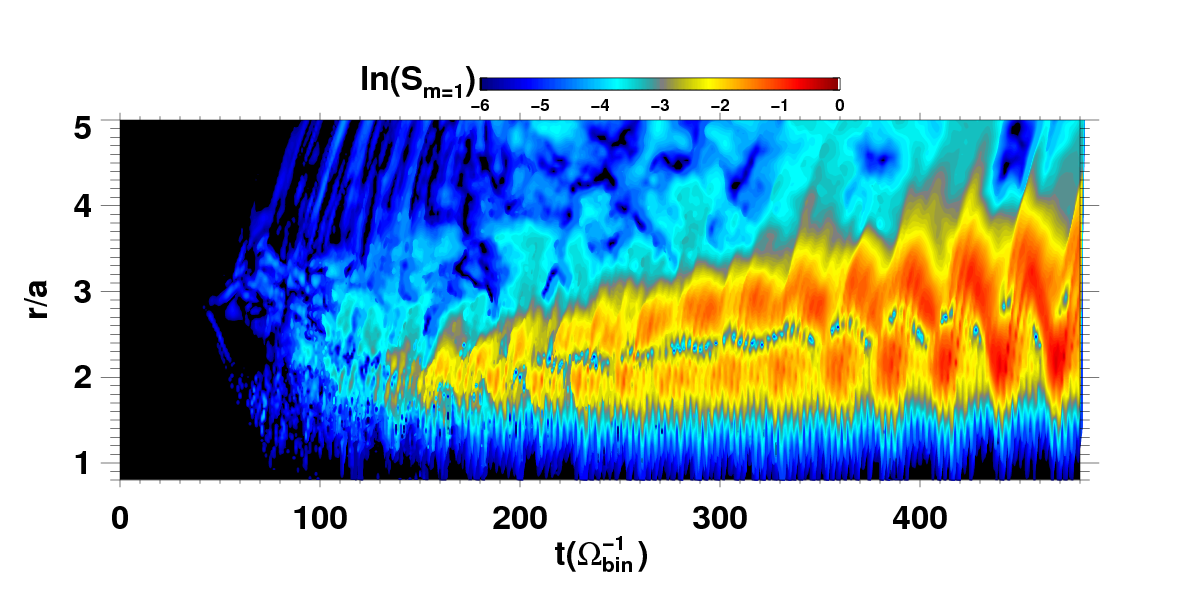}
\caption{\small{Space-time diagram of the m=1 Fourier component of the surface density.
Color contours are logarithmic (see color bar).}}
\label{fig:m1_diagram}
\end{figure}

There is another disk characteristic closely connected to the eccentricity: the density
concentration that appears near the inner disk edge at late time. We call it the `lump'
(see Figure~\ref{fig:4snapshots}). The lump orbits
around the binary at nearly the same orbital speed as the eccentric disk. To quantitatively study
the asymmetric feature we introduce the $m=1$ mode strength component
\beq
S_{\rm m=1}(r,t) \equiv \left |\int \rho e^{im\phi} rd\theta d\phi \right |~~~(m=1),
\label{eq:s1_local}
\enq
which is basically the $m=1$ component of the Fourier transform of the surface density. We
plot the space-time diagram of $S_{\rm m=1}$ in Figure~\ref{fig:m1_diagram}.  The orbiting lump
appears as a zig-zag pattern whose peak moves in and out between $r\sim 2a$ and $4a$ after
about $300$ time units. The pattern period, which is also the orbital period of the eccentric
disk, is $\sim 30$ time units.  We will discuss the lump in $\S$~\ref{sec:discussion_lump} and
explain it as due to a combination of disk eccentricity and the action of
periodic streams impacting upon the edge of the disk.

\subsection{Streams in the Gap \label{sec:stream}}

As shown in Figure~\ref{fig:4snapshots}, gas is not entirely absent from the gap. Consistent
gas flows launch from the inner edge of the disk and stream toward the binary. The streams not
only affect the accretion rate of the circumbinary disk, but also the structure of the inner
part of the disk.

\begin{figure}
\epsscale{1.2}\plottwo{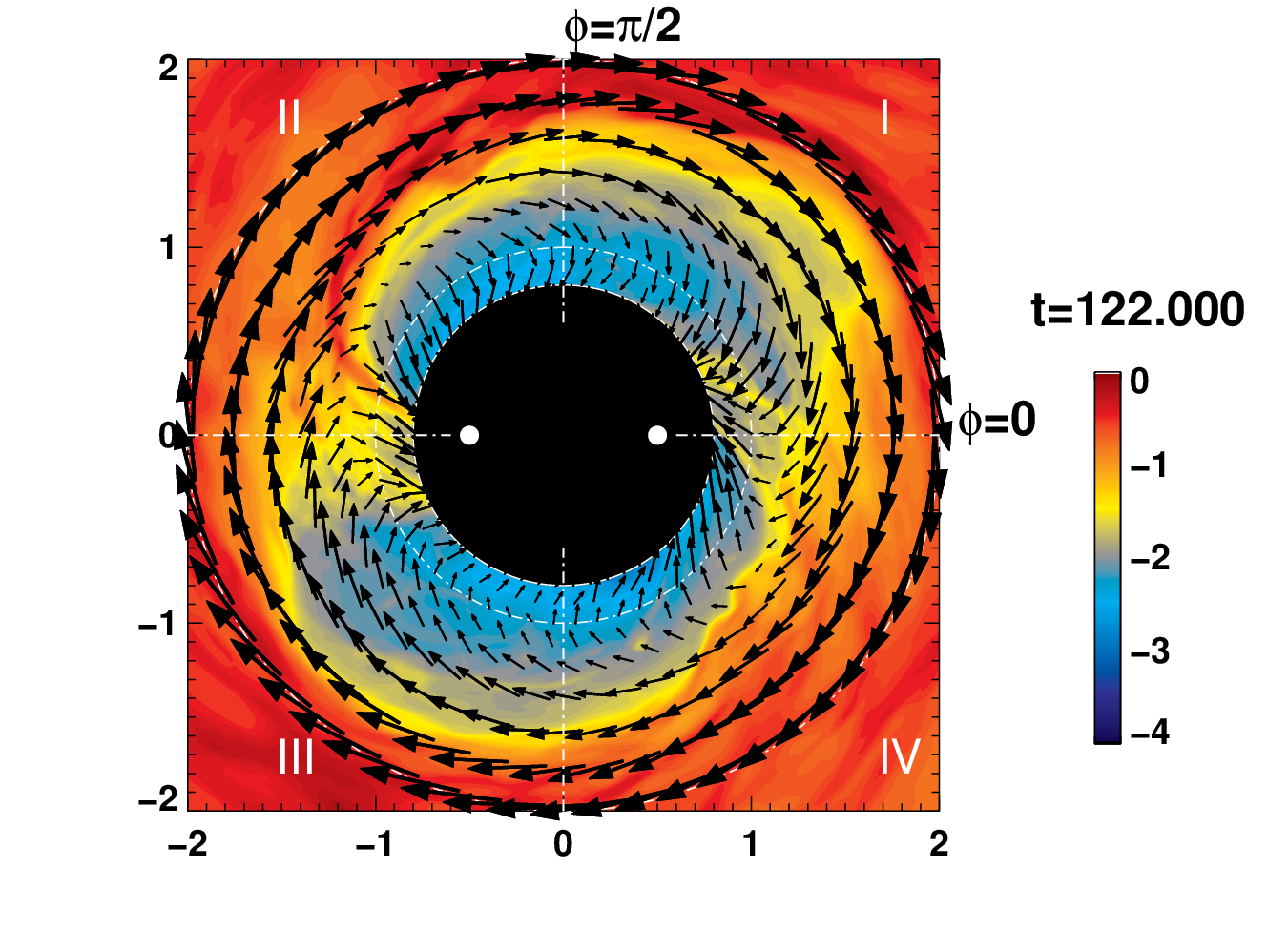}{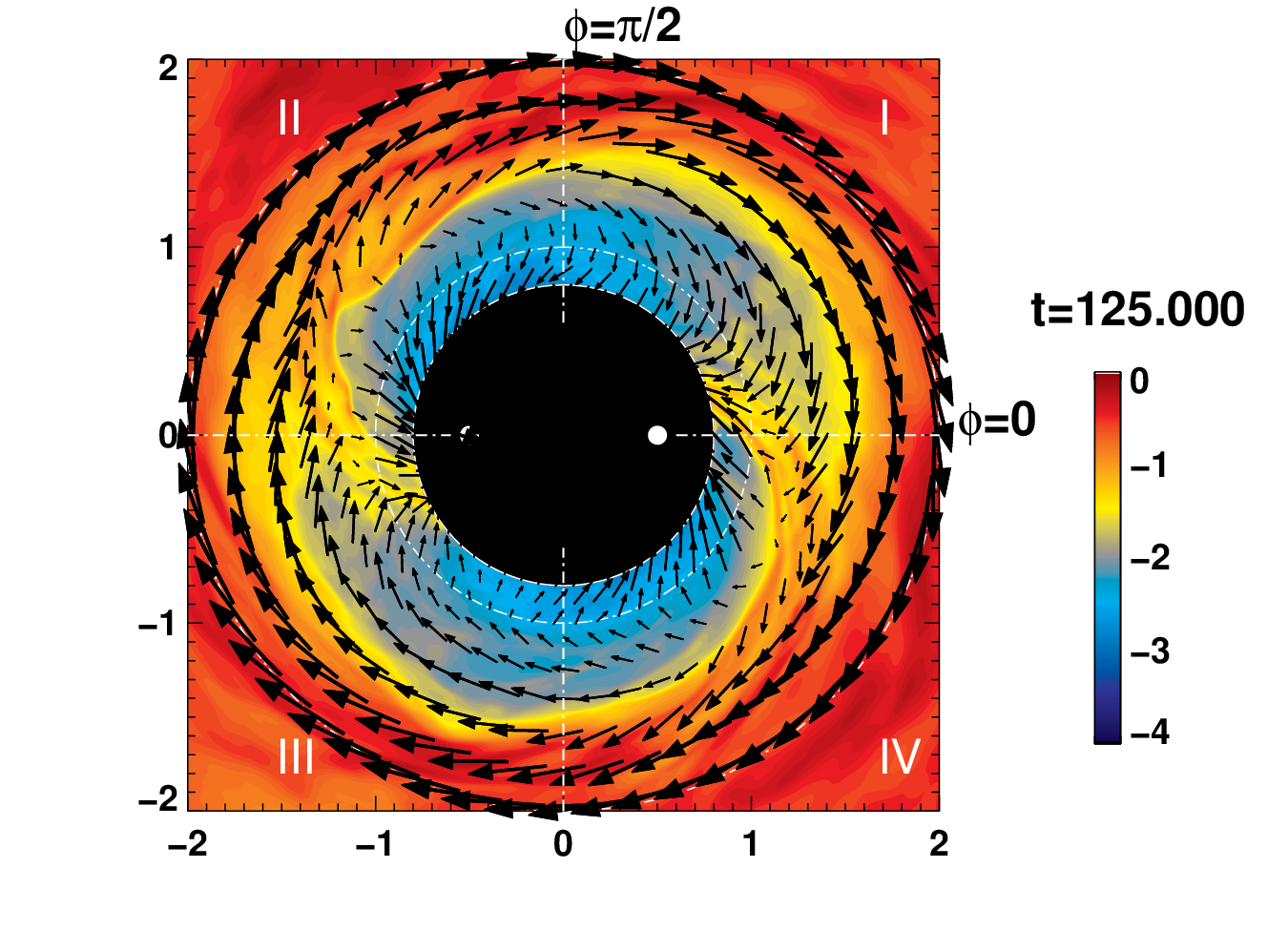}
\epsscale{1.2}\plottwo{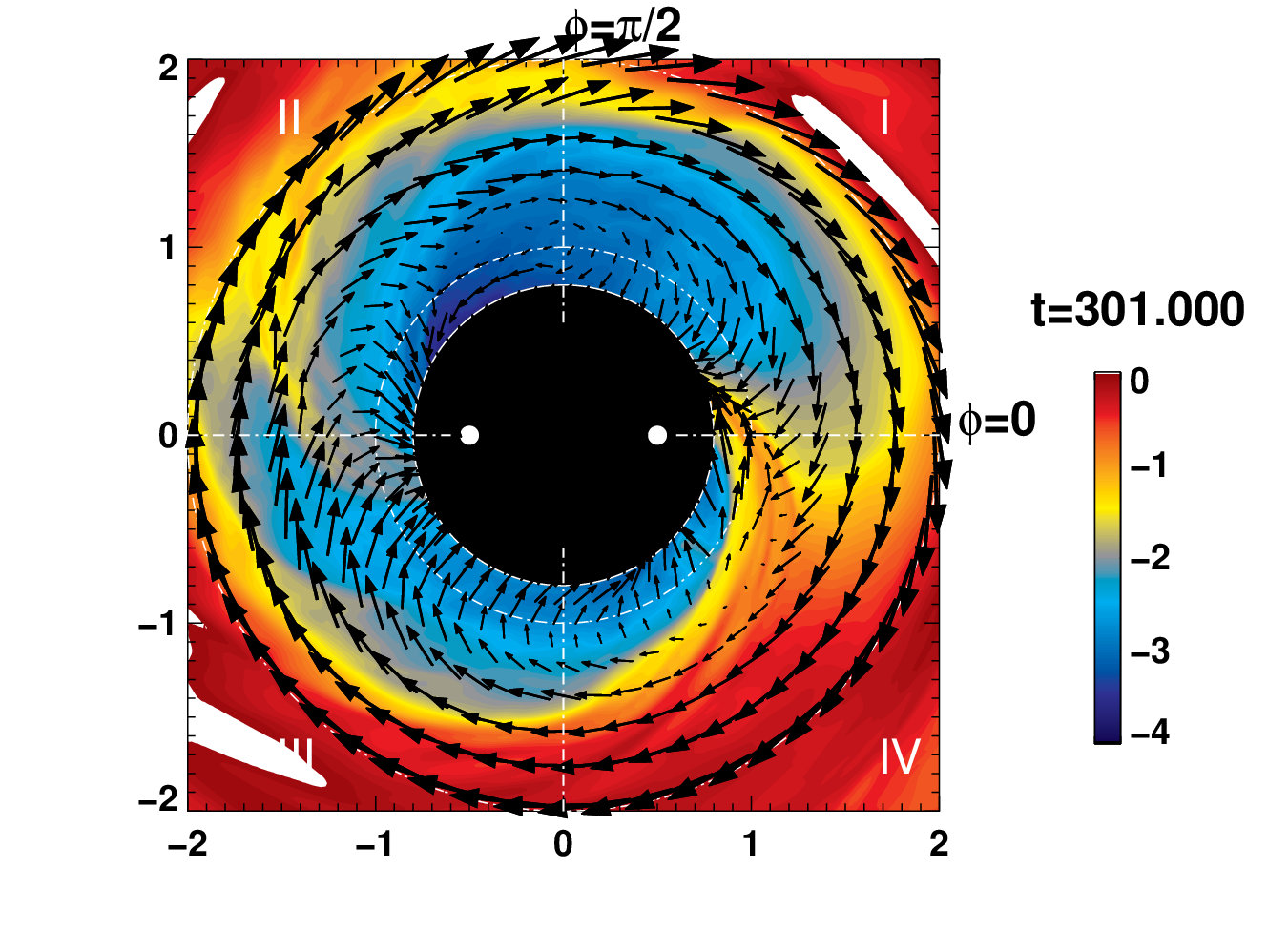}{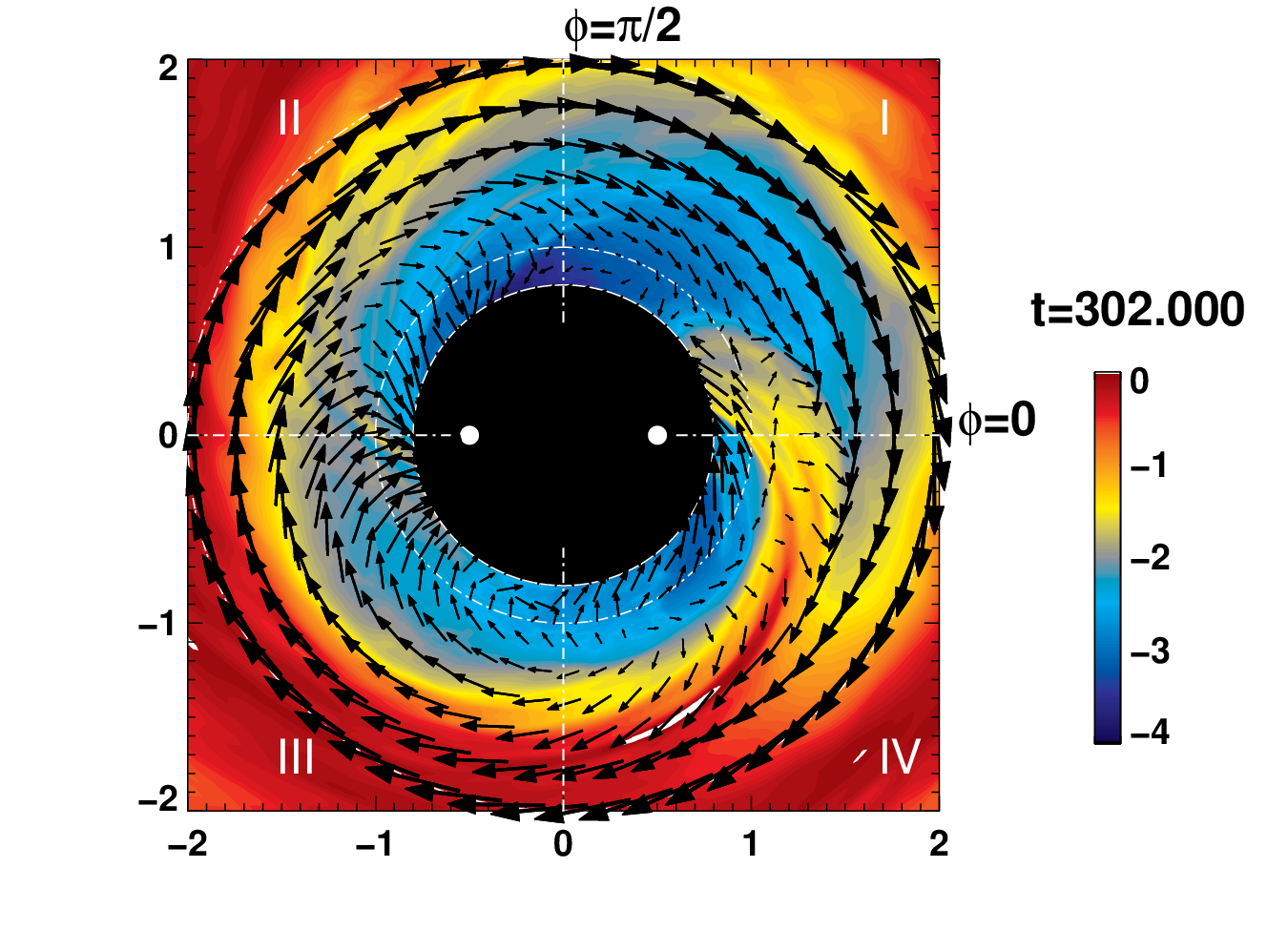}
\epsscale{1.2}\plottwo{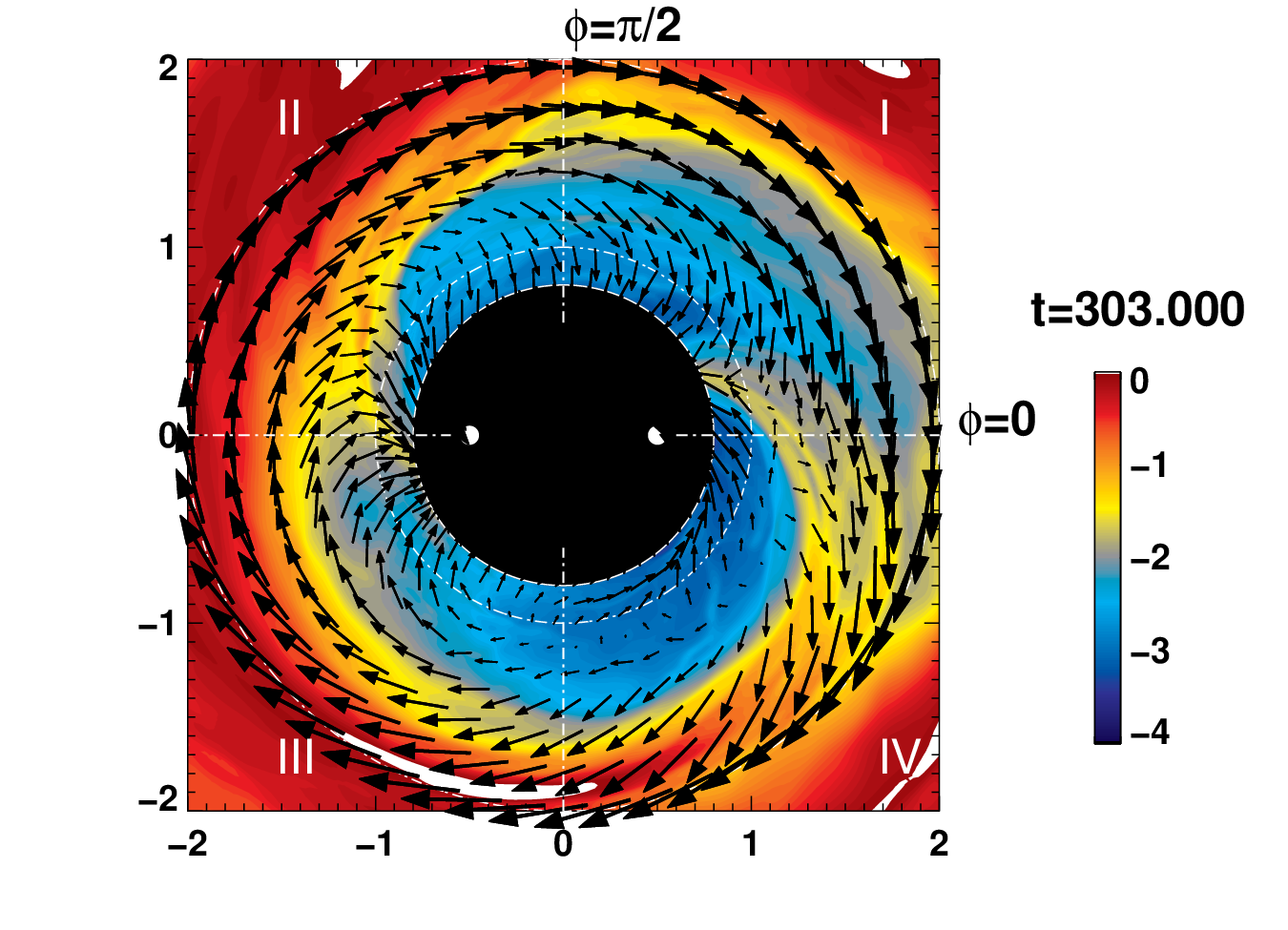}{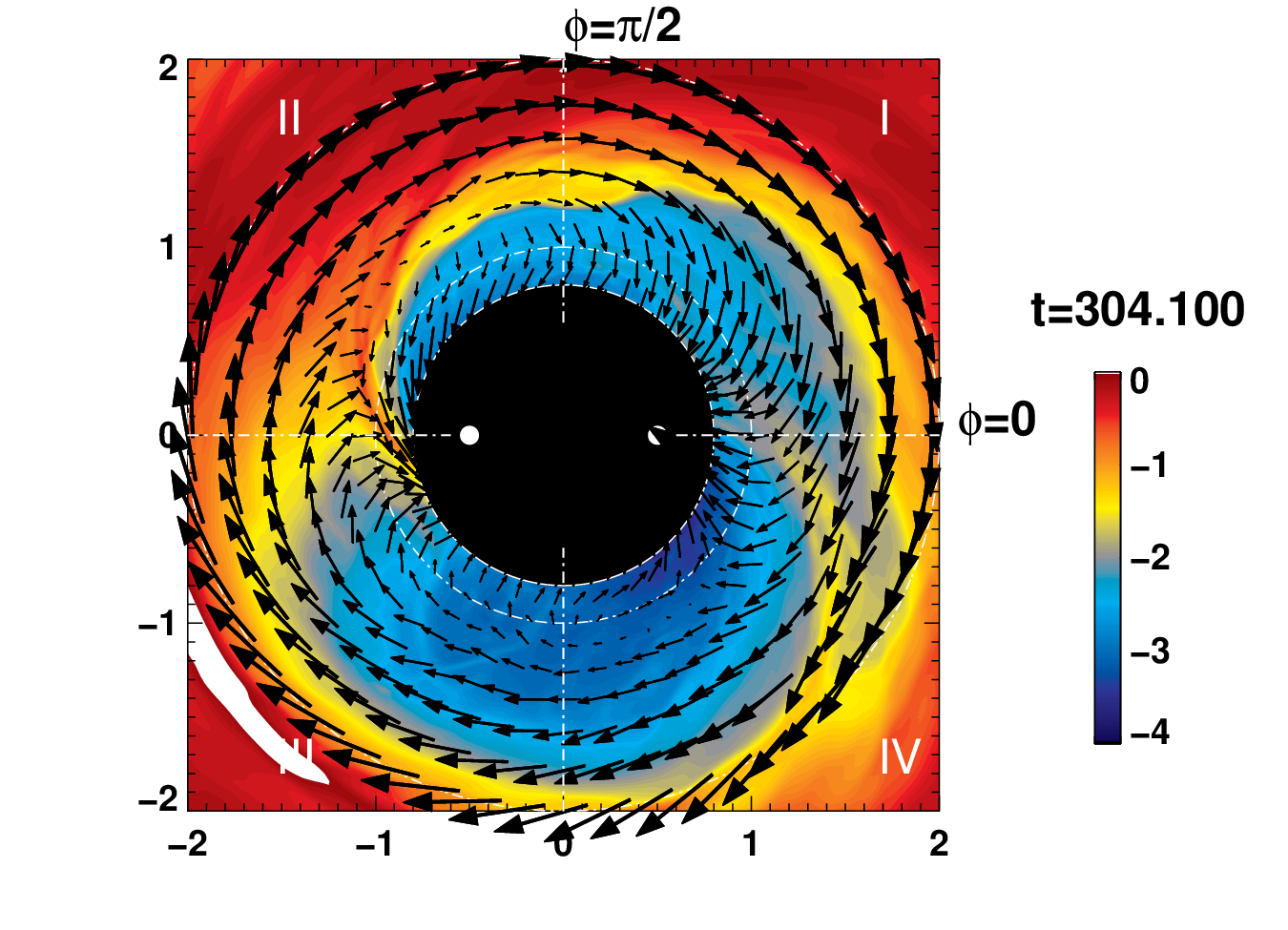}
\caption{\small{Three pairs of snapshots of the surface density in the binary's co-rotating
frame in logarithmic scale (white regions are off the scale).  
The density-weighted vertically averaged velocity is shown by black arrows.  
To show how phase-dependent stream effects change secularly over time, the time separation
between the right and left panels in each pair is a fraction of a binary orbit, while the intervals
from the first pair to the rest are many orbits.  The Roman numerals show the quadrants referred
in section~\ref{sec:stream}. 
}}\label{fig:stream}
\end{figure}

In a frame that co-rotates with the binary, we find the streams follow roughly the static
bi-symmetric potential of the binary and pass through the inner boundary about $0.3$--$0.4$
radians beyond the Lagrange $L2$ and $L3$ points (see Figure~\ref{fig:stream}). This phase
offset is consistent with the negative average torque at $r\lesssim a$ in \S~\ref{sec:torq}.
Early in the simulation, the two streams have roughly equal strength, i.e., similar density
contrast between the stream and the nearby gap region.  We show the surface density near
the inner edge at $t=122$ and $125$ in the first two panels of Figure~\ref{fig:stream}, and
draw the velocity field vectors measured in the co-moving frame on top of it. The snapshots
at $t=122$ show the inflowing gas is regulated by the binary torque as it leaves the inner disk
edge and streams towards the saddle
points. The sign of the angular speed near the inner boundary alternates from one quadrant to the
next, a direct signal that the binary torque changes its sign from one quadrant to the other.
In the co-rotating frame of the binary, the gas in the second (near $\phi=\pi$) and
fourth (near $\phi=0$) quadrants is pulled toward the closer component of the binary
(see $t=122$ snapshots). The tangential velocity results in a Coriolis force,
which tends to move matter away from the center. It takes about half the binary period
for the gas streams in those two quadrants to be kicked out and strongly interact with the
disk edge. As shown in the later snapshots at $t=125$, the velocity is mostly outward
at $r\sim 1.5a$ in those two quadrants, and there are strong
density enhancements near the regions where the velocity gradient is large.
There is also an equivalent way to think about the transition between $t=122$ and $t=125$. 
In the inertial frame, the binary almost always rotates faster than the disk; therefore,
the second and fourth quadrants have positive binary torque. As a result, gas in these
regions gains angular momentum from the binary and then drifts away from the center.

However, the disk inner edge changes its morphology once the eccentricity of the disk
becomes significant. Instead of a pair of streams with comparable size and strength, we find,
for instance at $t=301$, the stream on the right dominates the left. For an eccentric disk, the
apocenter side of the disk is far from both members of the binary. Thus, the gas stream is
strongly torqued, and therefore carries more mass, only when one of the members gets close to
the pericenter of the disk.  After half a binary period, the binary members switch their phases
(in the inertial frame) and a strong stream forms around the other member.
We illustrate this process with a time sequence of snapshots at
$t=301$,$302$, $303$ and $304.1$ in the corotating frame.
The stream on the right at $t=301$ is gradually pushed away at
$t=302$ and $303$ and joins the apocenter of the eccentric disk.
Meanwhile, the stream associated with the other member of the binary strengthens as that mass
approaches the pericenter at $t=304.1$.

We also measure the fluid and magnetic effects on the dynamics of the streams.  In the
inner disk and the gap, the force ratio $|\mathbf{f_{g}}+\mathbf{f_{L}}|/|\nabla\Phi|$ is
always $\lesssim 1/3$, where $\mathbf{f_{g}}=-\nabla P/\rho$ is the force density of the gas
pressure gradient and $\mathbf{f_{L}}=(\nabla\times\mathbf{B})\times\nabla\mathbf{B}/4\pi\rho$
is the Lorentz force density.  Because gas pressure and magnetic forces are smaller than
the gravitational force, the trajectory of the gas is essentially ballistic.  Consequently,
we neglect the fluid and magnetic effects when calculating the interaction between the streams
and the disk inner edge in \S~\ref{sec:discussion_mechanism1}. 

\subsection{Field Structure\label{sec:bfield}}
\begin{figure}
\epsscale{0.7}
\plotone{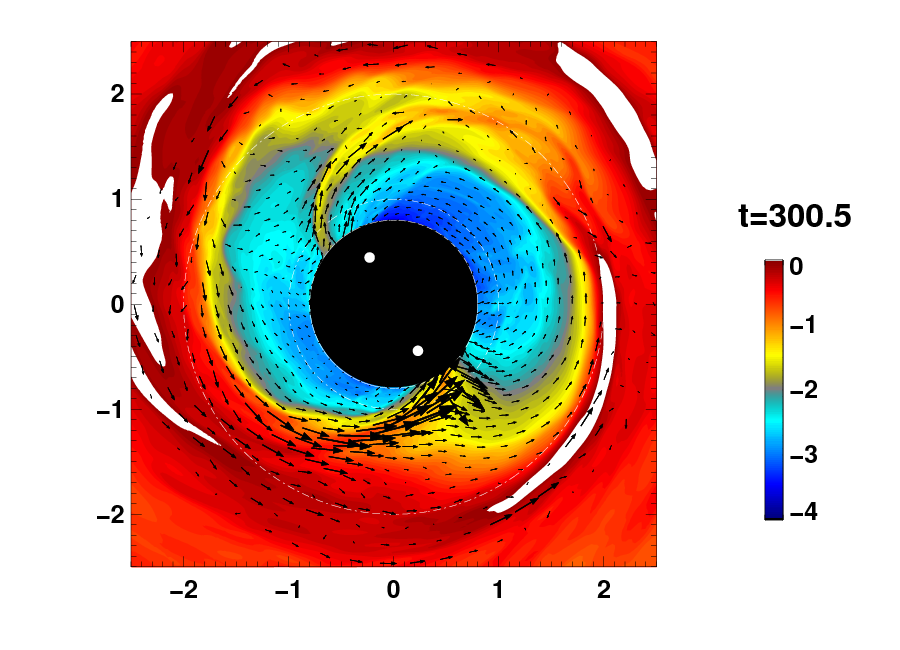}
\caption{\small{Vertically-averaged horizontal magnetic field superposed on surface density contours at
$t=300.5$.  Color surface density contours are logarithmic (see color bar), white are peak values
that is off the scale.
}}\label{fig:field}
\end{figure}

We found in the previous subsection that the Maxwell stress closely follows the gas streams in
the $x$--$y$ plane and is well confined near to the midplane. Now we show the
field structure in the gap region, especially in the stream, has similar features.
In Figure~\ref{fig:field}, we draw the vertical averaged magnetic field as vectors on
top of the logarithmic scaled surface density at $t=300.5$. The location and length of the
arrows show the position and relative strength of the field. Clearly, the field is well
collimated by the motion of the gas streams, with stronger inward field lines in the lower
stream (bottom half plane) and weaker outward field in the upper stream (top half plane).
The ordered field in the gap therefore produces greater Maxwell stress only within the
two streams.

\subsection{Time Dependent Accretion\label{sec:accretion}}

\begin{figure}
\epsscale{0.8}\plotone{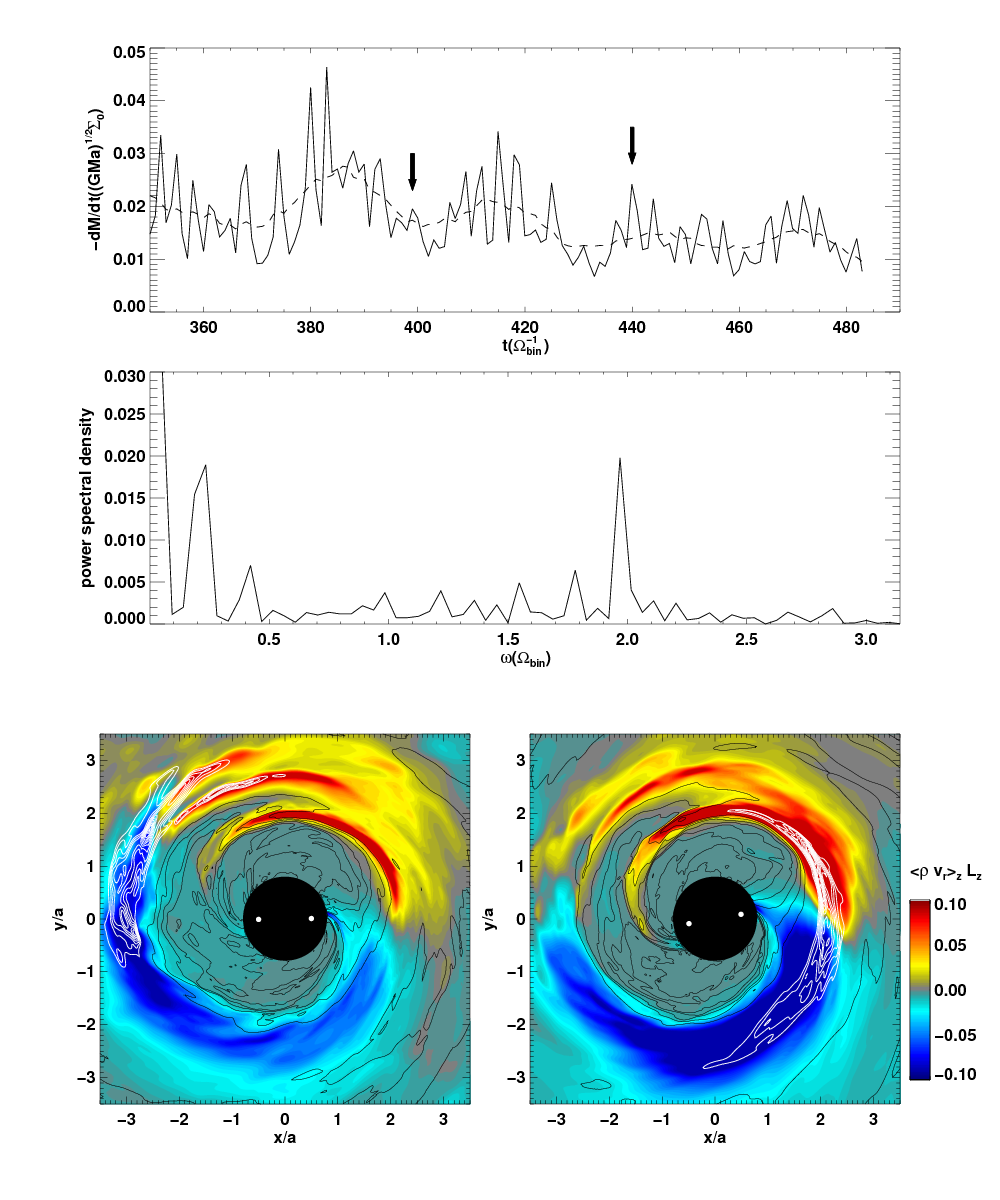}
\caption{\small{Top: Accretion rate as a function of time (in units of the binary orbital period
divided by $2\pi$) during the later stages of the simulation.
Middle: Power spectral density of the accretion rate as a function of frequency in units of
the binary orbital angular frequency $\Omega_{\rm bin}$, or $2\pi$ times the binary frequency.
Bottom: Snapshots of radial mass flux $\lan \rho v_r\ran_z L_z$ at the times shown by the arrows in
the top panel.  Color contours are in a linear scale (see color bar).  Contour lines represent
the surface density in two groups: black contours show low surface density
($10^{-4} < \Sigma < 10^{-0.5}$) on a logarithmic scale in order to highlight the streams;
white contours show high surface density ($1.4 < \Sigma < 3.0$ on a linear scale in order
to highlight the lump).}}\label{fig:mdot}
\end{figure}

We present the accretion rate through the inner boundary as a function of time in the top panel
of Figure~\ref{fig:mdot}. The sampling rate of the time sequence is determined by the output
rate into the history files, which was once per $\omgb^{-1}$.  There are high
frequency outbursts (the narrow spikes in the time sequence) and also a low frequency
modulation (the dashed curve, which is derived by smoothing the time sequence using a boxcar
average with width $15$ time units). These two time variations are the most significant
modes in the temporal profile of $\dot{M}$. By constructing the power spectral density of the
accretion rate using Fourier transforms (middle panel, Figure~\ref{fig:mdot}), we are able to
identify them: The higher frequency mode is at about twice the binary rotation rate, and it is
the same rate at which the streams are pulled inward by the binary
potential; the lower frequency mode is $\sim 0.2\omgb$, which is the same as the orbital
frequency of the lump during the later stages of the simulation (as shown in
\S~\ref{sec:ecc}). 
Our result in Figure~\ref{fig:mdot} is largely consistent with previous findings: a dominant
frequency associated with the binary orbital frequency
\citep[$2\omgb$ if $q=1$; MM08;][]{haya08,cuadra09,roe11,sesana11}, a low frequency component
due to the motion of the dense part of the disk \citep[MM08;][]{roe11}, and another component
(in our case at $1.8\Omega_{\rm bin}$) created by a beat between the binary orbital frequency
and the disk orbital frequency \citep[]{roe11}.  The principal contrast between our
results and this earlier work is that the dominant frequency is $2\Omega_{\rm bin}$ rather than
$\Omega_{\rm bin}$ because the members of the binary have exactly equal mass.

To further verify the cause of these two modes, we pick out two times ($t=400$ and
$441$ as marked in the top panel by black arrows) and plot their two-dimensional local
accretion rates in the inertial frame in the bottom panel of the same graph.  There were
accretion outbursts at both times, but the former is in the valley of a low
frequency oscillation, while the latter is at the crest. In the snapshots, the color
contours represent the vertically integrated accretion rate. The black contour lines show
the surface density in a logarithmic scale from $10^{-4}$ to $10^{-0.5}$.  These lines
pick out the location of the streams, while the white contour lines show the surface
density between $1.4$ to $3.0$ in a linear scale; these contours delineate the lump region.
We find the outbursts of $\dot{M}$ are indeed coming from the infalling streams
located in the fourth quadrants of both snapshots. However, the local accretion rate in the
stream from the left panel is much weaker than that from the right. In the right plot, the
lump is passing the pericenter region. The binary then pulls a gas stream with relatively
higher density than other times. Thus we conclude that the time dependent accretion shows
both high frequency outbursts corresponding to the periodic inflow from the streams
and low frequency modulation due to the orbiting lump. The low frequency mode becomes
important only when the lump forms and maintains itself after about $t=350$.

The time averaged accretion rate at the inner edge, normalized with the peak density, is
$\sim 0.028\sqrt{GMa}\sigmap$, a factor $\sim 40$ greater than found in MM08. 
Interestingly, during the quasi-steady period, the accretion rate in the gap is nearly
twice the rate measured at $\rp$, where the surface density peaks. This contrast
explains the slow mass depletion inside the gap shown in Figure~\ref{fig:mdisk}.
One of the most important questions about accretion onto a binary is the ratio of the
accretion rate accepted by the binary to the rate at which mass is fed into the disk
from the outside.   If the latter rate can be estimated by the {\it maximum} accretion
rate in the disk, we find that accretion through the inner boundary of the simulation,
and therefore presumably onto the binary, is about 1/3 of the accretion rate supplied.
This ratio is a factor of two greater than previously found by MM08.  However, in
evaluating these numbers one should be aware that the disk as a whole never reaches
equilibrium over the course of the simulation.  That the time-averaged accretion rate
as a function of radius is far from constant is a symptom of this fact.  Thus, our estimate
(and for the same reasons, that of MM08 as well) should be taken as no better than an
order of magnitude estimate of this fraction. We plan a more rigorous approach to this
problem in future work.
Lastly, we note that the effects of
the excision on the accretion rate are not explored in this MHD simulation.  However, our
hydrodynamic simulations with different cut-outs suggest that the rate hardly changes
as long as the cut-out size is smaller than $\lesssim a$.

\section{DISCUSSION\label{sec:discussion}}

%

\subsection{Binary Torque and Linear Resonance Theory \label{sec:discussion_torq}}

Using the results we have on the disk stresses and the binary torque, we want to further
elucidate the angular momentum transport mechanisms within a MHD turbulent
circumbinary disk. The binary strongly torques the surrounding gas in the gap as the gas
forms streams; the outward-moving portions of the streams deliver that angular momentum to the
inner part of the circumbinary disk.  By contrast, the torque directly exerted on the disk
body is much weaker and diminishes rapidly toward larger distance.  In that sense, we can
say that the binary mainly torques the gas in the low density gap; moreover, the magnitude
of that torque is proportional to the mass of gas in the gap.
This fact suggests that the binary torque is very sensitive to the basic
physics coupling the streams to the disk such as the stresses and the equation of state. For
instance, scaled with the peak surface density, our MHD turbulent disk
provides a factor of $14$ times stronger binary torque than that of MM08.

This fact also leads to a strong contrast with the commonly-used linear resonance theory of
interaction between disks and planets or other orbiting satelites \citep{gt82}.  To illustrate
that contrast, we calculate both the torque density and total torque as functions of radius,
assuming most of the torque comes from the $\omgb:\Omega=3:2$ outer Lindblad
resonance at $r_2/a=(3/2)^{2/3}\simeq 1.3$ \citep[e.g.,][]{ms87}. The estimate is shown in the
top panel of Figure~\ref{fig:torq}, with the red lines representing $1/4$ of the predicted values.
The linear theory prediction must be reduced by roughly that amount to match the measured total
torque.  In addition to having a larger amplitude than the simulation found, the linear theory
also predicts a much shorter wavelength for its radial oscillations.

Three distinctions between the circumstances of the MHD circumbinary disk we simulate and
the assumptions made in linear theory may explain these contrasts.
The first is the surface density profile. Standard linear theory assumes that any surface density
gradients have length scales much longer than the wavelength of the density wave excited by the
resonance.  However, the surface density profile in the gap is steep
enough that $(d\ln\Sigma/dr)^{-1} \lesssim 0.3 a$, which is comparable to the 
first wavelength
$5[(m+1)/(3m^2)]^{1/3}(H/r)^{2/3}a\sim 0.5 a$, where $H/r$ is evaluated at the resonance radius,
and $m=2$ is used here (see Figure~1 and the pressure-dominated case in Table~$1$ of \citet[][]{ms87}). 
Because the surface density at $r_2$ is so much smaller than at $r\sim 2$--$3a$, 
one might expect the maximum torque to move away from the resonance point toward the inner disk.
Secondly, linear theory assumes nearly circular orbits because it also assumes $q \ll 1$. However,
in our simulation $q=1$, and, largely for this reason, the orbits of gas within the inner edge of
the disk body are significantly non-Keplerian: they form streams.  Because the torque depends
so strongly on whatever gas mass is in the gap, this distinction is qualitatively important. 
Lastly, as the mass ratio of the binary of our simulation is unity, the forcing term, which is roughly
controlled by the ratio of the perturbing gravitational potential to the sound speed,
becomes large enough to drive nonlinear density disturbances. 
The longer wavelength oscillation found in our torque density might be partly explained
as a result of these nonlinear effects \citep{yc94}.

\subsection{Binary Orbital Evolution\label{sec:discussion_shrink}}
The time-averaged rate at which the binary loses angular momentum to the disk is
$T(\infty) \simeq 0.012 GMa\Sigma_0$. Meanwhile, the time-averaged angular
momentum flux across the inner boundary due to accretion is $\dot{M} j_{\rm in} \simeq 0.017
GMa\Sigma_0$, where $j_{\rm in} \simeq 0.94 \sqrt{GMa}$ is the averaged specific angular momentum at
$r_{\rm in}$.   On average, therefore, the binary actually gains angular momentum in net.

However, even though the binary gains angular momentum, it does not automatically follow
that the binary separation grows.  The evolution of the binary also depends on its growth in mass.
If the orbit remains circular, the shrinkage rate for an equal mass binary is
\beq
\frac{\dot a}{a} = 2\frac{\dot J}{J} - 3 \frac{\dot M}{M},
\label{eq:adota}
\enq
where $J=M j_{\rm bin} = 0.25 M\sqrt{GMa}$ denotes the angular momentum of the binary, and its
rate of change is $\dot{J}=\dot{M} j_{\rm in} - T(\infty)$.
 
After separating the accretion term of $\dot{J}$ and combining it with the second term in
equation~(\ref{eq:adota}), we have
\beq
\frac{\dot a}{a} = -2\frac{T(\infty)}{J} + 2 \frac{\dot{M}j_{\rm in}}{J}\left( 1-\frac{3}{2}\frac{j_{\rm
bin}}{j_{\rm in}}\right).
\label{eq:adota2}
\enq
If $j_{\rm in}/j_{\rm bin}>3/2$ (here it is $\simeq 3.75$), the shrinkage rate is controlled by the
competition between the first and second terms. Plugging in our numbers, we find the two terms are
surprisingly comparable in magnitude:
$-0.096\sqrt{Ga/M}\Sigma_0$ and $0.082\sqrt{Ga/M}\Sigma_0$ respectively. 
The net result is orbital compression, but at a rate much smaller than either of the two terms, 
$\dot{a}/a\simeq -0.014 \sqrt{Ga/M}\Sigma_0$.

However, it is also possible that interaction with the disk may induce eccentricity.  If so,
an extra term on the right hand side of equation~(\ref{eq:adota}) should be included, making it
\beq
\frac{\dot a}{a} = 2\frac{\dot J}{J} - 3 \frac{\dot M}{M} +
     \left(\frac{2\ebin^2}{1-\ebin^2}\right)\frac{\dot{e}_{\rm bin}}{\ebin},
\label{eq:adota_ecc}
\enq
where $\ebin$ denotes the binary's orbital eccentricity.  In addition, the angular momentum
of the binary should also be adjusted to account for the eccentricity:
$J = M j_{\rm bin}\sqrt{1-\ebin^2}$.  In Appendix~\ref{app0}, we show that the rate of eccentricity
growth depends on how the accretion flow interacts with the interior disk of each individual binary
member, an interaction our simulation did not treat.  However, we also show that the best estimate
we can make from the data we do have is that there is little evidence for any significant
${\dot e}_{\rm bin}/\ebin$.  Appendix~\ref{app0} also briefly discusses the possibility of unstable
eccentricity growth raised by linear theory; unfortunately, we can say little about it
because our simulation data do not bear on it, and linear theory may not apply in this regime. 
%
%
%
If, as we did in $\S$~\ref{sec:torq}, we define a characteristic disk mass
$\md = \pi \sigmap (3a)^2$, and further assume that the binary orbit remains circular,
then $\dot{a}/a\simeq - 8.0 \times 10^{-4} ({\md}/{M})\omgb$. 
For comparison, MM08 found a rate $\simeq - 0.003 \sqrt{Ga/M}{\Sigma_0}$, which can be scaled
used the definition of $\md$ to $\simeq - 3 \times 10^{-4} ({\md}/{M})\omgb$, somewhat
slower than ours. 
In other words, although we find a torque an order of magnitude larger than that of an
$\alpha$--viscosity hydrodynamic model, we also find an accretion rate so much larger that
the net effect on the binary orbit is reduced to an increase by only a factor of 2.7.

Another way to understand this result is to note that our $T(\infty)/\dot M \simeq 0.67 (GMa)^{1/2}$,
a factor of $3$ smaller than in MM08.  That is, the greater internal stresses produced by MHD effects,
especially in the gap region, both introduce more mass to that region (thereby magnifying the torque)
{\it and} lead to even greater accretion, which returns angular momentum to the binary.
Unfortunately, because the cancellation is so close, our calculation cannot be definitive in
regard to the shrinkage rate of comparable-mass binaries in Nature.  Further uncertainty comes
from the fact that we cannot track the gas flowing into the cutoff region with the present model,
so we do not know exactly how much energy and angular momentum accretion might bring to the binary,
yet our sizable accretion rate makes these contributions significant.  The net outcome in any
particular case may therefore depend on the specific details (gas equation of state, binary mass ratio,
etc.), as all of these can influence $j_{\rm in}$, $\dot{M}$, $T(\infty)$, as well as the detailed
mechanics of how the stream joins the interior disks associated with the members of the binary.

\subsection{Disk Eccentricity Growth \label{sec:discussion_ecc}}

\subsubsection{Eccentricity Distribution \label{sec:edistr}}

The distribution of eccentricity in radius plays an important role in the 
evolution of the disk. For the eccentricity to grow substantially
over time in a large circumbinary disk, it needs to be confined in radius.
Otherwise, the energy and angular momentum (more precisely, angular momentum
deficit) associated with an eccentric disturbance are spread over a large radius,
resulting in a small eccentricity in any one location.
We consider here a linear model for the eccentricity distribution and compare
it with the results of the simulations. 

The linear equation governing the eccentricity distribution and evolution is given by
\citet[][hereafter, GO06]{go06}. 
This also allows generation and damping of the disk eccentricity. 
The basic equation and boundary conditions are given in Appendix \ref{app1}. 
For our circumbinary disk, we adopt input parameters taken from the simulations while
the eccentricity was small and growing exponentially, i.e., the linear regime.
We define the late stage of the exponential growth phase as the period
$t=200-250$ time units (see Figure \ref{fig:ms}).  Surface density $\Sigma(r)$ is
taken directly from the simulation by averaging over this time interval. 
The disk inner radius is taken to be $r_{\rm i}=2 a$ and the outer radius
$r_{\rm o}=10 a$.  We assume the same equation of state as in the simulation: isothermal,
with sound speed $0.05 \omgb a$. The adiabatic index is taken to be $\gamma = 1- 0.1 i$.
The small imaginary part (which introduces slow damping) assures mathematical consistency.
We model the forces driving the initiation of eccentricty as a Gaussian in radius
(see equation~(\ref{sinj})), centered on $r_{\rm c} = 2.1a$ and having a width $w=0.05 a$.
These parameters are not well constrained by the simulation.  It follows that the
the results are also insensitive to their values within certain limits.
Parameter $s_{\rm c} $, the peak value of the eccentricity injection distribution in equation (\ref{sinj}),
is determined by the condition that the growth rate equal that in the simulation,
$\gamma_{\rm e} = -\rm{Im}(\omega) \simeq 0.018 \omgb$, where $\omega$ is the eigenvalue defined
in equation~(\ref{omeq}). The value is found to be $s_{\rm c}=0.0145 a \omgb$.  

\begin{figure}
\plotone{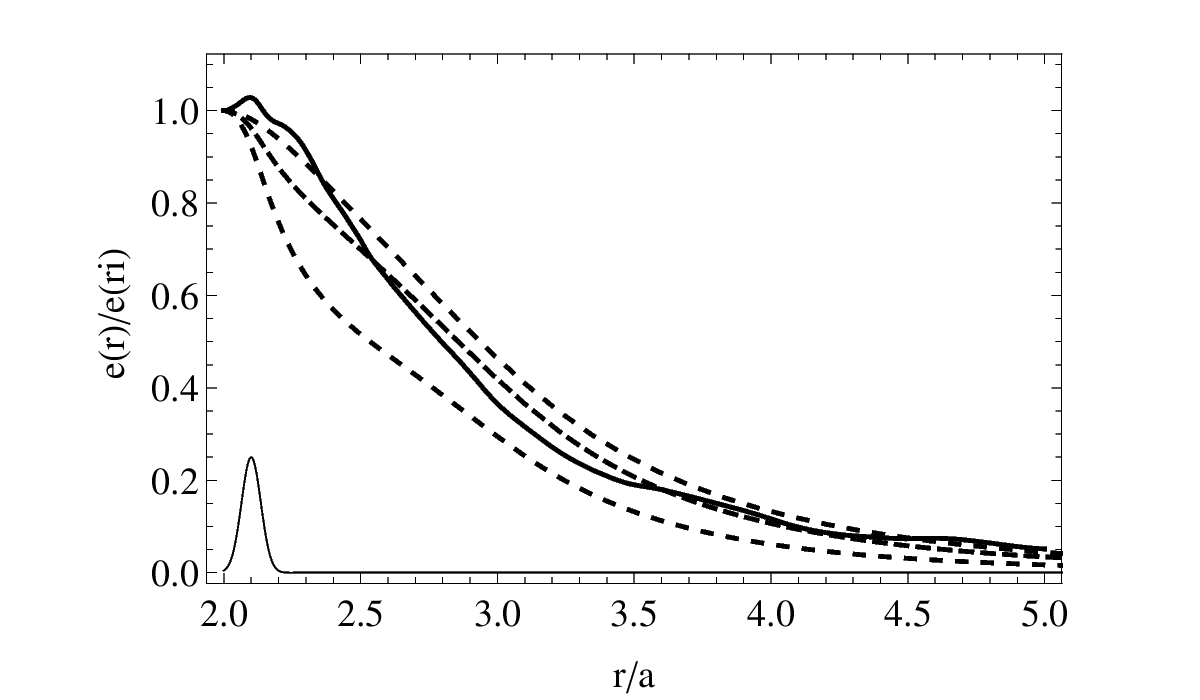} 
\caption{\small{Disk eccentricity as a function of radius from $r=2a$ to $r=5a$, normalized
by its value at the inner edge.  The heavy solid line is the eccentricity distribution obtained
from the simulations when the eccentricity evolution is in the linear regime. The middle dashed
line is the eccentricity distribution determined by linear theory (equation~(\ref{Eeq}),
with the input parameters appropriate for the simulation (see parameter details in
\S~\ref{sec:edistr}). The thin solid line is the normalized eccentricity injection rate distribution,
$0.25 s(r)/s(r_{\rm c})$.  The upper dashed curve is the eccentricity distribution for the fundamental
free eccentric mode. The lower dashed curve shows the result of a higher eccentricity growth rate,
about twice as high as in the middle dashed curve. 
}}
\label{fig:edistr}
\end{figure}

The physically relevant solution on long timescales corresponds to
the fastest growing mode. This solution was determined by the methods
described in \citet[][hereafter L10]{lubow10}.  
Figure~\ref{fig:edistr} plots the eccentricity distribution for this mode,
along with the eccentricity distribution obtained from the simulation.
The middle dashed line corresponds to the eccentricity distribution based on linear theory
with the set of parameters described above. The distribution matches well the one
obtained from the simulation (the solid line).  This distribution closely
resembles the distribution for the fundamental free precession mode in the disk
(the uppermost dashed line).  That mode has an eccentricity injection of zero, $s_{\rm c}=0$.
The confinement of eccentricity is due to the effects of differential precession of the binary
$\partial_r \dot{\varpi}_{g}$ (see equation~(\ref{prec_g}) in Appendix~\ref{app1}), while
pressure effects act to spread the mode. Differential precession effectively creates a
potential well that leads to trapping of the fundamental mode. 
Further localization is due to the injection of eccentricity, which is concentrated near the
disk inner edge (solid curve on lower left).  The faster eccentricity is injected into a small
region, the more the eccentricity distribution narrows.  For somewhat higher eccentricity injection
rates, the distribution is further confined (lowest dashed curve). Therefore, we see that the
eccentricity distribution is a consequence of the excitation of the fundamental free eccentric
mode in the circumbinary disk and that the eccentricity is confined within a radial width $\sim a$. 

The excitation of the fundamental mode occurs because that mode overlaps well with the eccentricity
injection distribution.  As discussed above, the location and width of the eccentricity injection
are not well determined by the simulations.
On the other hand, the linear equation has solutions that match the simulation results only if
$r_{\rm c}\lesssim 2.6a$, which indicates that the source of the eccentricity should be located near
the inner edge of our MHD disk.  
For larger values of $r_{\rm c}$, there is no value of $s_{\rm c}$ that could provide the growth
rate found in the simulation for an eccentricity distribution that resembles the simulated one.
This radius limit has a weak dependence on the width of the injected eccentricity
$w$. If the width is doubled to $0.1 a$, then the limiting radius increases by about 1\%.

The outer regions of the simulation exhibit an exponential falloff of eccentricity
with an e-folding length of about $0.8 a$ (see Figure~\ref{fig:ecc_distr} in
$\S$~\ref{sec:ecc_evolution}).  In Appendix \ref{app2}, we derive an asymptotic form for
the eccentricity distribution (valid for $r \gg a$) using linear theory. 
It is dominated by an exponential falloff in $r$ with an e-folding length also of about 0.8a.

To sum up the analysis in this subsection, we find the linear model can explain the eccentricity
distribution of our simulation. It shows the confinement of the disk eccentricity is mainly due
to the disk precession and eccentricity growth. The linear model also puts constraints on the
location of the source for eccentricity growth. Finally, it explains the exponential falloff of 
eccentricity profile.

\subsubsection{Eccentricity Growth Through Tidal Instability \label{sec:egrunst}}

One possible mechanism for eccentricity growth is a tidal instability.
This instability is believed to explain the superhump phenomenon
in cataclysmic binaries \citep[see][]{os96}. The instability involves the action
of the 3:1 eccentric Lindblad resonance in a disk that surrounds a star \citep[][hereafter
L91]{lubow91b}. As pointed out in L91, this instability can also be at work in circumbinary disks
at radial locations that satisfy the condition
\begin{equation}
\Omega(r) = \frac{m \omgb}{m+2},
\end{equation}
where $m$ is an integer that corresponds to the azimuthal wavenumber
of the tidal component of the potential involved in the instability.
In the case analyzed here of an equal mass binary, $m=1$ is excluded because
that component of the tidal potential is absent. For $m=2$, the instability
is associated with the $\omgb:\Omega = 2:1$ resonance that occurs at $r \simeq 1.6 a$.
In this model, the $m=2$ tidal field  interacts with
the $m=1$ eccentric motions of the disk to produce disturbances of
the form $\exp{( 3 i \phi - 2 i \omgb t)}$. These disturbances launch waves
of that form at the $2:1$ resonance. The waves in turn interact with the
tidal field to produce a stress that increases the disk eccentricity exponentially in time.

However, the $2:1$ resonance lies at $r \simeq 1.6a$, inside the circumbinary disk gap.
Furthermore, the gas motions are far from circular there.  Even if the gas followed circular
orbits in this region, the model of L91 predicts in this case an eccentricity growth rate
$\sim  a \omgb/ w_{\rm e}$, where $w_{\rm e}$ is the width of the eccentricity distribution
and is $\sim a$. This very rapid growth rate is a consequence of the powerful $m=2$
tidal potential for an equal mass binary, the dominant component of the tidal field. However,
this rate is more than an order of magnitude faster than the growth rate measured in the simulation.
It may be possible that the effects of the resonance are still felt, but at a reduced level, near the disk
inner edge as a consequence of finite width of the resonance. 
Thus, tidal instability may play a role in the eccentricity growth, but the evidence
suggests it is at most a limited role.

\subsubsection{Eccentricity Growth through Stream Impact\label{sec:discussion_mechanism1}}

As discussed in $\S$ \ref{sec:ecc_evolution}, the space-time diagram reveals a special feature of
the growth: the propagation of eccentricity from radii close to $\sim 2a$ toward larger
radius. In this section we examine the possibility that the impact of streams in the gap
striking the inner edge of the disk is the mechanism of eccentricity injection into the disk.

During the exponential growth phase of eccentricity, a pair of gas streams are pulled in from the
disk inner edge by the binary, and then flung back to the disk after half binary period. It is easy
to show that if the disk is on a circular orbit and free of eccentricity, the two streams are
not capable of breaking the bisymmetry that is intrinsic for an equal mass binary potential. 
We speculate that a small eccentricity breaks the symmetry by inducing unequal strength stream
pairs (in terms of both the density and the velocity) and/or asynchronous stream-disk interaction.
Once the symmetry of stream impact is broken, stream impact could amplify the initially small
disk eccentricity, in turn increasing the asymmetry of the streams themselves. This process could
then lead to sustained disk eccentricity growth.

We find two sorts of evidence that support this mechanism of eccentricity growth. The first
is based on a special set of simulations we performed.  In these, we eliminated part or all of the
returning streams by enlarging the central cut-off.  If the removal of outward streams halts eccentricity
growth, we may take that as evidence in support of a stream impact origin for eccentricity
excitation.  We do not consider cutoff radii large enough to affect the circular orbit region of the
disk because otherwise, the result might be equally well interpreted in terms of a resonance model
(\S~\ref{sec:egrunst}).  Because these are essentially hydrodynamic effects and involve no vertical
dynamics, these special simulations were performed in 2-d with no magnetic field (see $\S$~\ref{sec:hd_run} and
Table~\ref{tab1} for details of the simulations).

\begin{figure}
\plotone{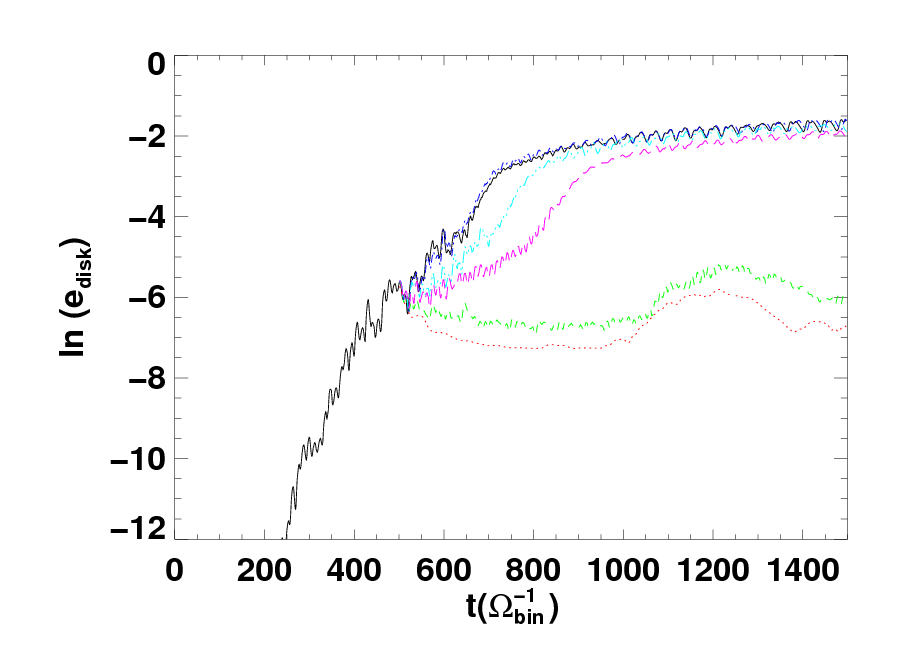}
\caption{\small{The evolution of the disk eccentricity with varying cut-off sizes: $r_{\rm in}=0.8a$
(black solid curve), $1.0a$ (blue dash-dotted curve), $1.1a$ (cyan dash-dot-dotted curve), $1.2a$
(magenta long dashed curve), $1.3a$ (green dashed curve) and $1.7a$ (red dotted curve).  There is
a qualitative change when the cut-off becomes larger than $\simeq 1.2a$}}
\label{fig:rin_ecc}
\end{figure}

In the control run (B2D.rin=0.8), the disk is truncated around $2a$, similar to B3D.
In the gap region, there is a pair of streams at $t=300-700$ while the disk eccentricity grows
exponentially. A single strong stream appears when the disk eccentricity becomes significant.
The simulation is terminated at $t=1500$, and at that time the
disk eccentricity is saturated. We find the stream dynamics and the eccentricity growth in the
control run are very similar to what were found in B3D, demonstrating that hydrodynamic effects
alone can yield eccentricity qualitatively similar to MHD. 
Thus, the results based on hydrodynamic simulations can provide a clear indication of the role of the
streams in eccentricity generation.  We display the growth history of the disk eccentricity in
Figure~\ref{fig:rin_ecc} (black solid).  Here the disk eccentricity is calculated using a two
dimensional definition of equation~(\ref{eq:edisk}). Similar to the MHD simulation, we find
exponential growth from $t\sim 300$--$700$. The growth rate in that period is $\simeq 0.017\omgb$,
very close to that of the MHD simulation ($\simeq0.018\omgb$).
We then restart from $t=500$ with five different values of $r_{\rm in}$ while keeping all other
parameters fixed. We terminate the reruns at $t=1500$, the stopping time for B2D.ri=0.8. 

In Figure~\ref{fig:rin_ecc}, we plot the evolution of disk eccentricities for cases with different $r_{\rm in}$.
The case with $r_{\rm in}=1.0a$ (blue dash-dotted) loses only the tips of the
streams, and its eccentricity evolution closely follows the control run, with
exponential growth continuing without any interruptions until it becomes saturated.
The disks with larger cut-offs ($r_{\rm in}=1.3a$: green dashed curve; $1.7a$: red dotted curve), however,
cease exponential growth of their eccentricities right after the restart because most of the
streams forming in the gap are lost in the hole and never get a chance to interact with the disk edge. 
There are also cases in between.  When $r_{\rm in}=1.1a$ (cyan dash dot dot curve) and $r_{\rm in} = 1.2a$
(magenta long dash curve), the growth rate of the eccentricity diminishes after the restart. It takes
longer times for both disks to reach the eccentricity of the control run.  We attribute these results
to partial loss of outward stream motion.

When $r_{\rm in}=1.3a$, the simulation domain includes the region well inside the gap where gas motions
are noncircular and are dominated by the streams.  In this case, the simulation region also
extends well inside the 2:1 resonance at $r \simeq 1.6 a$ whose presence is required for the mechanism
of \S~\ref{sec:egrunst}.  Consequently, any resonant instability should
not be seriously affected by this inner boundary change. However,
the eccentricity growth rate did change. This evidence then favors the stream impact mechanism.

The other evidence favoring a stream impact origin of the eccentricity comes from directly
measuring the rate of change of eccentricity due to the stream-disk interaction in the MHD simulation.
While the eccentricity grows exponentially, any asymmetry in stream impacts is necessarily small,
so the effects of the gas stream are mild and difficult to directly measure. 
By the time $t=300$, when the eccentricity growth is slow, the inner portion of the disk forms
an eccentric ring between $r\sim 2$--$3a$ with $e\sim 0.08$. 
Its pericenter is at $r\sim 2a$ and $\phi \sim 3\pi/2$, and the apocenter is at $\sim 3a$ and
$\phi\sim \pi/2$ in inertial frame. The eccentric ring slowly precesses around the binary.
In this state there is only a single dominant stream.  In this configuration, the binary
interacts strongly only with the periastron side of the disk. At this stage, the effects of the
stream are strong enough for us to reliably measure its effects on the disk.

The Gauss equations of celestial mechanics \citep{bc61} provide a way to relate the
perturbing effects of the stream to the evolution of disk eccentricity. This method has been
previously used to calculate the effects of an external perturbing force on a ring in the
context of dwarf-nova accretion disks \citep{lubow94}. The evolution of the disk eccentricity is
described by
\beq
\frac{de}{dt} = \frac{\sqrt{1-e^2}}{a_{\rm d} n}\left[R\sin f + S\frac{4\cos f + 3e +e \cos(2f)}{2(1+e\cos
f)}\right ],
\label{eq:dedt}
\enq
where $a_{\rm d}$ is the semi-major axis of the eccentric ring of disk and
$n=\sqrt{GM/a_{\rm d}^3}$ is the mean motion of the disk. Quantity $f$ is the true
anomaly where the stream impact occurs. $R$ and $S$ are the radial and angular components of the
disturbance force density, and are defined as
\beq
(R,S) = \frac{\dot{m_s}(\mathbf{v_s} - \mathbf{u_e})}{m_e},
\label{eq:disturbance}
\enq
where $\dot{m_s}$ denotes the mass injection rate of the stream, $m_e$ is the total mass of the
eccentric ring, $\mathbf{v_s}$ and $\mathbf{u_e}$ are the stream and disk velocities near the
impact, respectively.

Since the impact takes place over a short time, we must use the higher time resolution
simulation ($t=300$--$322$) in order to estimate the disk perturbation.
We carry out our analysis using the vertically-integrated two-dimensional data. 
%
%
For the Gauss equation to apply, the properties of the eccentric ring, such as $e$,
$a_{\rm d}$, $n$, and the longitude of pericenter $\varpi$, need to be nearly constant over the
orbital period of the ring. This assumption holds well because the timescale for the ring
properties to change is much longer. In addition, the mass transferred to the ring by the stream
impact over an orbital period is considerably smaller than the mass of the eccentric ring. We use
time-averaged values for those parameters. The values are: $e=0.08$, $a_{\rm d}=2.5a$ and
$\varpi=1.48$~radians.
The stream-disk impact is localized in space. We analyze the stream-disk impact over an arc that
is defined by $r=2a$ and $\phi\in[-\pi/4, \pi/4]$.  The quantities
$f$ and $\mathbf{v_s}$ are directly measured on the arc by taking the
density weighted averages at each time. The mass injection rate is calculated by integrating the
mass flux along the arc.
The instantaneous velocity of the eccentric edge $\mathbf{u_e}$ is obtained from the unaffected
matter at a slightly greater radius, which we take to be an arc at $r=2.1a$.
In fact, the disk velocity is not very sensitive to the location we choose, so long as it is taken
within the eccentric ring.

\begin{figure}
\plotone{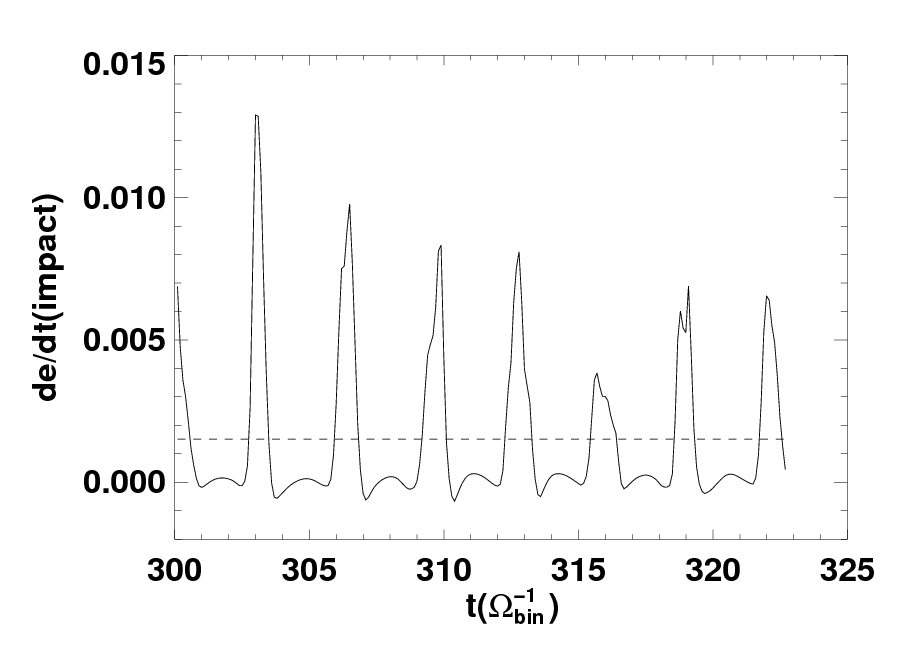}
\caption{\small{Rate of change of the disk eccentricity due to the stream impact.  Solid line
shows the instantaneous rate; dashed line shows the time-averaged rate.
}}\label{fig:impact}
\end{figure}

Using these parameters, we calculate the rate of change of the eccentricity from
equation~(\ref{eq:dedt}). The result is presented in Figure ~\ref{fig:impact}. We find that each impact lasts for less than one time unit. The impacts induce
peak values of $de/dt$ that range between $4.0\times 10^{-3}\omgb$ and $0.013 \omgb$ at each
half binary period. Most of the time, $de/dt$ remains close to zero because stream-disk impacts
are so brief. Taking time averages of the curve, we obtain an estimate of the impact-induced rate
of change of eccentricity
$\lan de/dt\ran_{t}\simeq 1.5\times 10^{-3}\omgb$. The corresponding growth
rate $\gamma_{\rm e}\simeq 0.019\omgb$ is consistent with the growth rate at earlier times
in the simulation ($\simeq 0.018\omgb$).
The growth rate contribution predicted by the stream impact at this later time 
may not be quite the same as the impact contribution at the earlier stage.  Nonetheless,
the approximate correspondence is encouraging.  Indeed, one question that arises
is why, given the estimate just made, the growth rate is smaller at this later time.
We can only speculate that nonlinear effects, significant at this time but not earlier,
may limit eccentricity growth.

%

\subsection{Interpretation of the Lump\label{sec:discussion_lump}}
The disk after $t=400$ shows a significant asymmetry which we call the ``density lump". We
believe the lump is due to the combined effect of stream impact and disk
eccentricity.
As discussed in \S~\ref{sec:stream}, only one stream strongly interacts with
the binary at any given time once the disk is noticeably eccentric. The stream moves outward after
gaining angular momentum from the binary. Once it reaches the disk inner edge, the stream is shocked
and its fluid compressed.  Its mass is added to the nearby gas mass, increasing the local density.
This region of enhanced density then moves around the binary at about the local orbital
speed, decaying over roughly one orbital period due to pressure forces and shearing.  One
might not therefore expect such a small scale density enhancement to become a large scale,
long-lasting lump.

\begin{figure}
\epsscale{1.2}\plottwo{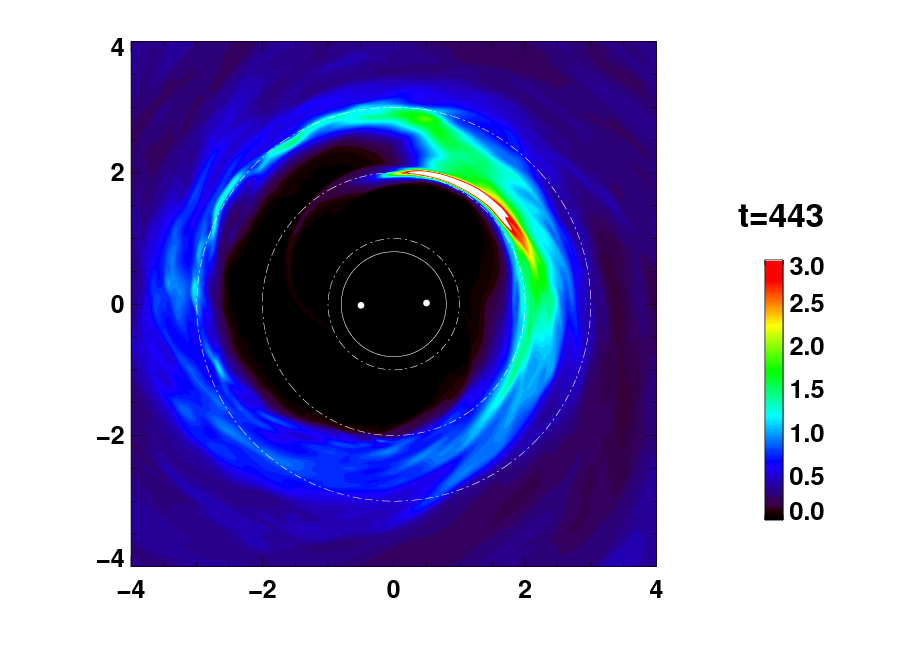}{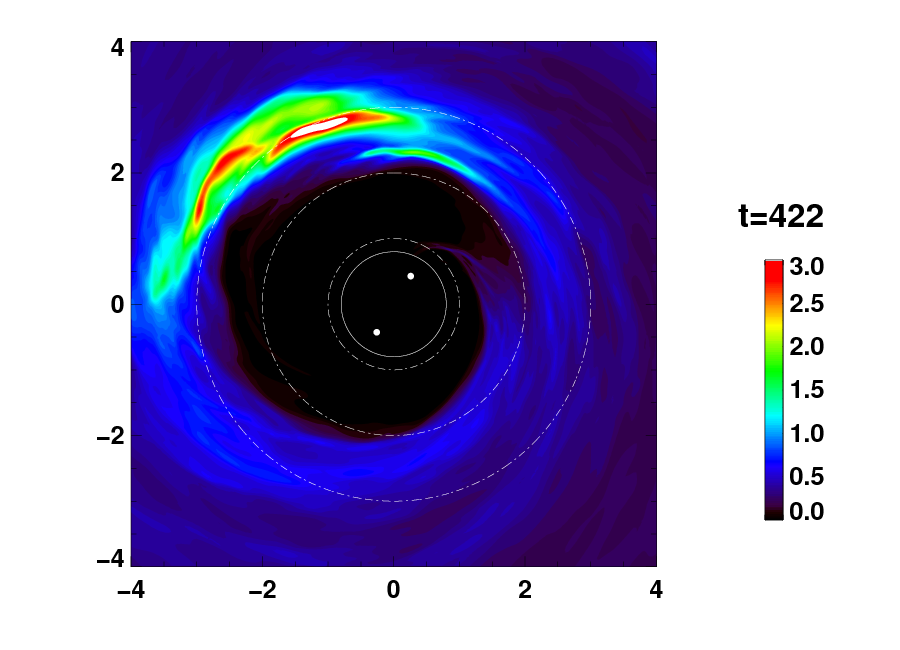}
\caption{\small{Snapshots of the lump concentration and feeding mechanisms at two specific times,
$t=443$ (left) and $t=422$ (right), in linear scale (white represents the peak density which is off
the scale).}}\label{fig:lump}
\end{figure}

We have, however, identified two mechanisms that sustain the lump and foster its growth. 
First, the lump can grow more concentrated via streams coming from the lump itself. We name this
mechanism \textsl{stream reabsorption}.  When the lump approaches pericenter, the binary peels
off a gas stream that is denser than streams drawn from lower density regions. This relatively
dense stream is then kicked back out by the binary torque.  Although the azimuthal location
reached by the returning stream may not be exactly where its starting point has arrived at
this time, given the relatively large azimuthal extent of the lump, the chances that the
returning stream strikes somewhere in the lump are relatively good.  Moreover, the forward shock
driven by the returning stream into the lump gas compresses it, restoring the density loss
incurred by shearing during the time the stream passed near the binary.
The left panel in Figure~\ref{fig:lump} shows the moment an episode of stream reabsorption taking place.

The second mechanism makes use of streams drawn from regions of the inner disk other than the
lump.  We call it \textsl{lump feeding}, as this channel of development not only enhances the
concentration of the lump, but also increases its mass. After $t=400$, the density enhancement orbits
eccentrically, and as it follows its orbit toward apocenter, both $v_{r}$ and $v_{\phi}$ diminish.
Meanwhile, streams traversing the gap continue to strike the disk inner edge, creating a new,
smaller density enhancement at orbital phase angles behind the lump. The orbital speed of this
new density concentration is much greater than that of the main lump because it is at rather
smaller radius.  As a result, the newly-created and smaller lump can catch up with the main
lump and join it before the lump returns from apocenter and accelerates.
In the right panel of Figure~\ref{fig:lump}, we show an example of this process, pictured at
a time ($t=422$) shortly before an outgoing stream reaches the lump.   Three time units later,
the small enhancement (the green region close to $r=2a$ at $\phi\lesssim \pi/2$) joins the
main lump (green and red colored area in the second quadrant).
The density-weighted velocity in the enhancement is $(v_r,v_{\phi}) \sim (0.17, 0.66)\omgb a$
(we average the stream gas where the surface density is greater than
$1.5\Sigma_0$). However, the lump has a slower velocity $\sim (0.07, 0.52)\omgb a$ (averaging
over locations where $\Sigma > 2\Sigma_0$). 
As a result of both the lump feeding and stream reabsorption, the density lump grows both in mass
and density contrast.

Another way to illustrate the connection between streams and the lump is to look at the space-time
diagram in Figure~\ref{fig:m1_diagram}.
In the diagram, the stream clearly stands out as the radially stretched features between $r\lesssim 1.5a$
and $r\gtrsim 2a$ which propagate away from the binary with a pattern speed of about $0.2 \omgb a$. 
This pattern repeats itself at a rate about twice the binary frequency. Meanwhile, the
zigzag oscillation at late time ($t>400$) between radius $\sim 2a$ and $4a$ shows the movement of the
density lump as it orbits around the binary on an eccentric orbit with a period of $\sim 30$ time
units. Clearly, when the lump passes the pericenter, the stream reabsorption process
promotes growth in the density contrast, driving an increase of the mode strength at
$r=2$--$2.5a$ in the ascending part of the zig-zag. The ascending part at even larger radius is
able to maintain its strength via lump feeding.  However, once the lump passes apocenter, the lump
feeding is limited as the lump is now much further away from the stream in azimuthal angle, and thus we find the
descending part at $r\gtrsim2.5a$ is less affected by the stream features. As the lump approaches
pericenter again, another cycle of stream reabsoprtion and lump feeding begins.
Note that the blue dots (low mode strength regions) around $2.5a$ between the zig-zags actually show
the moments that the newly kicked out stream reaches the radial coordinate (but is distant in
azimuth) of the lump, so that it smooths the $m=1$ mode at that radius. 
The zig-zag feature's growth in both strength and radial range demonstrates that the density of
the lump increases gradually, and as the disk becomes more eccentric the semi-major axis of the
lump's orbit grows.

\section{CONCLUSIONS \label{sec:conclusion}}

\subsection{Specific Results}

In this work, we have performed the first three-dimensional MHD simulation of a circumbinary disk
around an equal-mass binary, which in this case follows a circular orbit.  The main results are:
\begin{enumerate}
\item The disk exhibits a number of nearly steady features. There is 
a low-density gap within $r\leq 2a$ , an eccentric inner disk at $(2$--$3)~a$.
At early times, there is a pair of gas streams that flow into the gap from disk inner edge. They are
nearly steady in the corotating frame of the binary. At later times, there is a single dominant
stream whose mass flux is time varying with a period of half the binary orbital period.
Parts of these streams are torqued so strongly by the binary that they return and impact on the disk inner edge. 

\item Some aspects of the disk evolve secularly. At late time,
the disk inner edge develops an asymmetric density concentration (`the lump') whose mass, density
contrast, and orbital eccentricity grow steadily. We find the lump is due to a combination of
stream impact and disk eccentricity. 

\item The disk eccentricity grows exponentially during the time $t=100$--$250$ (16--40 binary
orbits) with a growth rate of $\sim 0.018\omgb$. The growth rate then slows down significantly.
By the end of the simulation, the disk body at $2a\lesssim r \lesssim 3a$ has reached an
eccentricity $\sim 0.1$.  Its pericenter precesses slowly due to the quadrupolar component
of the binary potential. Stream impact largely accounts for the eccentricity growth.

\item Reynolds stress associated with the streams is the leading transport mechanism in the gap
region, while Maxwell stress dominates in the disk body. 

\item Although the profile of the binary torque is similar to that of previous hydrodynamic
simulations, its magnitude depends strongly on the magnitude of internal disk stresses.
Normalized to the disk mass near the density peak, we find a measured total torque $T(\infty)$
$\simeq 14$ times greater than found by MM08.  Because the torque at any particular radius is
proportional to the gas mass there, the integrated torque is directly proportional
to the gas mass in the gap region, where the binary torques are strongest.  This mass, always
a small fraction of the disk mass, reaches the gap through the action of internal stresses
in the accretion flow.  Relative to the gas pressure, MHD stresses in the disk body
are about an order of magnitude larger than the ``viscous" stresses estimated
phenomenologically in previous work; in the gap, this ratio increases by another
order of magnitude.  As a result, the density of matter in the gap is also
about an order of magnitude larger than found in hydrodynamic calculations, and this
contrast leads directly to the larger torque.

\item After time averaging, the accretion rate at the inner boundary is $\sim 30\%$ of the
peak rate in the disk body. Compared with MM08 and normalized to the peak surface density, the
accretion rate at the inner boundary in this simulation is $\sim 40$ times greater.  This
contrast, like the contrast in the total torque, can be largely attributed to the stronger
internal stresses due to MHD effects.  The time-dependent accretion rate is strongly
modulated on both the binary orbital frequency (by the stream) and the inner-disk orbital
frequency (by the lump).

\item %
Previous work on the angular momentum budget of the binary has focussed on the torque it
exerts on the disk.   Because we find a substantially larger accretion rate than previous
calculations, we also find that the binary's gain in angular momentum due to accretion
can substantially offset its loss by torque.  As a result, the estimated binary contraction
rate $\dot{a}/a \sim -8\times 10^{-4}(M_{\rm d}/M)(GM)^{1/2}a^{-3/2}$ is only slightly larger
than the rate estimated by MM08.  

\end{enumerate}

\subsection{Consequences for Orbital Evolution: $M_d < M$ vs. $M_d > M$}

These detailed results have a number of more general implications for the evolution
of circumbinary disks, particularly in the context of supermassive black hole binaries.
It is generally believed that stellar dynamical effects become relatively ineffective at
driving the evolution of this sort of binary when its separation falls much below
$\sim 1$~pc \citep{bbr80}.
Recently, there have been several attempts to solve this `final--parsec problem' with stellar
dynamics.  Although they are potential solutions to this problem, they require either special
non-axisymmetric stellar distributions \citep[][]{berczik06,khan11,preto11} or extra perturbers
such as giant molecular clouds \citep[][]{perets08}.
Given the uncertainty about whether these candidate mechanisms suffice to solve the problem,
the prospect that angular momentum loss
to a surrounding disk may push the binary through this barrier is an attractive
one \citep[e.g.,][]{ivan99,gr2000,an2002,escala05,kl08,sk08,cuadra09,dotti09,chm10,tkz11,fls11}.

We have previously described this process in terms of the orbital shrinkage rate $\dot a/a$.
Defining the orbital shrinkage time $t_{\rm shrink}\equiv |a/\dot{a}|$, the time taken by
torque alone to change the binary orbital angular momentum
$t_{\rm torque} \equiv Mj_{\rm bin}/T(\infty)$, and the disk accretion time
$t_{\rm acc} \equiv \md/\dot M$, we can rewrite this rate as
\begin{equation}
\frac{t_{\rm acc}}{t_{\rm shrink}} = 2\left|-\frac{t_{\rm acc}}{t_{\rm torque}} + 
                \frac{\md}{M}\left(\frac{j_{\rm in}}{j_{\rm bin}} - \frac{3}{2}\right)\right|.
\end{equation}
%
Using the values found in the simulation makes $t_{\rm acc}/t_{\rm shrink} \simeq 0.6 \md/M$, i.e.,
the ratio of the accretion time to the orbital shrinkage time
is roughly the same as the ratio of the disk mass to the binary mass.  Indeed, \cite{lodato09}
argued that, unless the disk mass were at least comparable to the binary's secondary mass,
it would be ineffective in driving binary evolution.  Although it is true that a mass
comparable to the secondary's mass must {\it pass through} the inner region of the disk
over an orbital evolution time $t_{\rm shrink}$, it is not necessarily true that this
mass must be there all at once.

Suppose, for example, that $\md \ll M$.  Because $t_{\rm acc} \ll t_{\rm shrink}$ in
this case, if the disk mass is put in place once and for all, it would certainly be drained
long before the binary has significantly evolved.   On the other hand, one could also
imagine situations in which the disk is continuously replenished.  In that case, the
binary orbit could change substantially due to this interaction even though the instantaneous
disk mass is always much smaller than the binary mass; it's just that this process takes
longer.  Before determining how much time is required, it is
important to recall that the close subtraction between the torque term and the accretion
term may make the numerical value of $(t_{\rm acc}/t_{\rm shrink})/(\md/M)$ rather sensitive
to specific assumptions of our simulation, e.g., the binary mass ratio and the disk gas's
equation of state.  If the mass accretion rate were reduced by a factor of two
while the torque was held fixed, the shrinkage rate would increase by a factor
$\simeq 4$; conversely, if it were increased relative to a constant torque
by a ratio $\gtrsim 1.2$, the binary orbit would actually expand over time.

These processes could also be affected by the fact that the accretion rate onto the binary
is a substantial fraction of the accretion rate in the outer disk.  The specific
number we found ($\simeq 30\%$) is not terribly well-defined because the lack of inflow
equilibrium in the outer disk makes the denominator in that ratio very uncertain.  However,
a reinterpretation of that figure as, more vaguely, tens of percent, nonetheless has
significant implications.   If the accretion rate in the outer disk were sufficient to
supply an AGN, even the fraction leaking through the disk's inner edge would still
be large enough to fuel AGN activity, even if somewhat weaker.  As a result, the
inner edge of the disk would be illuminated in much the same way AGN are known to
illuminate the inner edge of their ``obscuring tori" \citep{ski93}.  Low density gas
in the gap region would then be strongly heated by absorption of the AGN continuum,
and much of it driven outward in a wind \citep{kb86,bk93}.  Such a wind would lead
to a reduction of the rate at which mass is captured by the members of the binary.  As a result,
the AGN luminosity would be smaller, and an equilibrium in which only a fraction of
the accretion rate through the disk inner edge is captured by one or the other of
the black holes might be achieved.  An immediate consequence would be an increase
in the ratio of torque to angular momentum captured by the binary, and therefore
a factor of several increase in the shrinkage rate.  Of course, a qualitative change
in the equation of state of gas in the gap might well lead to an order unity change
in the torque, as well.  At the same time, radiation forces associated with
illumination of the disk body can thicken it \citep[][]{pk92,krolik07,shi08}.
It is unclear how such a situation (effective vertical gravity weaker than $z\Omega^2$)
would alter saturation of the magneto-rotational instability, but more rapid
inflow is one plausible consequence.

The opposite case is also worth considering, in which $\md \gg M$, so that
$t_{\rm acc} \gg t_{\rm shrink}$.  If this is so, then even a one-time deposit
of mass in the disk might drive strong evolution in the binary orbit.  However,
when the binary separation decreases, the position of the strong torques moves
to smaller distance from the center of mass as well, raising the question of
how much mass might be there to receive those torques.   
We can estimate that time within the disk body of our simulation via 
$t_{\rm inflow}(r)\equiv \int^r dr'/\lan v_r\ran_{\rho,t}$, where $\lan v_r\ran_{\rho,t}$ 
represents the time- and density-weighted shell-averaged inflow velocity. 
In a steady-state disk around a point-mass, $t_{\rm acc}$ would match $t_{\rm inflow}$ at 
$r = \rp$. However, to the degree that binary torques retard accretion,  $t_{\rm acc}$
exceeds the inflow time measured at the surface density peak, which is 
$t_{\rm inflow}(\rp)\sim 5\times 10^2 \omgb^{-1}$. 
At sufficiently
large radius, $t_{\rm inflow}(r)$ will nearly always be greater than $t_{\rm acc}$ because
$t_{\rm inflow}(r) \propto r^{3/2} (r/H)^2$; only a rapid outward flare in the disk
thickness could alter this conclusion.  It has long been thought that if
$t_{\rm inflow}(r_p) \sim t_{\rm acc} \gg t_{\rm shrink}$, the disk and the binary
would decouple once the binary shrinks by a factor of a few because too little gas
would be found close enough to the binary to feel the torques; if orbital compression
depends solely upon interaction with a circumbinary disk, the shrinkage time could therefore
never be much shorter than
$t_{\rm inflow}(r_p)$ \citep[e.g.,][]{gr2000,an2002}.

However, the dynamical behavior seen both in our simulation and in others hints that
this assumption might not be valid.  The streams occurring in our simulation travel
inward on a timescale $\sim \omgb^{-1}$, which is $\ll t_{\rm inflow}$.  Although a
smaller binary separation would make the trajectories of the streams at radii not
far inside the disk edge more nearly angular momentum-conserving, and therefore make
it more difficult for them to penetrate to much smaller radii, the internal magnetic
stresses within the streams might be strong enough to replace the binary torques.
Indeed, we found that the nominal ``$\alpha$" in the gap region in our simulation
was one or two orders of magnitude larger than in the disk body.  There is also
another mechanism, commonly seen in global MHD simulations of disks around point-masses,
that might contribute in this way.  As emphasized by \cite{ghk11}, it is easy
for disks far from inflow equilibrium to drive mass-inflow rates much greater than
those expected on the basis of inflow equilibrium estimates.  For example, when
an annular disk with a seed poloidal magnetic field is created, orbital shear rapidly
creates toroidal field from radial field.  The $-B_r B_\phi$ stress acting on the
low density matter near the disk's inner edge rapidly removes its angular momentum,
propelling it inward.  As the binary separation becomes small compared to $r_p$, the
gravitational potential comes to resemble a point-mass, and similar dynamics might
be expected.  Because the exertion of a sizable torque requires only a small fraction
of the disk mass to orbit in the gap region, where the torque per unit mass is greatest,
a strong torque might continue even though the bulk of the disk mass remains far from the binary.

A massive disk is also likely to be subject to self-gravity, a mechanism not treated
in our simulation.  If $\md/M \gtrsim (c_s/v_{\rm orb})$,
self-gravitational instability can be triggered \citep[][]{gl65}.  The
evolution of this instability depends on the ratio of the cooling time to the dynamical
timescale. If the ratio is greater than unity, the disk can maintain marginal stability,
and the self-gravity would add extra stress to facilitate the accretion
\citep[][]{gammie2001,lr04,lr05}. If the cooling time is short, which is likely to be the
case for disks around massive black holes, the disk is likely to fragment and much of
its mass may be transformed into stars \citep[][]{sb89,gammie2001,naya06}, reducing
the gas surface density \citep[][]{lodato09}.  The consequences for interaction with
the binary remain somewhat unclear.  In sufficiently massive disks, stars might form
with orbits taking them sufficiently close to the binary that the summed torques from
many individual star-binary encounters could continue to shrink the binary; in effect,
the stellar loss-cone is repopulated by local star formation
\citep[][]{cuadra09}.  In less massive disks, stars
might form only in the disk body.  If most of the disk mass is converted into stars,
the gas mass available to fill the gap would be reduced, possibly leading
to both a smaller torque and a smaller accretion rate.  On the other hand, gravitational
torques due to non-axisymmetric fluctuations in the stellar density might create
stresses strong enough to overcome the reduction in gaseous disk mass.
\citep[][]{bt08}.  Given all these complicated possibilities, it is clear that further
study of self-gravitating circumbinary disks will be necessary before any strong
conclusions can be drawn on the fate of massive disks.

\subsection{Heating From Stream Impact\label{sec:heating}}

The work done by the binary on the streams can heat the inner edge of the disk through
stream impact.  The total energy per unit time delivered by the binary torque can be estimated as
\beq
L_{\rm torq}  \equiv \omgb T(\infty)\simeq 6.5\times 10^{-4} \omgb^3 a^2 \md.
\label{eq:ltorq}
\enq
For disks around massive binary black holes,
\begin{equation}
L_{\rm torq} \sim 1.9 \times 10^{38} M_6^{3/2} a_{0.1}^{-1/2} N_{24} \hbox{~ergs~s$^{-1}$},
\end{equation}
where $M_6 = M/10^6 $M$_{\odot}$ is the binary mass, $a_{0.1}= a/0.1$ pc the separation, and
$N_{24} = N_{\rm H}/1.7\times 10^{24}\rm{cm}^{-2}$ is the column density of the disk.
Thus, the expected luminosity is well below a typical AGN luminosity, and would therefore
be difficult to detect unless its power were emitted in a small number of lines. 
Similarly, we can get the heating rate for disks around stellar binaries: 
\beq
L_{\rm torq} \sim 7.9 \times 10^{33} M_0^{3/2} a_{10}^{-5/2} M_{\rm d,-1} ~\rm{ergs}~\rm{s}^{-1},
\label{eq:ltorq_stellar}
\enq
where $M_0 = M/$M$_{\odot}$ is the mass of the binary in solar masses, $a_{10}= a/10$ AU is the binary
separation in units of $10$~AU, and $M_{\rm d,-1} = \md/0.1$~M$_{\odot}$ is the disk mass in
units of $0.1$~M$_{\odot}$. We note that, with our ideal MHD model, this luminosity only sets the
upper limit of the possible heating rate due to the binary torque. In reality, disks outside 
$10$~AU might possess low-ionization dead zones \citep[][]{gammie96} which suppress the MHD turbulence 
and diminish the mean effective $\alpha_{\rm SS}$. Based on the accretion rate observed in disks
around young stars, which is typically $\sim
10^{-8}$~M$_{\odot}$~yr$^{-1}$\citep[e.g.,][]{heg95,g98} for T Tauri stars, we expect the actual
luminosity will be two to three orders of magnitude smaller than estimated in
equation~(\ref{eq:ltorq_stellar}).
Because stream impact is periodic with the period of the binary, the resulting luminosity
might be modulated with a period $\simeq 16 M_0^{-1/2} a_{10}^{3/2}$ yrs.

\subsection{Implications for the  ``Last Parsec Problem"}

In the black hole binary context, circumbinary disk studies have been in part motivated by a search
for mechanisms to solve the ``last parsec problem", the expected slow-down in orbital evolution
by stellar encounters when the binary separation becomes $\lesssim 1$~pc.  Our results have, in
one sense, quantitatively weakened the constraints on this proposed solution: per unit disk mass,
our nominal result is an orbital shrinkage rate a few times faster than previously thought.  However,
in a number of other ways, we have raised potential complications.  One is that the nominal
shrinkage rate is very sensitive to parameters because the larger accretion rate per unit disk
mass we find leads to a near cancellation between the binary's loss of angular momentum through
torques on the circumbinary disk and acquisition of angular momentum through accretion.  Another
stems from the likely back-reaction on both the disk proper and the matter in the gap caused by AGN
illumination as a result of accretion onto the binary.  Still another is the question of how
the asymmetry in the disk (the ``lump") affects dynamics driven by disk self-gravity when the
disk is sufficiently massive.   For all these reasons, our work has enriched, rather than settled,
the question of whether, or to what degree, circumbinary disks alleviate the ``last parsec problem".

\subsection{Future Directions}

Finally, we acknowledge that, as the first MHD simulation of a circumbinary disk with
order-unity mass ratio, this work is limited in several aspects.

Most of all, it would be
interesting to know the long-term behavior of the lump.  The reason why we ended the
simulation when we did was that matter accumulates more rapidly than magnetic field in
the lump.  Consequently, its internal Alfven speed diminished, and the ability of the
simulation code to resolve adequately the MHD turbulence along with it.  Such a situation
leads to an artificial weakening of the magnetic field.  To follow properly what actually
happens in the lump will require significantly better spatial resolution.

There are also numerous different parameters of the disks that need to be explored, such
as the mass ratio and the disk thickness. For example, a mass ratio different from unity
might result in noticeable changes in both the structure of the disk and the leakage rate
because the secondary would come closer to the disk inner edge.  A smaller thickness
would likely reduce the inflow rate in the disk body.  Such a change might diminish the
ratio of gas mass in the gap to disk mass, but the considerations discussed above about
MHD dynamics specific to the gap might counteract this effect.

We have also assumed that the disk orbits precisely in the binary orbital plane, and in a
prograde sense.  Inclined orbits might well occur, and their interaction with the binary's
quadrupole potential will lead to precession of the inner disk's orbital plane.   Coplanar,
but retrograde, gas orbits eliminate the resonance structure characterizing the linear
theory of binary--disk interaction and could lead to other qualitative effects \citep{nixon11}.

\acknowledgments
We thank an anonymous referee for many useful comments and suggestions. 
We also thank Milos Milosavljevic for useful email exchanges.
This research was supported by an allocation of advanced computing resources provided by the
National Science Foundation. The computations were performed on Kraken at the
National Institute for Computational Sciences (http://www.nics.tennessee.edu/).
Part of the work was also performed on the Johns Hopkins Homewood High-Performance
Computing Center cluster.
This work was partially supported by NSF grants AST-0908336 and AST-1028111.

\appendix
\section*{APPENDIX}

\section{Binary Eccentricity Evolution \label{app0}}
In this section, we estimate the growth rate of the binary's eccentricity based on a
simple model of accretion and energy exchange between the disk and the binary.
The orbital eccentricity $\ebin$ can be written in terms of the total orbital energy $E$, angular
momentum $J$, and total mass $M$:
\beq
\ebin = \sqrt{1+\frac{2EJ^2}{\mu G^2M^4}},
\label{eq:ebin}
\enq
where $\mu = m_1m_2/(m_1+m_2)$ is the binary's reduced mass, $m_1$ and $m_2$ are the masses of
the binary components, and $M=m_1+m_2$.  
Taking the time derivative of equation~(\ref{eq:ebin}), we find
$\dot{e}_{\rm bin}$:
\beq
\frac{\dot{e}_{\rm bin}}{\ebin} =
\frac{\ebin^2-1}{2\ebin^2}\left(\frac{\dot E}{E}+2\frac{\dot{J}}{J}-5\frac{\dot M}{M} \right).
\label{eq:edote}
\enq
Note that if the mass ratio $m_1/m_2 = 1$ (as in our simulation), $\dot{e}_{\rm bin}$ is independent
of any change in $m_1/m_2$.

In order to estimate the rate of change of the orbital energy $E$, we must understand how the binary
and circumbinary disk exchange energy. The binary can lose orbital energy by doing work on the
surrounding disk through its time-averaged torque.  In addition, the accretion flow penetrating
the gap ultimately becomes attached
to the binary, bringing its energy to the binary.  However, when the accreting matter begins
to orbit around an individual member of the binary, we must distinguish how much of its energy
becomes associated with that motion as opposed to the binary orbit.  In addition, there may also be
radiation losses.  Both of these effects could be computed if the binary members were on the
grid of the simulation, but they were not in the calculation we performed.  Consequently, the
best we can do here is to bound the range of possibilities.

For the greatest physical insight, it is convenient to group the contributions according to
whether they are associated with the binary-disk torque or with accretion,
i.e., writing the terms in the parentheses of equation~(\ref{eq:edote}) as $({\dot E}/E+2{\dot J}/J)_{\rm
torq} +({\dot E}/E+2{\dot J}/J)_{\rm acc} -5{\dot M}/M $, where the subscript `torq' denotes torque related
contributions, and `acc' the accretion related quantities. 

Consider the
binary-disk interaction first.  When the binary's orbit is exactly circular, the work done by
the disk on the binary can be written in terms of the azimuthal component of the binary's gravity
acting on the disk:
\beq
W= \int \Sigma(r,\phi)\left(\frac{\partial{\Phi}}{r\partial{\phi}}\right)\omgb rdr d\phi =
-T(\infty)\omgb.
\label{eq:work}
\enq
In our simulation, this contribution (in terms of the change in binary orbital energy)
is $\simeq -0.012GMa\Sigma_0\omgb$ (also see section~\ref{sec:heating}).  Because the
orbital energy of the binary is $-GM^2/(8a)$, the ratio $W/E = 0.096 (a^2\Sigma_0/M)\omgb$.
The rate at which the orbital angular momentum changes due to torque on the disk is
$-0.012 GMa\Sigma_0$; in ratio to the binary angular momentum, $(1/4)M\sqrt{GMa}$,
this contribution to $\dot J/J$ is $-0.048 (a^2\Sigma_0/M)\omgb$.  The combination
$(\dot E/E + 2\dot J/J)_{\rm torq}$ for these two pieces is therefore zero to within the accuracy
with which we can do the calculation.

To describe the energy and angular momentum brought to the binary orbit by accretion, we employ the
following toy model.  Around each of the binary member, the accreting gas would presumably form an
interior disk:
a disk extending out to a distance $r_{\rm d}$ from its central mass.  We imagine each binary
component together with its disk as a `hard sphere' of that radius orbiting around the
center of mass of the binary system.  When the stream of accreting matter reaches the sphere's surface,
it sticks to the sphere, adding its momentum to the sphere's.  It also delivers its potential
energy.  For simplicity, we will deal only in quantities averaged over times much longer
than the binary's orbital period, but shorter than the timescale of the binary's orbital evolution. 

In this model, the rate of change of kinetic energy in the binary orbit due to accretion is
\beq
\dot{E}_k = \dot{M} \vec{v}_{\rm s}\cdot \vec{v}_{\rm bin},
\label{eq:dotek}
\enq
where $E_k$ is the binary's kinetic energy, $\vec{v}_{\rm s}$ the velocity of the stream, and
$\vec{v}_{\rm bin}$ the velocity of the sphere, i.e., the orbital velocity of the individual
binary component. For an equal mass binary, the time averaged
$|\vec{v}_{\rm bin}|=(1/2)\sqrt{GM/a}$ with respect to the center of mass of the binary.
From energy conservation, we see that
$|\vec{v}_{\rm s}|=\sqrt{2\epsilon_{\rm in}^{\prime}+GM/a+ GM/r_{\rm d}}$, 
where $\epsilon_{\rm in}^{\prime}$ is the specific energy of the stream when it hits the `hard
sphere'.  This energy could be different from $\epsilon_{\rm in}$, the one measured when the
stream crosses the simulation boundary.  We approximate $r_{\rm d}=0.3a$ as
the radius of the tidally truncated interior disk for an equal-mass binary
\citep[][]{p77}.  Mass addition from the stream diminishes the binary's potential energy by
$\dot{E}_{p}=-\dot{M}(GM/2a)$.  This can be viewed alternatively as the potential energy of
the mass stream with respect to the {\it other} binary member or as
twice the potential energy per unit mass of the binary as a whole because there is a contribution
both from the arriving matter and from the deepening of the potential due to the increase
in total mass associated with accretion.  The potential energy of the stream with
respect to the binary member it is joining is associated entirely with the orbit of the stream
around that member and does not contribute to the orbital energy of the binary.
Summing the kinetic and potential energy brought to the binary by accretion and forming
its ratio to the orbital energy, we find
\beq
\left(\frac{\dot E}{E}\right)_{\rm acc} = 4\frac{\dot M}{M}\left[1 - 
         \cost \sqrt{\frac{2\epsilon_{\rm in}^{\prime}}{GM/a}+1+\frac{a}{r_{\rm d}}}\right],
\enq
where $\lan \cos\theta\ran$ is the time averaged angle between $\vec{v}_{\rm s}$ and
$\vec{v}_{\rm bin}$.  Assuming $\epsilon_{\rm in}^{\prime}\simeq \epsilon_{\rm in} = -0.81 GM/a$
(its value measured at the simulation inner boundary $r=r_{\rm in}$ by taking mass flux
weighted shell averages of the kinetic and potential energy), the accretion
contribution to the change in binary energy is
\beq
\left(\frac{\dot E}{E}\right)_{\rm acc} = 4\frac{\dot M}{M}\left(1 - 1.65\cost\right).
\enq
Angular momentum is brought to the binary at the fractional rate
\beq
\left(\frac{\dot J}{J}\right)_{\rm acc} = 3.76 \frac{\dot M}{M}
\enq
because the specific angular momentum of the binary is $(1/4)\sqrt{GMa}$ and the specific
angular momentum of the stream when it crosses the simulation boundary is 
$j_{\rm in} = 0.94\sqrt{GMa}$.

The total rate of change of eccentricity (see equation~(\ref{eq:edote}))is then the sum of
the accretion contributions due to energy, angular momentum, and mass acquisition
\beq
\frac{\dot{e}_{\rm bin}}{\ebin} = \frac{1-e^2}{2e^2}\left(6.59\cost - 6.52\right)\frac{\dot M}{M}.
\enq
We have previously found that $\dot M/M = 0.018 (a^2\Sigma_0/M)\omgb$, so that
\beq
\frac{\dot{e}_{\rm bin}}{\ebin} = \frac{1-e^2}{2e^2}\left(6.44\cost - 6.36\right)\times 10^{-3}
\left(\frac{\md}{M}\right)\omgb.
\enq
In other words, the eccentricity {\it decreases} unless $\cost$ is very nearly unity.
Because our simulation assumes a circular orbit, this result seems surprising because
there is no obvious reason why $\cost$ cannot be significantly smaller than one.
In fact, however, the orbital mechanics {\it require} $\cost \simeq 1$, as can be
seen from the following argument.

When measuring the specific orbital energy of the stream as it crosses the inner problem
boundary, we find its azimuthal velocity is $\simeq 1.21 \sqrt{GM/a}$, almost twice the
magnitude of its radial
velocity ($\simeq -0.66 \sqrt{GM/a}$). Thus, the stream's velocity is nearly in the
azimuthal direction. 
Moreover, this azimuthal speed is rather greater than the binary orbital
speed $0.5\sqrt{GM/a}$, and, as shown by the six panels of Figure~\ref{fig:stream}, most
matter crosses the inner boundary at an orbital phase slightly {\it ahead} of the nearest
member of the binary.  Therefore, when the stream crosses the boundary, it does so at a
small angle and then rapidly catches up with the other member of the binary, traveling
at almost constant radius.  In other words, when it strikes that disk, it does so with
$\cost \simeq 1$.

The net result is that the accretion contribution to $\dot{e}_{\rm bin}/\ebin$, as well as
the disk interaction contribution, almost exactly cancels.  To within the accuracy
of this simulation, these effects appear to have at most a weak effect on the
binary eccentricity.

There is, however, a resonant interaction between binaries and their surrounding
disks that has the potential to drive a linear instability in the binary eccentricity.
The resonance involves forcing by a binary potential component of the form
$\Phi_{m,l}(r,\phi, t)= \phi_{m,l}(r) \cos{(m \phi - \omgb t)}$.
It is strongest at the 1:3 resonance, where $m=2$ and $l=1$ \citep{arty91}.
Using the same formalism as previously, the rate of growth of eccentricity
can be written as
\begin{equation}
\frac{d \ln{e}}{dt}  = -\frac{\dot{E} - \omgb \dot{J} }{2 e^2 |E|}
\end{equation}
when $e \ll 1$.  Using $\dot{E} = \Omega_p \dot{J}$ with $\Omega_p = \omgb l/m= \omgb/2$
for the 1:3  resonance, it  then follows that
\begin{equation}
\frac{d \ln{e}}{dt}  = -\frac{\dot{J} }{2 e^2 \mu \omgb a^2},
\label{dedtr}
\end{equation}
where $\mu$ is the reduced mass of the binary.
To evaluate this eccentricity growth rate for the resonance,
we determine the torque $\dot{J}=-T_{2,1}$ as defined in equation~11 of \cite{al94}.
This torque is determined through the use of equations~(12)-(14) and (21) in that paper.
We obtain
\begin{equation}
\frac{d \ln{e}}{dt} = \frac{49 \pi^2}{16} \frac{m_1 m_2}{M^2} \frac{a^2 \Sigma}{M} \Omega_b,
\end{equation}
where surface density $\Sigma$ is evaluated near the resonance
at $r \simeq 3^{2/3} a$ and for an equal mass binary $m_1=m_2=M/2$. If the density varies near this radius, then it should be
suitably averaged over the resonance width of order $\sim (H/r)^{2/3} r$.
There is a caveat regarding this result, however: it is derived in the context
of linear theory and assumes an axisymmetric disk whose fluid follows circular
orbits.  It is uncertain how much the growth rate might change in a highly
non-axisymmetric, eccentric disk like the one found in our simulation.  Moreover,
once the binary eccentricity begins to grow, new resonances appear whose
effect tends to counteract eccentricity growth \citep{lubow92}.  Thus, it is hard
to evaluate the ultimate impact of this possible instability.

\section{Linear Equation for Eccentricity Distribution \label{app1}}

We apply the linear equation for eccentricity evolution given by GO06
that can be written as
\begin{align}
i \partial_r ( f_1(r) \partial_r E(r,t) ) + i f_2(r) E(r,t) +  J(r) s(r)  E(r,t) \nonumber & \\ = J(r) \partial_t E(r,t),
\label{Eeq}
\end{align}
where 
$E(r,t) = e(r,t) \exp{(i \varpi(r,t))}$ is the complex eccentricity, for real
eccentricity $e$ and periapse angle $\varpi$. 

$E$ is related to the linear perturbations from the axisymmetric circular
velocity. For the velocity expressed cylindrical coordinates as
$(u' (r,t) \exp{(-i \phi)}, v'(r,t)  \exp{(-i \phi)}),$ we have that
\begin{equation}
u'(r,t) = i r \Omega(r) E(r,t)
\label{u}
\end{equation}
and
\begin{equation}
v'(r,t) = \frac{1}{2} r \Omega(r) E(r,t),
\label{v}
\end{equation} 
where $\Omega(r)$ is the Keplerian orbital frequency 
about the binary of mass $M$ given by
\begin{equation}
\Omega(r) = \sqrt{\frac{G M}{r^3}}.
\end{equation}
Quantity  $J(r)$ is the disk angular momentum per unit radius divided by $\pi$ and is given by
\begin{equation}
J(r) = 2 r^3 \Omega(r) \Sigma(r).
\end{equation}
Functions $f_1(r)$ and $f_2(r)$ are given by
\begin{eqnarray}
f_1(r) &=& \gamma P(r) r^3, \\
f_2(r) &=&  \frac{d P}{dr} r^2 +  J(r)  \dot{\varpi}_{\rm g}(r),
\label{ab}
\end{eqnarray}
where
$P(r)$ is the two-dimensional (vertically integrated) disk pressure, $\Sigma(r)$ is the disk surface density.
Quantity $\dot{\varpi}_g$ is the gravitational precession rate 
of a free particle on an eccentric orbit about the equal mass binary, which is given by
\begin{equation}
\dot{\varpi}_g(r) = -\frac{1}{2 r^2 \Omega} \partial_r (r^2 \partial_r \Phi_0(r))
\label{prec_g}
\end{equation}
where
\begin{equation}
\Phi_0(r) = -\frac{G M}{\pi r} \, \chi  K(\chi)
\end{equation}
with
\begin{equation}
\chi=\frac{2 r a}{(0.5 a +r)^2}
\end{equation}
and $K$ is the complete elliptic integral of the first kind.
The terms involving quantities $f_1$ and $f_2$ describe the 
eccentricity propagation and precession, respectively.
Quantity $\gamma$ is the gas adiabatic index.
Real function $s(r)$ is the eccentricity injection rate due to some source
that is assumed to be distributed as a gaussian of width $w$ centered at radius
$r_{\rm c}$ and central value $ s_{\rm c}/(\sqrt{\pi} w)$
\begin{equation}
s(r)=\frac{  s_{\rm c}  \exp{(-(r-r_{\rm c})^2/w^2)}}{\sqrt{\pi} w}.
\label{sinj}
\end{equation}

Following along the lines G06 and L10, 
we adopt the inner boundary condition 
\begin{equation}
\partial_r E(r_{\rm i},t) = 0.
\label{iBC}
\end{equation}
For a Keplerian disk, the divergence of the velocity is proportional to $\partial_r E$.
Consequently, this boundary condition is equivalent to requiring that the
Lagrangian density perturbation near the disk inner edge vanishes.
The outer boundary condition is taken to be
\begin{equation}
E(r_{\rm o},t) = 0.
\label{oBC}
\end{equation}
To ensure that the trapping of the eccentricity is not
an artifact of the outer boundary condition, we test that the resulting
mode structure is independent of the outer radius $r_{\rm o}$.

To obtain modes, we write the eccentricity as
\begin{equation}
E(r,t) = E(r) \exp{(i \omega t)}
\label{omeq}
\end{equation}
where $\omega$ is a complex eigenfrequency.
The eccentricity distribution equation (\ref{Eeq}) has infinitely many modes
and eigenfrequencies.

\section{Eccentricity Distribution Far From Disk Inner Edge \label{app2}}

We determine here the analytic form of the eccentricity distribution
far from the inner edge of the circumbinary disk, based on the linear
theory of Appendix \ref{app1}.
In that limit, we assume that the local precession rate is small
compared with the eigenvalue $\omega$ defined in equation (\ref{omeq}). 
This assumption is expected to hold, since the gravitational precession
rate $\dot{\varpi}_{\rm g}$ declines rapidly with $r$ . We also expect the
pressure contribution to the precession rate to decline, since the pressure
is expected to decline with $r$.
We assume power law dependences for the gas sound speed and surface density
\begin{eqnarray}
c_{s}^2(r) &=& c_{s0}^2 \, x^{-b_1}, \\
\Sigma(r) &=& \Sigma_0 \, x^{-b_2},
\end{eqnarray}
where $x=r/a$ and $c_{s0}, \Sigma_0, b_1,$ and $b_2$ are constants.
Equation (\ref{Eeq}) then simplifies to 
\begin{align}
x^{-b_1+3/2} E''(x) + (-b_1-b_2 + 3) x^{-b_1+1/2}E'(x) = \tilde{\omega} E(x),
\label{Eeq_bigr}
\end{align}
where we ignored the eccentricity source term that is assumed to
be localized near the disk inner edge. Dimensionless
frequency $\tilde{\omega}$ is a constant given by
\begin{equation}
\tilde{\omega} = \frac{2}{\gamma} \left( \frac{ \omgb a}{c_{s0}} \right)^2 \frac{\omega}{\omgb}.
\end{equation}

For $x \gg 1$, equation (\ref{Eeq_bigr}) has an asymptotic solution in lowest
order of the form
\begin{equation}
E(x) \sim \frac{\exp{(-c_1 x^{c_2})}}{x^{c_3}},
\end{equation}
where
\begin{eqnarray}
c_1 &=& \frac{4 \sqrt{\tilde{\omega}}}{1+ 2 b_1},\\
c_2 &=& \frac{1}{4} (1 + 2 b_1), \\
c_3 &=& \frac{1}{8} (9- 2 b_1 - 4 b_2).
\end{eqnarray}
In this relation we ignore the overall scale factor for the eccentricity distribution, since 
we are interested in its function form. The appropriate root for $\sqrt{\tilde{\omega}}$
is taken such that $Re(\sqrt{\tilde{\omega}}) >0$.

For the case of the simulation, we have that
$\gamma=1, b_1=0$ (isothermal), 
$c_{s0} = 0.05 \omgb a$, and $\omega \simeq - 0.02 \omgb i $.
Quantity $b_2 = 1$ for an isothermal constant $\alpha$ decretion disk, although
$b_2$ is closer to -1 in the outer simulated region during much of the simulation. 
In any case, quantity $b_2$ has no influence on the dominant factor, the exponential.
We adopt $b_2 = 1$ and obtain
\begin{equation}
e(x) = |E(x)| \sim\frac{ \exp{(- 11.3 \, x^{1/4})}}{x^{5/8}}.
\label{Eiso}
\end{equation}

We approximate the
exponent of equation (\ref{Eiso}) by means of a Taylor series about $x = 3$ (typical of the simulated outer disk region) and obtain in leading order 
\begin{equation}
e(x)  \sim \frac{ \exp{(- 1.24 x)}}{x^{5/8}}
\end{equation}
and the length scale for the exponential decay is then $\sim 0.8 a$.
\clearpage

\clearpage

%

%


\begin{deluxetable}{cccc}
\tablecolumns{4} \tablewidth{0pc}

\tablecaption{Properties of Simulations of Circumbinary Accretion Disks}

\tablehead{\colhead{Label} & \colhead{Type of Simulation} & \colhead{Resolution\tablenotemark{a}
} & \colhead{Radial
Extent\tablenotemark{b}}}

\startdata
B3D         & MHD           & $400\times160\times512$ & $(0.8,16) $  \\
B2D.rin=0.8 & Hydrodynamics & $512\times1024$         & $(0.8,16) $  \\
B2D.rin=1.0 & Hydrodynamics & $472\times1024$         & $(1.01,16)$  \\
B2D.rin=1.1 & Hydrodynamics & $456\times1024$         & $(1.11,16)$  \\
B2D.rin=1.2 & Hydrodynamics & $440\times1024$         & $(1.22,16)$  \\
B2D.rin=1.3 & Hydrodynamics & $428\times1024$         & $(1.31,16)$  \\
B2D.rin=1.7 & Hydrodynamics & $380\times1024$         & $(1.73,16)$
\enddata 
\small{\tablenotetext{a}{Resolution is listed as $r\times\theta\times\phi$ in MHD simulation and
$r\times\phi$ in Hydrodynamic simulations.}}
\small{\tablenotetext{b}{The azimuthal extent is $(0,2\pi)$ for all simulations.}}
\label{tab1}
\end{deluxetable}

\clearpage

\end{document}